\documentclass[usenatbib]{mn2e}
\usepackage{bm}
\usepackage{morefloats}
\usepackage{hyperref}
\usepackage{tabularx}
\usepackage{graphicx}
\usepackage{multirow}
\usepackage{pdflscape}
\usepackage{amsmath}

\newcommand {\gaia}{\textit{Gaia }}
\newcommand {\hst}{\textit{HST }}

\title[Accurate distances to Galactic globular clusters]
{Accurate distances to Galactic globular clusters through a combination of \gaia EDR3, \hst and literature data}
\author[Baumgardt \& Vasiliev]{H. Baumgardt$^{1}$\thanks{E-mail:
h.baumgardt@uq.edu.au}, E. Vasiliev$^{2,3}$\\
$^{1}$ School of Mathematics and Physics, The University of Queensland, St. Lucia, QLD 4072, Australia \\
$^{2}$ Institute of Astronomy, University of Cambridge, Madingley Road, Cambridge CB3 0HA, UK\\
$^{3}$ Lebedev Physical Institute, Leninsky prospekt 53, 119991 Moscow, Russia\\
}

\begin{document}

\date{Accepted 2021 xx xx. Received 2021 xx xx; in original form 2021 xx xx}

\pagerange{\pageref{firstpage}--\pageref{lastpage}} \pubyear{201x}

\maketitle

\label{firstpage}

\begin{abstract}
We have derived accurate distances to Galactic globular clusters by combining data from the \gaia Early Data Release 3 (EDR3) with distances based on Hubble Space telescope (\textit{HST}) data and literature based distances. We determine distances either directly
from the \gaia EDR3 parallaxes, or kinematically by combining line-of-sight velocity dispersion profiles with \gaia EDR3 and \hst based proper motion velocity dispersion profiles.
We furthermore calculate cluster distances from fitting nearby subdwarfs, whose absolute luminosities we determine from their \gaia EDR3 parallaxes, to globular cluster main-sequences. We finally
use \hst based stellar number counts to determine distances. We find good agreement in the average distances derived from the different methods down to a level of about 2\%.
Combining all available data, we are able to derive distances to 162 Galactic globular clusters, with the distances to about 20 nearby globular clusters determined with an accuracy of 1\% or better. 
We finally discuss the implications of our distances for the value of the local Hubble constant. 
\end{abstract}

\begin{keywords}
globular clusters: general -- stars: distances 
\end{keywords}

\section{Introduction} \label{sec:intro}

Galactic globular clusters constitute an important rung on the extragalactic distance ladder. They are nearby enough to be resolved into individual stars, making them ideal
objects to calibrate the brightnesses of physically interesting stars like RR Lyrae or Type II Cepheids. In addition they are also massive enough to contain statistically significant
samples of these stars. Hence globular clusters are useful to determine the slopes and zero-points of RR Lyrae period-luminosity (P/L) relations \citep[e.g.][]{bonoetal2007,dambisetal2014}
as well as a possible metallicity dependence of the zero points \citep[e.g][]{sollimaetal2006}. Globular clusters can also be used as calibrators for other distance methods, like the tip of the red giant branch distance method (TRGB), which
allows to determine distances to external galaxies without having to rely on P/L relations of variable stars \citep[e.g][]{cernyetal2020,freedmanetal2020,soltisetal2021}.

%%Most interesting is their use for the calibration of the distances to the Large (LMC) and Small Magellanic clouds (SMC),
%%since these galaxies are important stepping stones for the calibration of the Cepheid period-luminosity (P/L) relationship \citep[e.g][]{pietrzynskietal2019,riessetal2019}. 
%%Due to their relative faintness compared to Cepheids, accurate direct parallax determinations have so far not been available for the large majority of Galactic RR Lyrae stars
%%(but see \citet{neeleyetal2019} for recent distance determinations to RR Lyrae stars based on \gaia data), hence globular clusters are

Distances to globular clusters are determined either through fits of their color-magnitude diagrams (CMDs) with theoretical isochrones \citep[e.g][]{ferraroetal1999,dotteretal2010,gontcharovetal2019,valcinetal2020},
or by using variable stars that follow known relations between their periods and absolute luminosities like RR Lyrae stars \citep[e.g][]{bonoetal2007,hernitscheketal2019}, Type II Cepheids \citep{matsunagaetal2006} or Mira type variables \citep{feastetal2002}.
In order to avoid circularity, the absolute luminosities of these stars need to be determined independently from globular cluster distances by using for example theoretical models \citep[e.g][]{catelanetal2004} or \textit{Hipparcos} or
\gaia parallaxes \citep{fernleyetal1998,neeleyetal2019,ripepietal2019}.
Accurate distances have also been obtained for globular clusters by using eclipsing binaries \citep{kaluznyetal2007,thompsonetal2020}, however the faintness and scarcity of suitable binaries 
means that observations have so far been limited to a few globular clusters. Finally it is possible to determine distances by comparing the magnitudes of main-sequence stars with stars of similar metallicity in the solar
neighborhood, the so-called subdwarf method \citep[e.g][]{reid1998,cohensarajedini2012} or kinematically by comparing line-of-sight and proper motion velocity dispersion profiles in globular clusters
\citep[e.g.][]{mcnamaraetal2004,vandevenetal2006,watkinsetal2015b}. The latter method has the advantage that the derived distance is not influenced by the reddening of the cluster.

Typical uncertainties in the zero points of the P/L relations of RR Lyrae stars are thought to be of order 0.05 mag \citep{bhardwaj2020} and other methods like CMD fitting have
similar uncertainties. In addition, the reddening of many clusters presents an additional challenge for the determination of accurate distances since it can be variable even across small,
arcminute size fields \citep{bonattoetal2013,pallancaetal2019}. Furthermore there is evidence for non-standard reddening laws in the directions of several globular clusters like for example M4 \citep{dixonlongmore1993,hendricksetal2012}. 

It is therefore important to use independent methods
to verify ine distances to globular clusters, especially methods which are not effected by the reddening of stars. In this paper we use data from the \gaia EDR3 catalogue \citep{gaiamissionmain2016,gaiaedr3main}
to determine distances to Galactic globular clusters. In addition to using the \gaia EDR3 parallaxes directly, we also determine moving group distances and kinematic distances derived by comparing proper motion velocity dispersion
profiles with line-of-sight velocity dispersion ones. We finally use the \gaia EDR3 parallaxes of nearby subdwarfs as well as
Hubble Space Telescope (HST) star counts together with kinematic information to determine distances.

Our paper is organised as follows: In sec. 2 we present our derivation of cluster distances using the methods mentioned above. In sec.~3 we describe our survey of literature
distances. We compare the distances that we derive from the different methods and derive the average distance to each globular cluster in sec. 4. We finally draw our 
conclusions in sec. 5.

\section{Analysis}
\label{sec:analysis}

\subsection{Cluster sample and selection of member stars}

We take our target list of globular clusters from \citet{baumgardtetal2019a}. To this sample we add six additional Milky Way globular clusters that have been found in recent years:
VVV-CL001 \citep{minnitietal2011,ftetal2021}, BH~140 and FSR~1758 \citep{cantatgaudinetal2018}, Sagittarius II \citep{laevensetal2015,mutlupakdiletal2018}, RLGC~1 and RLGC~2 \citep{ryulee2018}, and
Laevens~3 \citep{longeardetal2019}. The status of Sagittarius II is still debated, while \citet{mutlupakdiletal2018} argue for it to be a star cluster based on its location in a size vs.
luminosity plane, \citet{longeardetal2020} argue, 
based on a small spread in [Fe/H] that they find among the cluster stars, for it to be a dwarf galaxy. We tentatively include the object in our list of clusters. The other systems are most likely globular clusters based
on their color-magnitude diagrams, kinematics and location in the Milky Way. Together with the 158 GCs studied in \citet{baumgardtetal2019a}, we therefore have a sample of 162 globular clusters. 

\subsection{Cluster parallax distances}
\label{sec:gedr3par}

Determining distances to globular clusters via the parallaxes of individual stars has the advantage that the distances are determined directly, without having to rely on another distance method as 
an intermediate step in the distance ladder. In addition, parallaxes are not influenced by the reddening of the clusters. 
In this paper, we take the cluster parallaxes from \citet{vasilievbaumgardt2021} (hereafter VB21). VB21 determined mean parallaxes for 170 globular and outer halo star clusters by averaging 
the cluster parallaxes of individual member stars from the \gaia EDR3 catalogue. Cluster members were selected based on the \gaia EDR3 proper motions and parallaxes.
%%individual stellar line-of-sight velocities from \citet{baumgardthilker2018} (if available) and the location of stars in a color-magnitude diagram. 
VB21 also applied multi-gaussian mixture modeling to classify stars into cluster members and background stars and determined the mean cluster parameters and their errors over multiple realizations
of statistically created member catalogues. In order to account for possible magnitude, color
and position dependent biases in the \gaia catalogue, VB21 applied the parallax corrections of \citet{lindegrenetal2021} to the individual stellar parallaxes. However when
testing the derived cluster parallaxes against the literature distances that we derive further below, VB21 found that even after applying the Lindegren et al. corrections,
there is still a mean offset in the \gaia parallaxes. In particular, they found that the parallax corrections suggested by
\citet{lindegrenetal2021} might have been overcorrecting the \gaia EDR3 parallaxes by $\Delta\varpi \sim 0.007$~mas. VB21 also found evidence for spatially correlated small scale 
systematic errors of order $\epsilon_\varpi\sim 0.01$~mas. We therefore subtract $\Delta\varpi = 0.007$~mas from the derived mean parallaxes to correct for the parallax bias. The 
small-scale correlated errors were already taken into account by the parallax averaging procedure of VB21. We list the median distances $D=1/\varpi$ for clusters with
$\varpi/\sigma_\varpi>10$ in Table~\ref{disresult} to allow a comparison of the \gaia EDR3 parallax distances with the other distances that we derive.
When calculating mean cluster distances we use the data for all clusters and the parallax values directly as will be described further below.

\subsection{Kinematic distances}

For clusters with accurately measured line-of-sight and proper motion velocity dispersion profiles one can determine cluster distances kinematically by varying the
cluster distance until the velocity dispersions are the same or a best match to a theoretical model of the cluster is achieved \citep[e.g.][]{henaultbrunetetal2019}. The advantage of such kinematic distances is that, like parallaxes, 
the kinematic distance method is a direct method which does not rely on other, more nearby methods in the distance ladder. Similar to parallax distances, kinematic distances 
are also not affected by cluster reddening, allowing in principle to derive accurate distances for highly reddened clusters. In order to calculate kinematic distances, one either needs
to know the anisotropy profile of the internal velocity dispersion of a cluster, or assume that the cluster is isotropic. The latter is most often assumed since observed
velocity dispersion profiles usually show that globular clusters are isotropic, at least in their inner parts \citep{vanleeuwenetal2000,watkinsetal2015a,rasoetal2020,cohenetal2021}, 
$N$-body simulations of star clusters also show that most globular clusters should have isotropic velocity dispersion profiles \citep{baumgardtmakino2003,lutzgendorfetal2011,tiongcoetal2016}.
\begin{figure*}
\begin{center}
\includegraphics[width=0.63\textwidth]{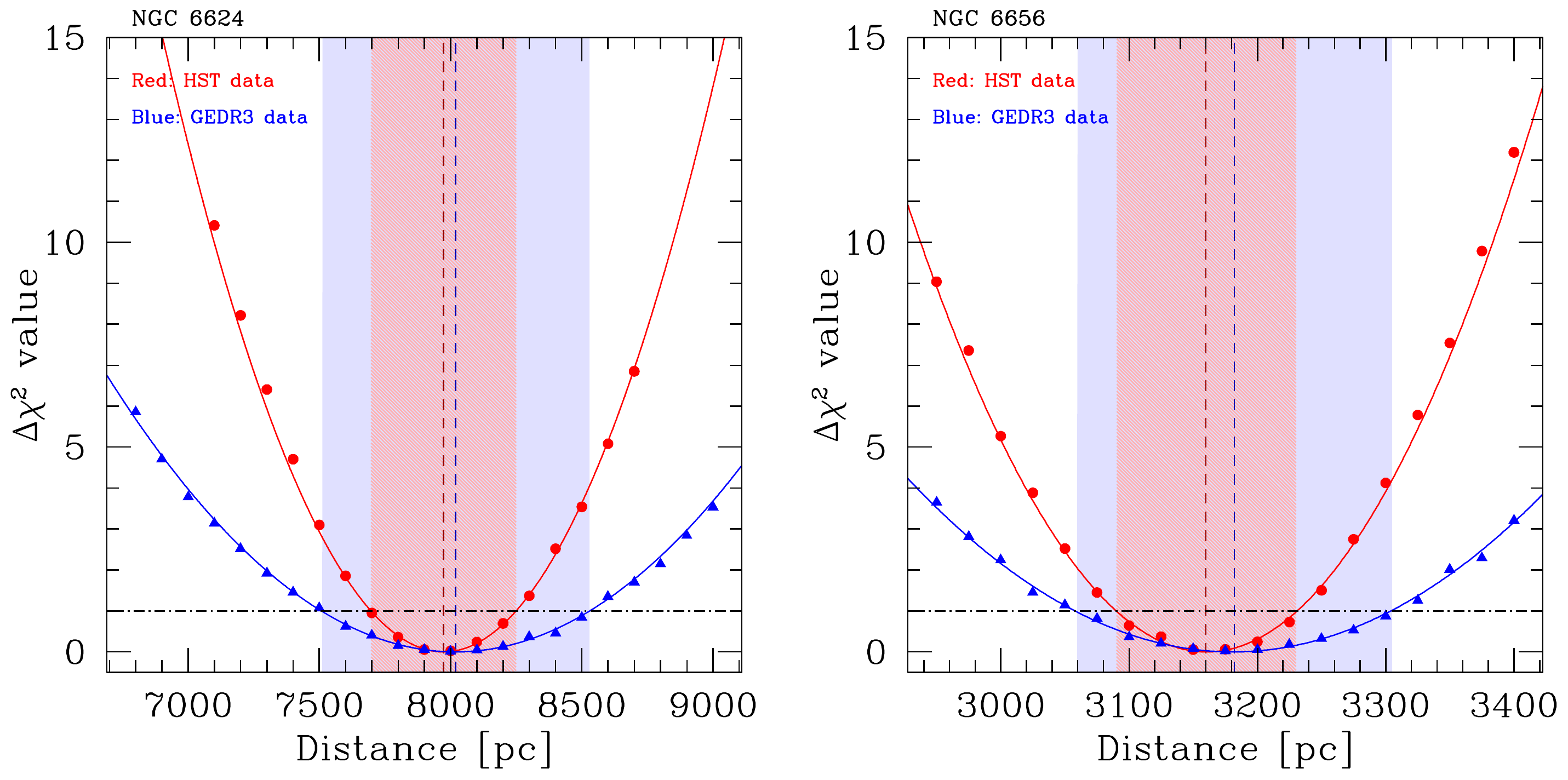}
\hfill
\includegraphics[width=0.30\textwidth]{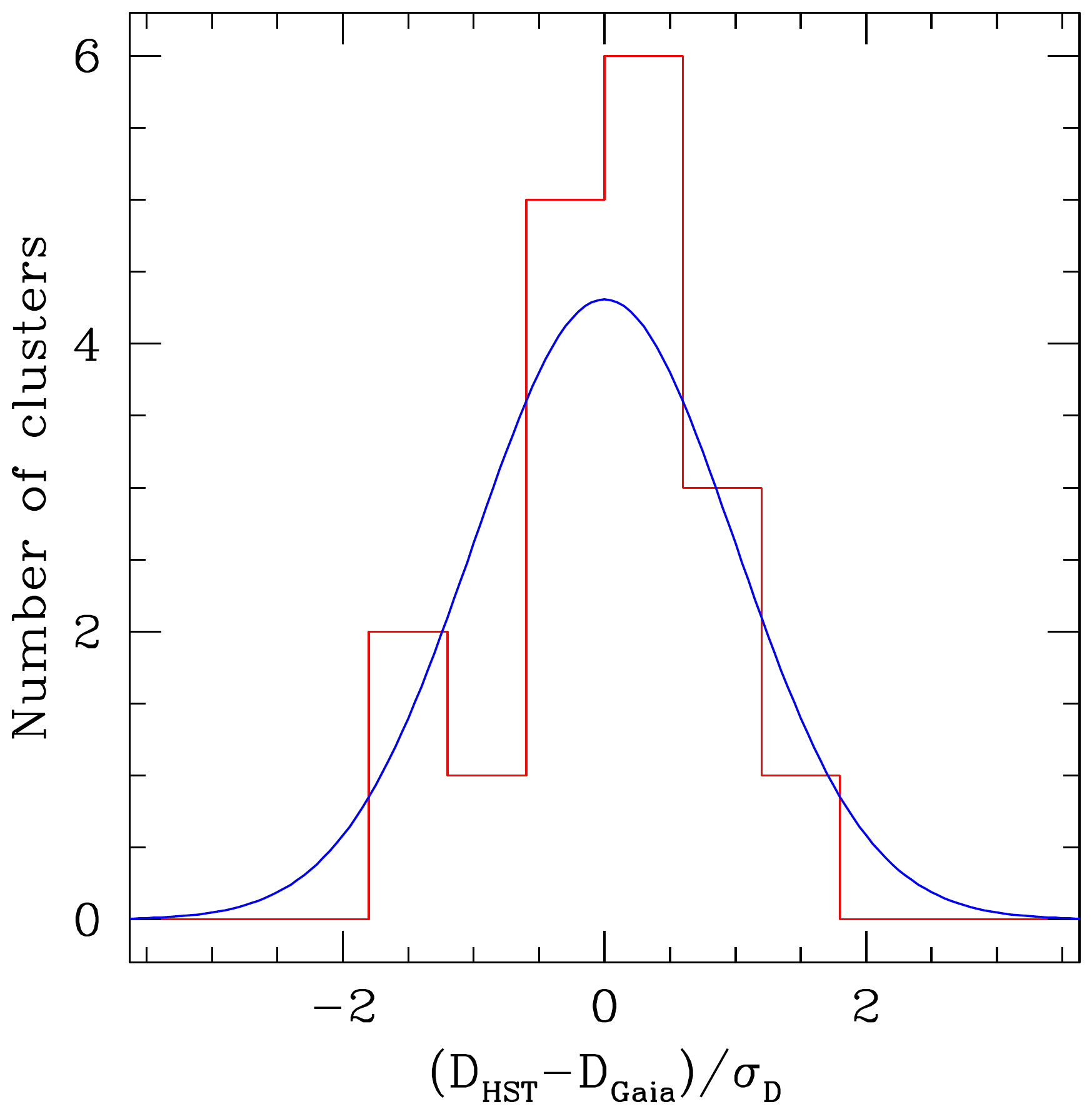}
\end{center}
\vspace*{-0.2cm}
\caption{Illustration of the kinematic distance determination for the globular clusters NGC~6624 (left panel) and NGC~6656 (middle panel). Shown are the $\chi^2$ values from the fit of our $N$-body models to the
velocity dispersion profiles as a function of distance for the \hst proper motions (red circles) and \gaia EDR3 proper motions (blue triangles). Solid lines show a quadratic fit to the $\chi^2$ values of each data 
set and the dashed lines and shaded areas mark the best-fitting distances and their 1$\sigma$ error. The literature distances are $D=8020$~pc (NGC~6624) and $D=3330$ pc (NGC~6656) and are in agreement with the 
kinematic ones. The right panel compare  the error weighted distance differences from \gaia and \hst data for clusters which have proper motion velocity dispersion profiles from both satellites. The resulting 
distribution is compatible with a normal distribution (shown by a blue solid line).
}
\label{kindis}
\end{figure*}

We take the line-of-sight velocity dispersion profiles for the kinematic distance fitting from \citet{baumgardthilker2018}, \citet{baumgardtetal2019b} and \citet{baumgardtetal2019a}. \citet{baumgardthilker2018} 
calculated line-of-sight velocity dispersion profiles by combining literature velocities with line-of-sight velocities derived from archival spectra from the ESO and Keck data archives. 
\citet{baumgardtetal2019b} and \citet{baumgardtetal2019a}  added line-of-sight velocities from \gaia DR2 and the Anglo-Australian Observatory (AAO) data archive to these profiles. In 
addition we use the line-of-sight
velocity dispersion profiles published by \citet{kamannetal2018} based on MUSE data. We take the proper motion dispersion
profiles either from VB21 based on \gaia EDR3 data, or from published profiles based on multi-epoch \hst measurements.  Table~\ref{tab:hstkin} lists
the \hst proper motion velocity dispersion profiles that we use in this paper together with the distances that we derive from these profiles. Unfortunately we could not use the recently 
published proper motion dispersion profiles for nine inner Milky Way globular clusters by \citet{cohenetal2021} since no accurate line-of-sight velocity dispersion profiles are available for these clusters.
\begin{table}
\caption{Kinematic distances based on \hst proper motion velocity dispersion profiles. The second last panel gives the kinematic distance
derived by the authors of the original \hst data set, the last column gives our distances for the same data set.}
\begin{tabular}{@{}l@{$\;\;$}l@{}c@{$\;\;$}c}
\hline
  \multirow{2}{*}{Cluster} & \multirow{2}{*}{HST Data set} & Lit. Dist. & Our Dist. \\
    & & [kpc] & [kpc] \\
\hline
NGC 104 & \citet{watkinsetal2015a}   & $\;\;4.45 \pm 0.50$ & $\;\;4.545 \pm  0.047$ \\
        & \citet{heyletal2017}       & $\;\;4.29 \pm 0.47$ & $\;\;4.347 \pm  0.236$ \\[+0.05cm]
NGC 288 & \citet{watkinsetal2015a}   & $\;\;9.98 \pm 0.37$ & $\;\;9.098 \pm  0.291$ \\[+0.05cm]
NGC 362 & \citet{watkinsetal2015a}   & $\;\;9.37 \pm 0.18$ & $\;\;9.202 \pm  0.280$ \\[+0.05cm]
NGC 1261 & \citet{rasoetal2020}      &     $-$             & $16.775 \pm  0.824$ \\[+0.05cm]
NGC 1851 & \citet{watkinsetal2015a}  & $11.41 \pm 0.20$    & $11.440 \pm  0.254$ \\[+0.05cm]
NGC 2808 & \citet{watkinsetal2015a}  & $10.18 \pm 0.12$    & $\;\;9.837 \pm  0.122$ \\[+0.05cm]
NGC 5139 & \citet{watkinsetal2015a}  & $\;\;5.22 \pm 0.05$ & $\;\;5.264 \pm  0.082$ \\[+0.05cm]
NGC 5904 & \citet{watkinsetal2015a}  & $\;\;8.77 \pm 0.15$ & $\;\;7.456 \pm  0.146$ \\[+0.05cm]
NGC 6266 & \citet{watkinsetal2015a}  & $\;\;5.67 \pm 0.07$ & $\;\;6.354 \pm  0.128$ \\
         & \citet{mcnamaraetal2012}  &     $-$             & $\;\;6.945 \pm  0.264$ \\[+0.05cm]
NGC 6341 & \citet{watkinsetal2015a}  & $\;\;8.43 \pm 0.34$ & $\;\;8.231 \pm  0.347$ \\[+0.05cm]
NGC 6352 & \citet{libralatoetal2019} &     $-$             & $\;\;6.260 \pm  0.762$\\[+0.05cm]
NGC 6362 & \citet{watkinsetal2015a}  & $\;\;7.34 \pm 0.24$ & $\;\;7.720 \pm  0.315$ \\[+0.05cm]
NGC 6388 & \citet{watkinsetal2015a}  & $10.44 \pm 0.12$    & $10.894 \pm  0.137$ \\[+0.05cm]
NGC 6397 & \citet{watkinsetal2015a}  & $\;\;2.54 \pm 0.05$ & $\;\;2.332 \pm  0.057$ \\
         & \citet{heyletal2012}      & $\;\;2.20 \pm 0.60$ & $\;\;2.416 \pm  0.062$\\[+0.05cm]
NGC 6441 & \citet{watkinsetal2015a}  & $11.77 \pm 0.20$    & $12.059 \pm  0.172$ \\
         & \citet{haeberleetal2021}  & $12.74 \pm 0.16$    & $12.364 \pm  0.226$ \\[+0.05cm]
NGC 6624 & \citet{watkinsetal2015a}  & $\;\;6.69 \pm 0.36$ & $\;\;7.972 \pm  0.277$\\[+0.05cm]
NGC 6656 & \citet{watkinsetal2015a}  & $\;\;3.18 \pm 0.07$ & $\;\;3.161 \pm  0.070$\\[+0.05cm]
NGC 6681 & \citet{watkinsetal2015a}  & $\;\;9.33 \pm 0.14$ & $\;\;9.260 \pm  0.165$ \\
NGC 6715 & \citet{watkinsetal2015a}  & $23.79 \pm 0.33$    & $25.019 \pm  0.646$\\[+0.05cm]
NGC 6752 & \citet{watkinsetal2015a}  & $\;\;4.28 \pm 0.03$ & $\;\;4.005 \pm  0.130$\\[+0.05cm]
NGC 7078 & \citet{watkinsetal2015a}  & $10.23 \pm 0.13$    & $10.375 \pm  0.237$\\[+0.05cm]
\hline
\end{tabular}
\label{tab:hstkin}
\end{table}

We determine a kinematic distance independently for each proper motion velocity dispersion profile that we have from either \gaia EDR3 or \hst data. The best-fitting distance is determined by
fitting each cluster with the grid of $N$-body models calculated by \citet{baumgardt2017} and 
\citet{baumgardthilker2018}. Their grid contains several thousand $N$-body models of star clusters, varying the size, initial density profile, metallicity  
and stellar mass function of each cluster. We interpolate within this grid and determine for each cluster the $N$-body model that simultaneously provides the best fit to the internal mass function,
velocity dispersion and surface density profile. More details about the $N$-body models can be found in \citet{baumgardt2017} and 
\citet{baumgardthilker2018} and we refer the reader to these papers for a full description of the fitting procedure. 

%%We varied the cluster distance until the we achieved the best fit to the combined line-of-sight velocity and proper motion velocity dispersion profiles.
We restrict the radial extent of the
line-of-sight velocity dispersion profiles to the same range for which we have proper motion dispersions in order to avoid a possible bias due to the $N$-body models not providing a
match to the full profile. In the $N$-body models we measure the line-of-sight velocity dispersion profile for stars brighter than the main-sequence turnoff. The magnitude limit 
for the proper motion velocity dispersion profiles was varied for each data set independently so we match the magnitude limit of the observed data. Hence, except for the nearest clusters like NGC~6121,
we chose a magnitude limit equal to the main-sequence turnoff for fits to the \gaia EDR3 proper motions, while we choose deeper limits for the \hst data. The left and middle panels of Fig.~\ref{kindis} 
show the resulting fits to two clusters that have both \gaia and \hst data
and Table~\ref{tab:hstkin} gives the derived distances from the \hst data together with the kinematic distances derived by the authors of the original papers. There is usually good agreement 
between the distances derived by us and the original kinematic distances. The only exceptions are NGC~5904 and NGC~6266. At least for NGC~5904, one the reason for different distances could be 
the lack of a radial overlap between the line-of-sight and proper motion velocity dispersion profiles used by \citet{watkinsetal2015a}, which makes the derived distance strongly dependent on the
dynamical model used for the cluster.

We give the full set of distances including the \gaia ED3 kinematic distances in the Appendix.
The right panel of Fig.~\ref{kindis} shows the error weighted distribution of differences in the kinematic distances for the \gaia EDR3 and \hst data. It can be seen that they roughly follow a normal distribution, 
indicating good agreement between the \hst and \gaia distances.

\subsection{Subdwarf distances}
\label{sec:sd}

The main idea of our subdwarf distances is to make use of the excellent accuracy of \gaia EDR3 parallaxes for nearby stars. As discussed in sec.~\ref{sec:gedr3par}, \gaia EDR3 parallaxes have small-scale, systematic 
uncertainties of $\sim 0.01$ mas that can't be removed through calibration of the parallaxes against objects with known parallaxes. This means that even the distances to the nearest globular clusters cannot be determined directly with
an accuracy better than a few percent. The accuracy of parallax distances also quickly deteriorates with increasing distance and drops below the accuracy of CMD fitting distances beyond distances of
about 5~kpc. In contrast, the distance to a nearby star at $d=100$~pc has a relative uncertainty of only 0.1\% using \gaia EDR3 parallaxes. 
\begin{figure*}
\begin{center}
\includegraphics[width=0.99\textwidth]{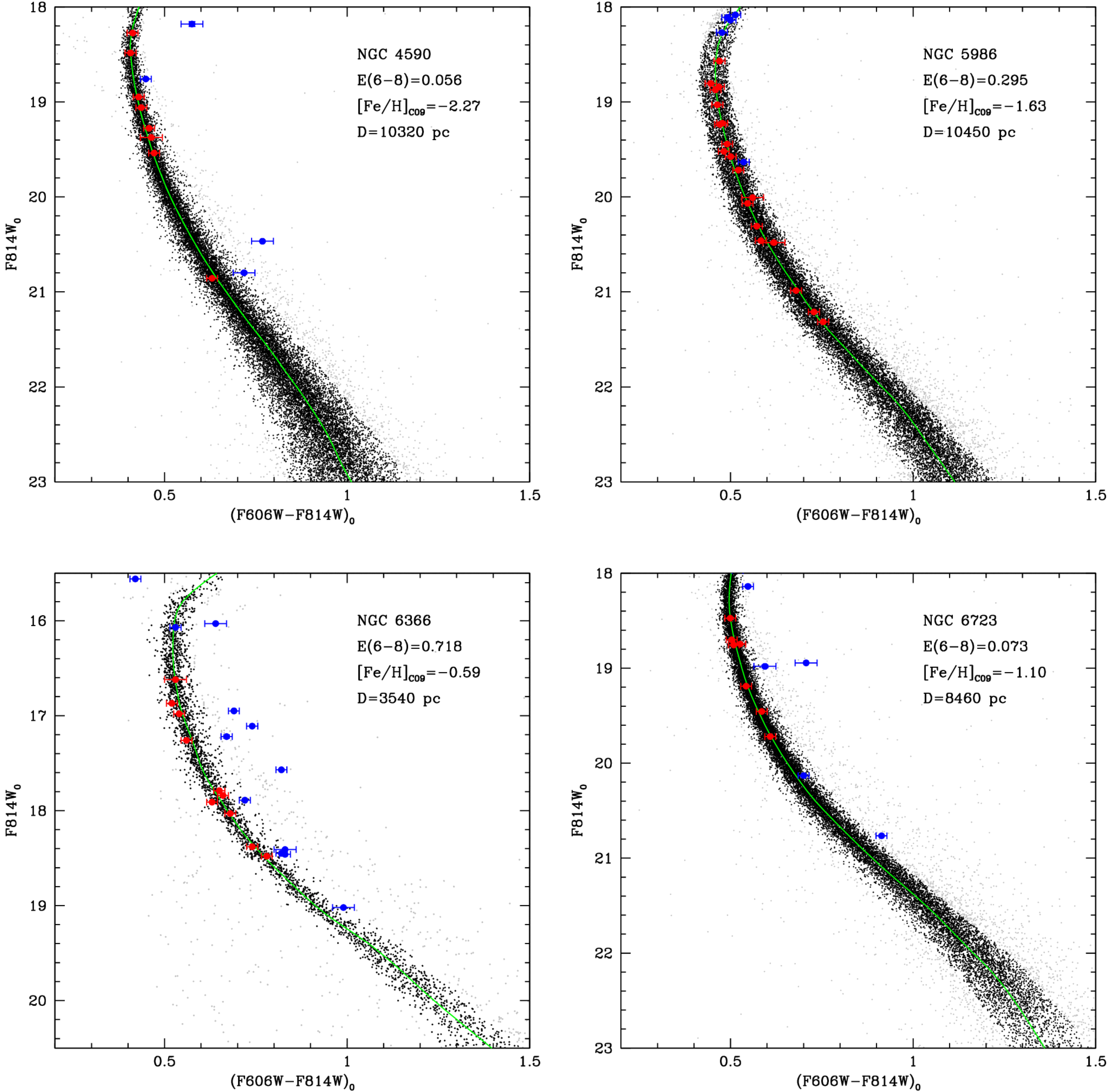}
\end{center}
\vspace*{-0.2cm}
\caption{Illustration of the subdwarf distance determination for four globular clusters with a range of metallicity and reddening values. Grey points mark stars considered to be background stars, black
points are cluster members. The green line shows our fitted ridge line for each cluster. Subdwarfs used in the distance fitting are shown by red circles, subdwarfs not used are shown by blue circles.
Photometric error bars are also shown. Note that the error in absolute magnitude due to the \gaia EDR3 parallax error is shown but usually too small to be seen.
}
\label{fig:subdwarf}
\end{figure*}

Our basic approach for deriving subdwarf distances is the same as the one used by \citet{cohensarajedini2012}, except that we replace the Hipparcos parallaxes by \gaia EDR3 ones. We start using the compilation of precise Johnson-Cousins UBVRI
photometry of nearby subdwarfs by \citet{casagrandeetal2010} and identify the counter-parts of these stars in the \gaia EDR3 catalogue. We then apply the \citet{lindegrenetal2021} parallax corrections to the stars as well as the
additional shift of  $\Delta\varpi = 0.007$~mas found by VB21.
Using only stars which have published $V$ and $I$ band photometry by \citet{casagrandeetal2010}, a reddening $E(B-V)<0.02$ and \gaia EDR3 parallaxes with a relative precision $\varpi/\epsilon_\varpi>20$
leaves us with a sample of 206 stars. Despite our more stringent limit on the parallax accuracy, our sample is almost a factor 10 larger than the one used by \citet{cohensarajedini2012}. Thanks to the superior accuracy 
of the \gaia EDR3 parallaxes, our sample also contains about 20 stars with metallicities $\mbox{[Fe/H]}<-2.0$, while no accurate Hipparcos parallaxes were available for any of these stars. We finally use the SIMBAD astronomical
data base to exclude known binaries from this sample, leaving us with a sample of 185 stars.

Our source for the globular cluster photometry is the compilation of \hst photometry published by \citet{sarajedinietal2007} based on observations made with the Advanced Camera for Surveys (ACS) onboard \textit{HST}.
Their survey contains deep photometry in the ACS/WFC F606W and F814W bands for 67 globular clusters. In addition to the photometry for 67 globular clusters from \citet{sarajedinietal2007}, we also use the ACS/WFC 
photometry published by \citet{dotteretal2011} for 6
globular clusters as well as the photometry for NGC~6528 from \citet{lagioiaetal2014} that was kindly provided to us by the authors. We finally use the photometry from \citet{baumgardtetal2019a} for an additional 6 globular clusters 
(NGC 6325, NGC 6342, NGC 6355, NGC 6380, NGC 6401 and NGC 6558). We skipped Pal~2 due to significant differential reddening and E~3 since the cluster has such a low mass that the location of its main sequence cannot be accurately determined.
This leaves us with a sample of 79 globular clusters with accurate, deep ACS/WFC photometry in the F606W and F814W bands.
\begin{figure*}
\begin{center}
\includegraphics[width=0.99\textwidth]{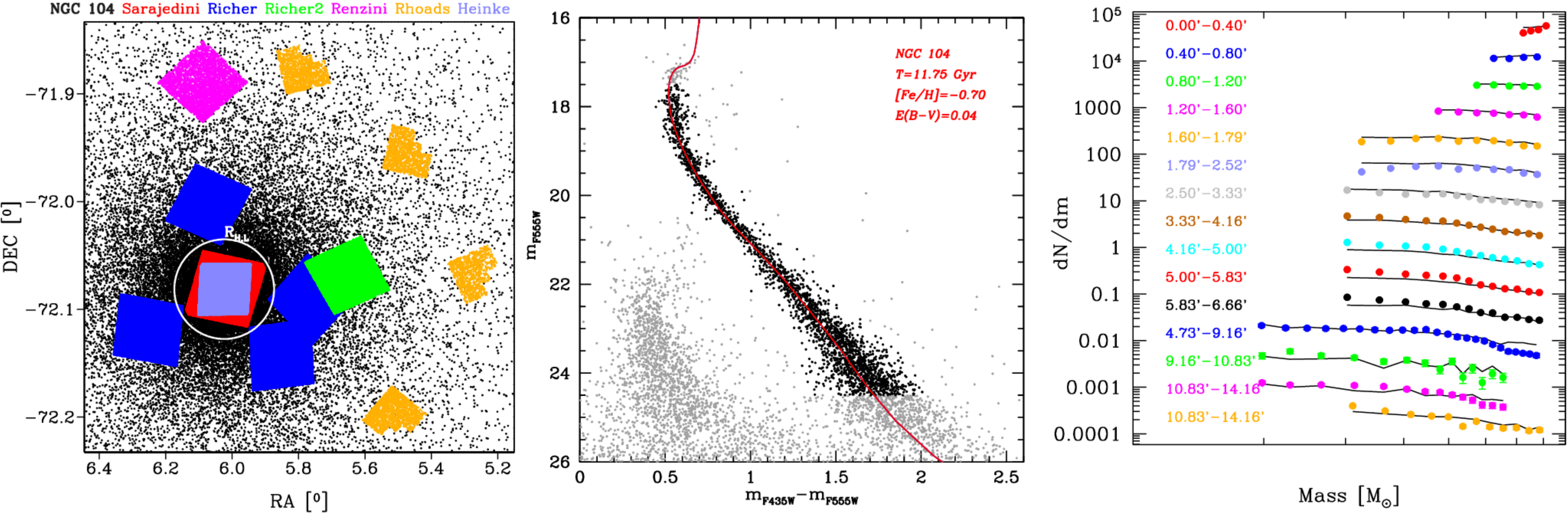}
\end{center}
\vspace*{-0.2cm}
\caption{Illustration of the fit of our $N$-body models to the globular cluster NGC~104. The left panel shows all fields in NGC~104 with available \hst photometry. The different fields are labeled by the name of the 
Principal Investigator of the \hst observations and the white circle marks the observed half-light radius of NGC~104. The middle panel shows the fit of a PARSEC isochrone (shown in red) to the 
combined photometry of the four HST/WFPC2 fields of Rhoads (HST Proposal ID: 9634) from the left panel. The parameters of the isochrone are given in the figure. Grey circles mark all stars in the field and black 
stars are the stars used for the determination of the stellar mass function of NGC~104. The group of stars visible in the lower left part of the middle panel is the SMC.
The right panel compares the derived mass functions as a function of radius against the mass function of the best-fitting $N$-body model (shown by solid lines). There is a clear change
in the derived mass function slope from increasing towards higher masses in the centre to increasing towards lower masses in the outermost parts, indicating that NGC~104 is mass segregated. This trend is reproduced by the 
best-fitting $N$-body model.
}
\label{fig:ngc104mf}
\end{figure*}

We dereddened the photometry of each globular cluster using the reddening values given by \citet{dotteretal2010} together with the extinction coefficients from \citet{siriannietal2005}. We then transformed the subdwarf photometry,
from the Johnson-Cousins UBVRI system to the ACS/WFC filter system using eqn.~12 of \citet{siriannietal2005} together with the transformation coefficients given in Table~18 of their paper. We then fitted PARSEC isochrones \citep{bressanetal2012}
to the color-magnitude diagram of each cluster, and used the isochrone to select the cluster members. We then calculated the average color of the cluster stars as a function 
of magnitude and define a main-sequence ridge line by cubic spline interpolation between the data points.

In order to compare the subdwarf photometry with the globular cluster photometry and derive the cluster distances, we select for each globular cluster all subdwarfs with metallicities $|$[Fe/H]$_{SD}$-[Fe/H]$_{GC}|<$0.2. We took the 
metallicities of the subdwarfs from \citet{casagrandeetal2010} and the metallicities of the globular clusters from \citet{carrettaetal2009b}. In order to correct the colors of the subdwarfs to the color they would have at the metallicity
of the globular cluster, we created a set of PARSEC isochrones equally spaced in metallicity by $\Delta \mbox{[Fe/H]}=0.5$ and calculate the colors of zero-age main sequence stars as a function of their absolute $V$-band luminosity. We then use this grid
of isochrones and the absolute magnitudes of the subdwarfs which we calculate using their \gaia EDR3 parallaxes to correct the subdwarf colors for the difference in metallicity. Due to the proximity of the subdwarf metallicities to the metallicity
of the clusters, the resulting shifts are always below 0.02 mag.

After transforming the subdwarf photometry to the ACS/WFC system and correcting the subdwarf metallicities, we fitted the cluster main sequences in a F814W$_0$ vs. (F606W-F814W)$_0$ CMD with the subdwarfs. To this end, we 
vary the assumed cluster distance modulus until the error-weighted difference between the subdwarfs and the previously determined main-sequence ridgeline are minimal. Due to the steepness of the cluster ridgelines and the accurate \gaia parallax distances, 
this essentially results in minimizing the error weighted color differences between the subdwarfs and the main sequence ridge lines. We exclude subdwarfs with $M_{814}<4$ from the fits since these could already have evolved away from the
main sequence. We also exclude subdwarfs with discrepant photometry since these could be undetected binaries. Fig.~\ref{fig:subdwarf} shows examples of our subdwarf fits for four globular clusters. The clusters shown are the same 
as the ones depicted in Fig.~2 of \citet{cohensarajedini2012}. We will derive an error bar for the subdwarf distance moduli in sec.~3 when we compare the subdwarf distance moduli against literature data.

\subsection{Star count distances}

For a given velocity dispersion profile, the derived mass of a cluster increases (decreases) with increasing (decreasing) cluster distance. If the velocity dispersion profile is based entirely on line-of-sight
velocities, the total cluster mass changes linearly with distance, while for a cluster with a velocity dispersion profile based only on proper motions, the mass changes with the distance to the third power.
For a cluster with measured line-of-sight and proper motion velocities, the scaling will be in between these limits. 
Since the predicted number of main-sequence stars changes linearly with the cluster mass (for a given mass function), one can use the observed number of cluster stars in a given magnitude interval and
a given field, together with the measured cluster kinematics and a theoretical model for the cluster, to determine the cluster distance.

We use as source for the globular cluster photometry the compilation of ACS/WFC F606W/F814W photometry published by \citet{sarajedinietal2007} as well the \hst photometry that was derived in \citet{baumgardtetal2019a},
\citet{baumgardtetal2020} and \citet{ebrahimietal2020}. We also use the photometry published by \citet{kerberetal2018} for NGC~6626 as input photometry.
We exclude NGC~5139 from this list due to the significant metallicity spread among the cluster stars. 
For each \hst observation, we again fit PARSEC isochrones \citep{bressanetal2012} to the cluster color-magnitude diagrams. To  create the isochrones, we use the cluster ages derived 
by \citet{vandenbergetal2013}, or, for clusters not studied by them, from available literature sources,
and take the cluster metallicities and reddenings from \citet{harris1996}. From the isochrones we then derive individual stellar masses for the main-sequence stars in the clusters.
We fit the $N$-body models of \citet{baumgardt2017} and \citet{baumgardthilker2018} to the observed surface density, velocity dispersion profiles and the observed mass function of main sequence stars in the
\hst fields and determine the best-fitting $N$-body model from the fit. For each \hst observation, we fit a power-law mass function $N(m) \sim m^{\alpha}$ to the observed stellar mass distribution as well as the
masses of stars in the $N$-body models in the same field. We use the mass function slopes $\alpha$ as fit parameters but not the absolute number of stars. Fig. \ref{fig:ngc104mf} shows as an example the distribution of 
\hst fields with measured photometry (left panel), the derived color-magnitude diagram (middle panel), and the measured stellar mass functions at different distances from the centre (right panel) for the globular cluster NGC~104. 

Varying the cluster distance, we then determine the distance that gives the best match between the number of main-sequence stars in the best-fitting $N$-body model and the observed number of main sequence stars 
in the different \hst fields. We restrict ourselves to well observed clusters in which the errors of the total mass (based on the cluster kinematics) are less than 5\%.
%% and the errors on the cluster distances based on the total cluster mass errors are similarly small.  

\begin{figure}
\begin{center}
\includegraphics[width=0.99\columnwidth]{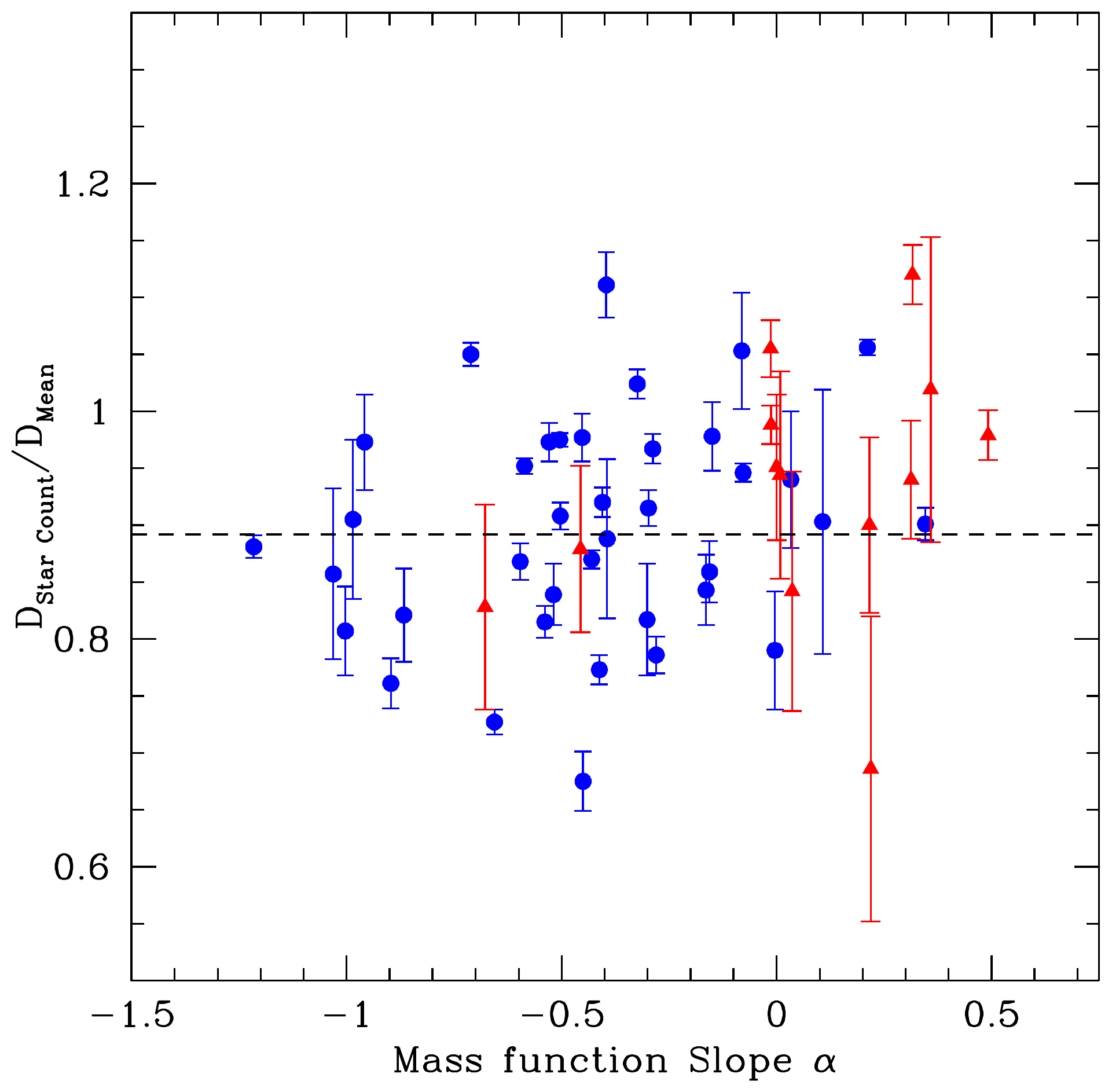}
\end{center}
\vspace*{-0.2cm}
\caption{Ratio of the best-fitting distance based on star counts to the average distance from the other methods as a function of the global mass function slope $\alpha$ of the clusters. The cluster sample is split into clusters with 
 10 or more distance measurements (blue circles) and those with less than 10 individual measurements (red triangles). The dashed line shows the average $D_{Star Count}/D_{Mean}$ ratio for the former group which has more secure
distance determinations. It can be seen that the average ratio is well below unity.
}
\label{fig:dscount}
\end{figure}

Fig.~\ref{fig:dscount} depicts the ratio between
the best-fitting cluster distances based on the star count method against the mean cluster distances that we derive from the other methods. We split the cluster sample into two groups, clusters with 10 or more
distance determinations from the other methods and clusters with less than 10 measurements. We expect that the distances of clusters in the first group are well determined so that
any deviation is mainly due to errors in the star count distances. It can be seen that the average distance ratio is well below unity. The reason is probably that the $N$-body models that we use to fit
the clusters do not contain primordial binaries. Since the average mass of binaries is larger than that of single stars, real globular clusters contain fewer stellar systems than models with only single stars 
for the same total mass, leading to an underprediction of the cluster distances with our method. The size of this underprediction is also in rough agreement with the observed binary fractions, which are around 
10\% \citep{sollimaetal2007,miloneetal2012,giesersetal2019}. Fig.~\ref{fig:dscount}  also indicates that the star count distances contain additional errors that are not accounted for by the uncertainties in the
cluster kinematics, since even when applying a constant shift, the deviations between the star count distances and the mean distances of the other methods are larger than what can be accounted for by the errors in either method. 
These additional errors could arise due to for example mismatches between the chosen mass functions of the $N$-body models and those of the real clusters or errors in the conversion from
luminosity to mass from the isochrones. We will fit for an additional shift in the star count distances as well as additional errors in sec.~4 when we compare all distances.

\subsection{Moving group distances}

Stars in star clusters move on very similar trajectories through the Milky Way since the internal motion of stars in the cluster is usually two orders of magnitudes smaller than the velocity with which the
cluster moves around the Galactic centre. As the distance and viewing angle to the stars changes along the orbit, the proper motion and line-of-sight velocity of the stars will change as well and one can
use this variation to measure the distance to the cluster. This is the so-called moving group or convergent point method \citep{brown1950,debruijne1999}. To first order, the change in proper motion $\mu$ 
due to these perspective effects can be approximated by
\begin{equation}
 \mu = -\frac{v_{Los} \alpha}{D A} \;\; \mbox{mas yr}^{-1}
\end{equation}
where $v_{Los}$ is the line-of-sight velocity of the cluster in km/sec, $\alpha$ the distance from the cluster centre in the direction of motion in radians, $D$ the distance of the cluster from the Sun in kpc, 
and $A=4.74$ \citep{brown1950,vandevenetal2006}.
Using the above equation, VB21 determined moving group distances to about 40 globular clusters based on \gaia EDR3 proper motions and the average line-of-sight velocities of the clusters from
\citet{baumgardthilker2018}. Here we use the two most
precisely determined distances $D=4.62 \pm 0.28$ kpc (NGC~3201) and $D=5.34 \pm 0.55$ kpc (NGC~5139). For all other clusters the derived distances have too large error bars to be useful.

\section{Literature survey of globular cluster distances}

In addition to our own distance determinations, we also performed a literature search of globular cluster distances. Limiting our search to the last 20 to 25 years, we were able to find about 1300 distance determinations
in total, with each globular cluster having between 1 to 35 individual measurements. Whenever a paper gave multiple measurements for the same cluster, we averaged these and took the scatter between the individual values as an indication
of the uncertainty. When we had multiple distance determinations from the same research group which were based on the same method and similar input data, we included only the newest value 
since we expect the different distances to not be independent from each other and the newest measurement to be the most accurate one. When a paper gave both a distance and a distance modulus to a cluster, we
used the distance modulus since we expect the errors to be Gaussian only in distance modulus. The resulting list of distance moduli and distances can be found in the Appendix.
\begin{figure*}
\begin{center}
\includegraphics[width=0.99\textwidth]{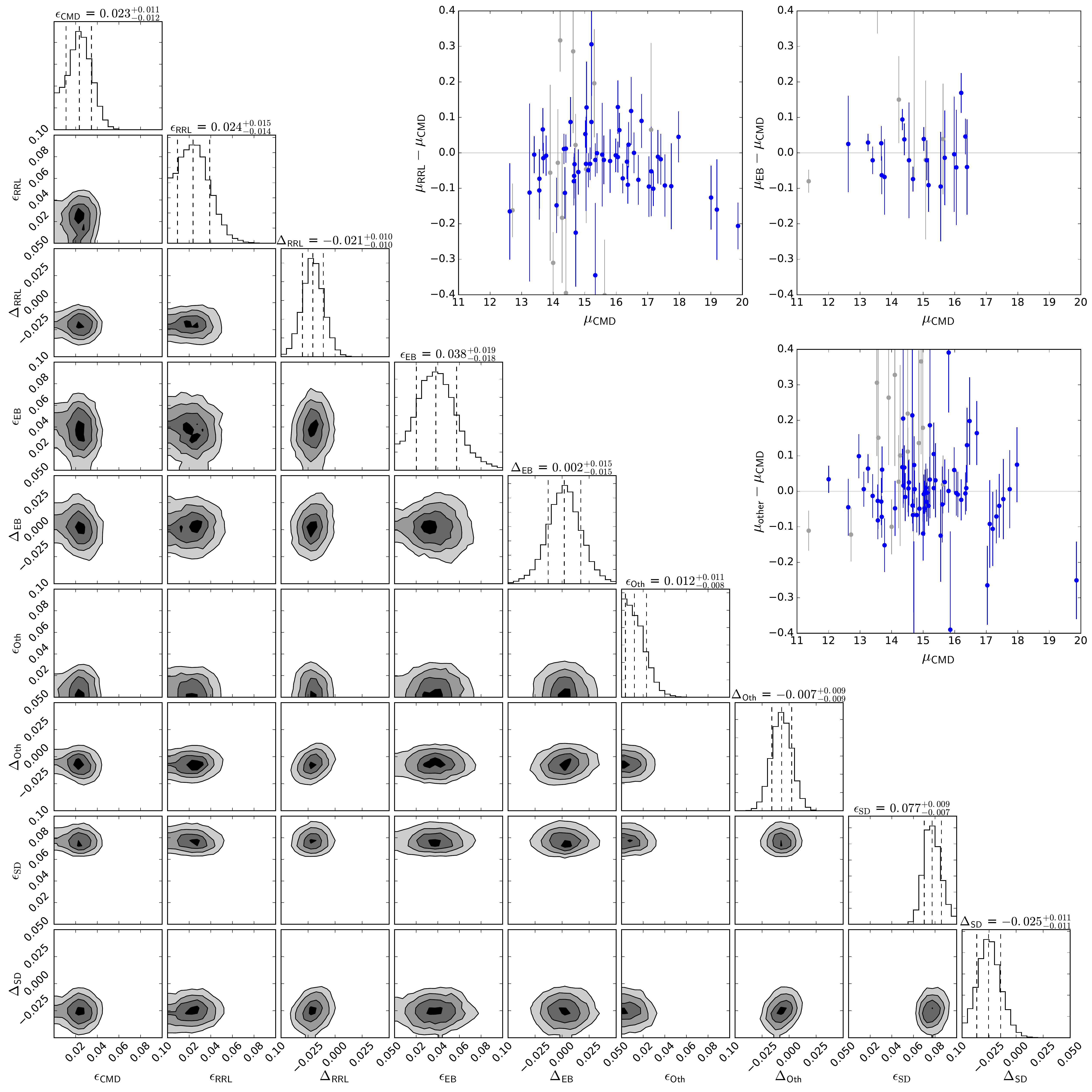}
\end{center}
\caption{Comparison of the distance moduli derived from isochrone fitting of CMDs (CMD), optical RR Lyrae P/L relations (RRL), eclipsing binaries and infrared RR Lyrae P/L relations (EB), and a combination 
  of all other literature methods (OTH) and our subdwarf distance moduli (SD). For each of the methods shown we use Monte Carlo Markov chain (MCMC) sampling to determine a maximum likelihood solution, solving for a constant
   shift $\Delta$ in distance modulus of each method against the CMD isochrone fitting distances as well as additional systematic errors $\epsilon$ of each method. The RR Lyrae distance moduli are on average 
    about $\Delta_{RRL} = 0.021$ mag smaller than the
   CMD fitting and our subdwarf distance moduli are smaller by about 0.025 mag.  The MCMC sampling also indicates additional errors of order 0.02 mag for the CMD and RR Lyrae distance moduli and 0.04 mag for the eclipsing binary ones. 
  It finally finds an error of 0.08 mag for our subdwarf distance moduli. The insets compare the distance moduli for individual clusters. Clusters with $E(B-V)<0.40$ that are used for the comparison are shown in blue, clusters with $E(B-V)>0.40$ are shown in grey.
}
\label{fig:litdiscomp}
\end{figure*}

In order to calculate a mean distance modulus for each cluster, we de-reddened distance moduli using the reddening values given in the papers, or, if no value was available, given in the 2010 edition of \citet{harris1996}. We also
transformed linear distances into distance moduli for those papers which did not give distance moduli. For distance moduli
without an error bar, we assumed an error equal to the mean error of the other distance determinations for the same cluster, with a minimum of 0.10 mag. A comparison of distance moduli for a few well observed clusters
shows that even in clusters with low reddening, individual distance moduli show a scatter of around 0.07 to 0.10 mag around the mean. This scatter is independent of the chosen method, with only eclipsing binary and RR Lyrae distances
based on infrared P/L relations showing smaller scatter. We therefore set the minimum error of any measurement except for EB or infrared RR Lyrae distances, for which we use the quoted errors, to 0.07 mag.
We neglected distances that deviate strongly from the other measurements.  In total we had to exclude about 25 measurements out of the 1300 measurements we found this way. Most of the excluded measurements were for highly 
reddened clusters with $E(B-V)>1.0$ where the individual 
measurements also often show a larger scatter. This scatter probably reflects an uncertainty in the exact amount of reddening and the proper selection of cluster and background stars. 

We also found that the subdwarf distance moduli
of \citet{cohensarajedini2012} were systematically smaller than the other literature values for low-metallicity clusters with $[Fe/H]<-1.2$. The average difference increased with decreasing metallicity and reached about 0.30 mag 
for the lowest-metallicity clusters. In order to correct this bias, we fitted a linear relation between the difference in distance modulus and metallicity to the data and shifted the \citet{cohensarajedini2012} distance moduli
upwards to bring them into agreement with the other literature data.
We finally shifted the distances that \citet{recioblancoetal2005} derived from CMD fitting down by 0.10 mag since we found a systematic bias of this size when comparing their distances with other literature distances.

In order to assess the accuracy of the derived distances and the reliability of the quoted errors, we compare in Fig.~\ref{fig:litdiscomp} the mean distance moduli calculated from various methods (isochrone fitting of CMDs,
visible light RR Lyrae P/L relations, eclipsing binaries and infrared RR Lyrae P/L relations and a combination of all other literature methods) plus the distance moduli we derived from subdwarfs with each other. For each
cluster we take a weighted average of all individual measurements of a given type. We split the RR Lyrae
distances into distances derived from visible or infrared P/L relations, since infrared P/L relations are usually thought to be more accurate \citep[e.g.][]{bonoetal2016}. We also group the infrared P/L distances together with the EB
distances to have a larger sample of clusters with very accurate distances. For each cluster we calculate average distance moduli for each method separately and then use Monte Carlo Markov chain (MCMC)
sampling and maximum likelihood optimisation to determine possible shifts $\Delta$ between the CMD distance moduli and the other distance moduli, as well as additional systematic errors $\epsilon$ that could still be present in the data.
We use only clusters with $E(B-V)<0.4$ mag for the comparison since more strongly reddened clusters show larger deviations between the individual measurements, 
and would therefore require a separate treatment. We compare all distance moduli against the CMD fitting ones since these are available for the largest number of clusters.

It can be seen that the RR Lyrae distance moduli are smaller by $\Delta_{RRL} = -0.021$ mag compared to the CMD fitting ones, which is significant at the 2$\sigma$ level. Our subdwarf distance moduli are also smaller, while the
other methods show no statistically significant offset to the CMD distances. Since, without independent information, it is not possible to determine which of these distance scales is closer to the true distances,
we do not apply shifts to any individual method and perform a straight averaging of the individual measurements instead.
Our MCMC sampling also finds evidence for additional systematic errors of $\epsilon \approx 0.023$ in the CMD fitting and
RR Lyrae distance moduli and additional systematic errors of  $\epsilon_{EB} \approx 0.038$ mag in the eclipsing binary/IR RR Lyrae distance moduli. The MCMC sampling finally shows that our subdwarf distance moduli have errors 
of about $\epsilon_{SD} = 0.077$ mag, which we use when calculating the mean distance moduli. 

Adding the above systematic errors in quadrature to the individual errors for the CMD, RR Lyrae and eclipsing binary distances, we then calculate a mean literature distance modulus and its formal error for each cluster from the mean
distance moduli of each individual method. We also calculated a reduced $\chi^2_r$ value of all distance determinations according to
\begin{equation}
\chi^2_r=\frac{1}{N-1} \sum_i (<\mu_{Lit}>-\mu_i)^2/\sigma_i^2
\end{equation}
where $<\!\mu_{Lit}\!>$ and $\mu_i$ are the average distance modulus and the individual distance moduli respectively and $\sigma_i$ is the error of the individual measurements. For clusters with $\chi^2_r>1$, i.e. a
larger than expected scatter of the individual measurements, we increase the final distance error by the square root of $\chi^2_r$.
Table~\ref{disresult} gives the average distance and its 1$\sigma$ upper and lower error that we obtain after averaging the literature distances for each cluster in this way. For easier comparison with the
other methods, we give the literature data in linear distances, however when calculating a final value for the distance we use the distance modulus as explained below. 
Plots depicting the individual measurements for each cluster can be found in the Appendix. 

\section{Results}
\label{sec:distances}

\subsection{Calculation of final distances}

\begin{figure*}
\begin{center}
\includegraphics[width=0.99\textwidth]{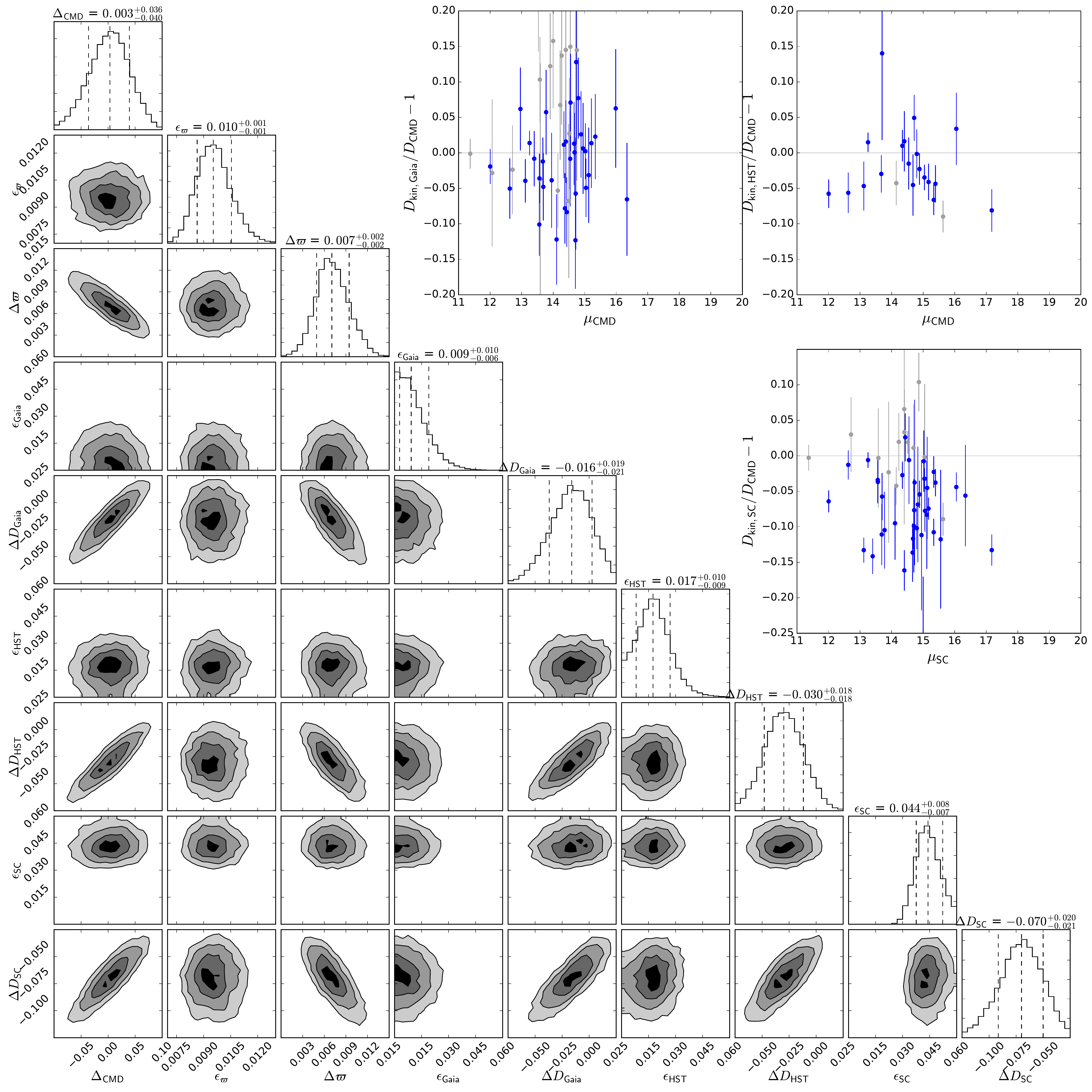}
\end{center}
\caption{Comparison of the CMD fitting distances with the Gaia parallax and kinematic distances and the \hst kinematic distances. We again use MCMC sampling and solve for constant scale factors $\Delta D$ in the distance scales of the various
methods as well as additional systematic errors $\epsilon$.
 $\Delta D_{Gaia}$ and $\Delta D_{HST}$ are relative shifts in the kinematic distances of the \gaia and \hst kinematic distances and $\Delta \varpi$ is a shift in the Gaia parallaxes. 
   We obtain a significant shift in the Gaia parallaxes of $\Delta \varpi \approx 0.007$ mas, in agreement with VB21. The \hst kinematic distances are on average also smaller than the literature distances by 
   about 3.0\%, while the Gaia kinematic distance shift $\Delta D_{Gaia}$ is compatible with zero. The \hst kinematic distances also show evidence of additional relative distance errors of about 1.7\%. The inset
     compares the kinematic and star count distances with the distances from isochrone fitting of the cluster CMDs.
}
\label{fig:distall}
\end{figure*}

Fig.~\ref{fig:distall} compares the distances that we derive from the \gaia EDR3 data (separated into parallax and kinematic distances), the \hst kinematic distances and the star count distances
with the mean literature distances determined in the previous section. We again use MCMC sampling and determine offsets between the distances as well as additional unaccounted for errors.
We restrict ourselves again to clusters with E(B-V)$<0.4$ mag for which the distances should be least influenced by reddening uncertainties. 

We find a constant shift of the Gaia parallax distances of $\Delta \varpi = 0.007$ mas, confirming the results of VB21. There is no
evidence for a significant shift of the CMD distances against the other methods. However, given that the error bar on $\Delta_{CMD}$ is about 0.04, the RR Lyrae distances would also be in statistical agreement with
the distances that we determined from the \gaia EDR3 data. The MCMC sampling finds $\Delta D_{Gaia} = -0.016 \pm 0.020$, implying that the Gaia kinematic distances are in statistical agreement with the other methods. However the \hst
kinematic distances are smaller by about 3.0\%. There is also some evidence for significant, additional unaccounted for errors in the \hst kinematic distances. We therefore add an error 
equal to 0.017 times the distance in quadrature to the formal distance error of the \hst distances for each cluster. We do not apply a shift to the \hst kinematic distances since it is first not clear which method gives 
the correct distance, and also because we found the \hst and \gaia kinematic distances to be in agreement with each other, meaning that the error might well be with the CMD distances. The MCMC analysis finally shows that 
the star count distances need to be increased by a factor $1/0.93=1.075$ and that their uncertainties have an additional systematic contribution of 4.4\% of the cluster distances. 

After increasing the star count distances and adding additional errors to the \hst kinematic and star count distances, we determine the final distance for each cluster by minimizing the combined likelihood of all measurements according to:
\begin{eqnarray}
\nonumber \ln L  & = & -\frac{1}{2} \frac{\left(\varpi_G-\frac{1}{<\!D\!>}\right)^2}{\epsilon^2_\varpi} -\frac{1}{2} \sum_i \frac{\left(D_i-<\!D\!>\right)^2}{\sigma_i^2} \\
   & & -\frac{1}{2} \sum_j \frac{\left(\mu_j- 5 \log{<\!D\!>}+5\right)^2}{\sigma_j^2} 
\end{eqnarray}

Here $\varpi_G$ and $\epsilon_\varpi$ are the cluster parallaxes and their $1\sigma$ errors, $D_i$ and $\sigma_i$ are the cluster distances and associated $1\sigma$ errors of the kinematic, star count and moving group distances.
$\mu_j$ and $\sigma_j$ are the distance moduli and error bars that we obtain from the averaging of the literature data and our subdwarf fits. $<\!D\!>$ is the mean cluster distance that we want to determine. We use 
the \gaia parallaxes here
instead of the distances that can be calculated from them since errors on the parallax distance are Gaussian only in parallax space. For the same reason we use distance moduli for the literature and subdwarf distances.
We calculate the mean distance from the point where $\ln L$ has a maximum and calculate
the associated upper and lower distance errors from the condition that $\Delta \ln L = 1.0$. The resulting mean distances along with the individual values from the various methods are
given in Table~\ref{disresult}. All data has been transformed to linear distances to make the different measurements easier comparable with each other. 

\begin{figure}
\begin{center}
\includegraphics[width=0.99\columnwidth]{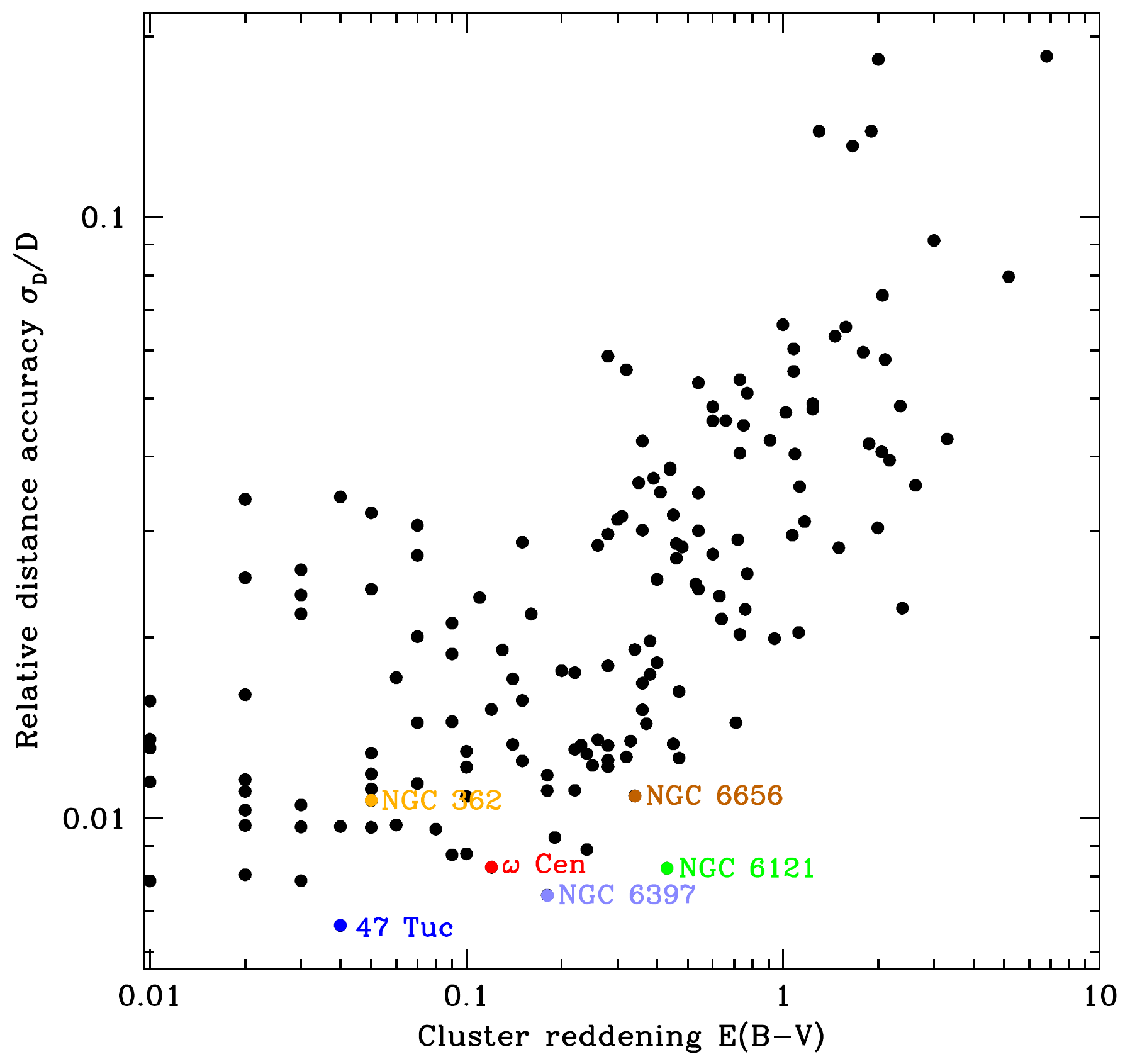}
\end{center}
\caption{Final distance error $\sigma_D/D$ as a function of the cluster reddening $E(B-V)$. Clusters with reddening values $E(B-V)<0.20$ have an average distance error of about 1\%. More reddened clusters show a continuous increase of the
distance error with reddening. The location of three of the clusters discussed in the text is marked in the plot.
}
\label{fig:disacc}
\end{figure}

Fig.\ref{fig:disacc} depicts the relative distance accuracy that we achieve by combining all data as a function of the cluster reddening. It can be seen that for clusters with low reddening values $E(B-V)<0.3$ we
achieve relative accuracies of about 1\%. For larger reddening values $E(B-V)>0.3$ there is a continuous rise of the distance errors and we achieve only 20\% accuracy for the most highly reddened clusters. This rise could partly
be driven by the fact that heavily reddened clusters have often only few individual distance determinations.

\subsection{Individual clusters}

\subsubsection{47 Tuc}

We derive a distance of $D=4.52 \pm 0.03$ kpc to 47~Tuc (NGC~104). For 47~Tuc, both literature distances and kinematic distances have small formal errors and are fully consistent with each other (see Fig.~\ref{fig:appfirst}).
Our mean distance also agrees very well with the value of $D=4.53 \pm 0.06$ kpc that \citet{maizapellanizetal2021} determined from the \gaia EDR3 parallaxes,
correcting for the small scale angular covariances in the \gaia data by comparing the parallaxes of 47~Tuc stars with those of background SMC stars.
It also agrees very well with the distance of $D=4.45 \pm 0.12$ kpc that \citet{chenetal2018} derived with the same method 
from \gaia DR2 data as well as the distance of $D=4.53 \pm 0.05$ kpc that \citet{thompsonetal2020} found from  the analysis of two detached eclipsing binaries in the cluster.
The distance to 47 Tuc therefore seems to be established with an accuracy of better than 1\%.

\subsubsection{NGC 362}

We correct the \gaia EDR3 parallaxes in the same way done by \citet{maizapellanizetal2021} for NGC~104 and \citet{chenetal2018} for NGC~104 and NGC~362. We again use SMC field stars in the 
direction of NGC~362 as reference, assuming a distance of $D=62.1 \pm 1.9$ kpc as distance to the SMC \citep{cionietal2000,graczyketal2014}. The resulting cluster parallax is $\varpi=109.9 \pm 3.7 \mu as$, corresponding to a 
distance of $9.10^{+0.32}_{-0.30}$ kpc. This distance is in agreement with the earlier value of \citet{chenetal2018} ($D=8.54 \pm 0.48$ kpc) as well as the
distance we derive from an averaging of literature values ($8.80 \pm 0.11$ kpc) and our own kinematic distance determinations (see Fig.~\ref{fig:appfirst}).

\subsubsection{$\omega$ Cen}

There is also good agreement in our distance determinations for $\omega$~Cen. The literature distances lead to an average distance of $D=5.47 \pm 0.06$ kpc (see Fig.~\ref{fig:appfirst}). The \gaia kinematic distance is 
in good agreement with the literature distance, however the \hst kinematic distance ($5.26 \pm 0.12$ kpc) is about 200 pc shorter.
One possible reason for the mismatch could be the intrinsic flattening of $\omega$~Cen. \citet{mackeygilmore2004} find
an ellipticity of $\epsilon=0.17$ for $\omega$~Cen based on its projected isophotes. The true flattening is likely even larger, with the exact value depending on the cluster inclination. \citet{vandevenetal2006}
found through axisymmetric Schwarzschild modeling that the resulting flattening of the velocity ellipsoid can decrease the kinematic cluster distance to $\omega$ Cen by several hundred pc if the flattening is not
taken into account in the modeling. Since the $N$-body models that we use to derive the kinematic distances are spherically symmetric, there is a chance that our kinematic distances underestimate the true distance
to $\omega$~Cen. This effect is much less likely for most other clusters which have much smaller relaxation times and should therefore be closer to being spherically symmetric. This is indeed also seen in observations
\citep[e.g.][]{whiteshawl1987}. We therefore regard the literature distance as more reliable for $\omega$ Cen. The \gaia EDR3 parallaxes after application of the systematic parallax offset
also favor a longer distance, however the error bar is too large to be decide between the two possibilities.

%%(see Fig.~\ref{fig:appfirst}). The \gaia kinematic distance is also in good agreement with the literature distance.  It should however be noted that the good agreement of the \hst kinematic distance is due to
%%the systematic increase that we applied to the \hst kinematic distances in sec.~4.1. Without this increase, the  \hst kinematic distance would be shorter by about 200~pc.  However even with a shorter \hst kinematic
%%distance, our final distance to $\omega$ Cen would not change by more than about 20~pc.

\subsubsection{NGC 6121}

We derive a distance of $D=1.851 \pm 0.015$ kpc to M4 (NGC~6121). This distance is slightly larger than the distance of $D=1.82 \pm 0.04$ kpc found by \citet{kaluznyetal2013} based on three eclipsing binary stars
as well as recent results from RR~Lyrae stars \citep[e.g][]{bragaetal2015,neeleyetal2019,bhardwajetal2020} (see Fig.~\ref{fig:appfirst}). Nevertheless their distances are compatible with our distance within the error bars.
The true distance to NGC~6121 is therefore likely somewhere in the range 1.83 to 1.85~kpc, making M4 another cluster whose distance is determined with an accuracy of about 1\%.

\subsubsection{NGC 6397 and NGC 6656}

NGC 6397 and NGC 6656 (M~22) are among the closest globular clusters to the Sun. Their small distances together with their low reddening make them ideal targets for globular cluster studies, so a large amount of observational
data exists for both clusters. For NGC~6397 we find a total of 27 individual distance determinations in the literature, from which we derive a mean cluster distance of $2.521 \pm 0.025$ kpc. Despite a relative distance error of only
about 1\%f or the literature distance and similarly small errors for most of the other methods, most distances are in good statistical agreement with each other, with the exception of the HST kinematic distances that are too short 
by about 100 pc. Taking the average over all methods, we derive a distance of $2.488 \pm 0.019$~kpc for NGC~6397, making this cluster the second closest cluster to the Sun. There is a similar good agreement in the individual distances 
values for NGC~6656, for which we derive a mean distance of $D=3.307 \pm 0.037$ kpc.

\subsection{Absolute luminosity of the TRGB and the value of $H_0$}

Globular clusters can be used to determine the absolute luminosity of the tip of the red giant branch (TRGB), which in turn can be used to measure the distance to nearby galaxies.
Using \gaia EDR3 parallaxes,
\citet{soltisetal2021} recently derived a distance of $5.24 \pm 0.11$ kpc to $\omega$~Cen. Using this distance, they derived an absolute I-band magnitude of the
TRGB of $M_{I,TRGB}=-3.97 \pm 0.06$, from which they deduced a value of the local Hubble constant of $H_0=71.2 \pm 2.0$ km/sec. In contrast,  \citet{freedmanetal2020} used stars in the LMC to derive
an an absolute TRGB magnitude of $M_{I,TRGB}=-4.05 \pm 0.06$, using eclipsing binaries in 47~Tuc and the SMC to calibrate the LMC distance.

The $\omega$ Cen distance of \citet{soltisetal2021} was based on a direct averaging of the \gaia EDR3 parallaxes of individual member stars, applying only the \citet{lindegrenetal2021} parallax corrections. Their parallax value
did not take systematic biases in the \gaia parallaxes or small scale correlated errors into account. Accounting for both, we derived a parallax of $\varpi=182.3 \pm 9.5 \mu as$ for $\omega$ Cen
\citep{vasilievbaumgardt2021}, corresponding to a distance
$D=5.48 \pm 0.24$ kpc, slightly larger than \citet{soltisetal2021} and also with a larger formal error than found by them. Combining this data with the other distances, we then derive a distance of $D=5.43 \pm 0.05$ to $\omega$ Cen. A similar analysis
as the one done by \citet{soltisetal2021} leads to
an absolute TRGB magnitude of $M_{I,TRGB}=-4.06 \pm 0.06$, confirming the value found by \citet{freedmanetal2020}. In addition, our results also confirm their adopted distance to NGC~104. The resulting value of the local Hubble constant 
of $H_0=69.4 \pm 2.0$ km/sec would be in better agreement with the expected value based on the Planck data of $67.4 \pm 0.5$ km/sec \citep{verdeetal2019}.

\section{Conclusions}

We have derived mean distances to globular clusters using a combination of our own measurements based on \gaia EDR3 data and published literature distances. We derived our own distances using six
different methods: \gaia ED3 parallaxes, kinematic distances using proper motion velocity dispersion profiles based on \gaia EDR3 proper motions, fitting nearby subdwarfs to the globular cluster
main sequences where the distances and absolute luminosities of the subdwarfs are calculated from their \gaia parallaxes and moving group distances based on  \gaia ED3 proper motions. In addition, 
we also derive kinematic distances from \hst proper motion velocity dispersion profiles and distances using \hst star counts in combination with the cluster kinematics. 

Apart from a bias in the \gaia EDR3 parallaxes, we find good internal agreement 
among the distances calculated with the different methods and also with published distances in the literature down to a level of about 2\%. There is a possibility that our \hst kinematic distances
could slightly underestimate the distances to globular clusters, however the difference that we see could also signify a problem in the isochrone fitting distances.
We also find some evidence that isochrone based distance moduli are larger than RR Lyrae based ones by about 0.02 mag.

Averaging over the various distance determinations, we are able to determine distances to a number of nearby globular clusters with an accuracy of 1\%, making these clusters valuable objects for the establishment
of a cosmic distance ladder. In particular, our results for the distances to 47 Tuc and $\omega$ Cen argue for an absolute TRGB magnitude of $M_{I,TRGB}=-4.05$ leading to a value of the local Hubble constant that is in agreement
with the expected value based on Planck data. They however do not remove the tension between the Cepheid based value of $H_0$ and the Planck data, which is currently significant at about the 4$\sigma$ level
\citep{riessetal2019}. Given that we find that the different literature distances to globular clusters are consistent down to a level of about 1-2\%, it is also unlikely that our results would lead to a significant
revision of the distance modulus to nearby galaxies like the LMC, and therefore a reduction of the tension in $H_0$.

Future releases of the \gaia catalogue will further increase the sample of globular clusters with highly accurate distances by reducing the random and systematic errors in the parallaxes and proper motions of 
stars. There is currently also still a substantial uncertainty in the distances of highly reddened
clusters, where individual distances can deviate from each other by up to a factor of two. At least for the more nearby clusters, \gaia parallaxes and proper motion velocity dispersion profiles based on future \gaia data releases
should bring a significant improvement to the distances to these clusters.
 
\section*{Acknowledgments}
EV acknowledges support from STFC via the Consolidated grant to the Institute of Astronomy. We thank Domenico Nardiello for providing us with the HST photometry for NGC~6626.
This work presents results from the European Space Agency (ESA) space mission \gaia. \gaia data are being processed by the Gaia Data Processing and Analysis Consortium (DPAC). Funding for the DPAC is provided by national institutions, in particular the institutions participating in the Gaia MultiLateral Agreement (MLA). This work is also based on observations made with the NASA/ESA Hubble Space Telescope, obtained from the data archive at the Space Telescope Science Institute. STScI is operated by the Association of Universities for Research in Astronomy, Inc. under NASA contract NAS 5-26555.

\section*{Data Availability}

The distances and the fit results of our $N$-body models to Galactic globular clusters using these distances can be obtained from the following webpage:
\href{https://people.smp.uq.edu.au/HolgerBaumgardt/globular/}{https://people.smp.uq.edu.au/HolgerBaumgardt/globular/}

\bibliographystyle{mn2e}
\bibliography{mybib}

\begin{table*}
\caption{Derived distances of the studied globular clusters. The final column lists the number of independent distance determinations used to calculate the mean distance.
The \gaia EDR3 parallax distances give the mean and $1\sigma$ uncertainty of the distance.  We have applied shifts to the star count distances and \gaia parallaxes to bring all distances to a common scale and also have added systematic errors as described in the text.}
\begin{tabular}{@{}l@{\hspace*{0.1cm}}c@{\hspace*{0.1cm}}c@{\hspace*{0.1cm}}c@{\hspace*{0.1cm}}c@{\hspace*{0.1cm}}c@{\hspace*{0.1cm}}c@{\hspace*{0.1cm}}c@{\hspace*{0.1cm}}c}
\hline
\multirow{2}{*}{Name} & GEDR3 parallax dist. &   GEDR3 kin. dist. & HST kin. dist. & Subdwarf dist. & Star Count dist. & Lit. dist. & Mean distance & \multirow{2}{*}{N$_{\mbox{Tot}}$}\\
 & [kpc] & [kpc] & [kpc] & [kpc] & [kpc] & [kpc] & [kpc]  & \\
\hline
2MASS-GC01  &  -   &  -   &  -   &  -   &  -   & $\,\,\,\,\,\,3.373^{+0.682}_{-0.567}$  & $\,\,\,\,\,\,3.373^{+0.682}_{-0.567}$  & 1 \\[+0.10cm]
2MASS-GC02  &  -   &  -   &  -   &  -   &  -   & $\,\,\,\,\,\,5.503^{+0.456}_{-0.421}$  & $\,\,\,\,\,\,5.503^{+0.456}_{-0.421}$  & 2 \\[+0.10cm]
AM 1  &  -   &  -   &  -   &  -   &  -   & $ 118.905^{+3.444}_{-3.347}$  & $ 118.905^{+3.444}_{-3.347}$  & 2 \\[+0.10cm]
AM 4  &  -   &  -   &  -   &  -   &  -   & $\,\,\,29.013^{+0.951}_{-0.920}$  & $\,\,\,29.013^{+0.951}_{-0.920}$  & 2 \\[+0.10cm]
Arp 2  &  -   &  -   &  -   & $\,\,\,28.184^{+1.058}_{-1.019}$  &  -   & $\,\,\,28.721^{+0.346}_{-0.342}$  & $\,\,\,28.726^{+0.346}_{-0.342}$  & 11 \\[+0.10cm]
BH 140  & $\,\,\,\,\,\,4.808^{+0.258}_{-0.233}$  &  -   &  -   &  -   &  -   &  -   & $\,\,\,\,\,\,4.808^{+0.258}_{-0.233}$  & 1 \\[+0.10cm]
BH 261  &  -   &  -   &  -   &  -   &  -   & $\,\,\,\,\,\,6.183^{+0.279}_{-0.267}$  & $\,\,\,\,\,\,6.115^{+0.265}_{-0.253}$  & 4 \\[+0.10cm]
Crater  &  -   &  -   &  -   &  -   &  -   & $ 147.231^{+4.334}_{-4.210}$  & $ 147.231^{+4.334}_{-4.210}$  & 3 \\[+0.10cm]
Djor 1  &  -   &  -   &  -   &  -   &  -   & $\,\,\,\,\,\,9.854^{+0.676}_{-0.632}$  & $\,\,\,\,\,\,9.879^{+0.671}_{-0.628}$  & 4 \\[+0.10cm]
Djor 2  &  -   &  -   &  -   &  -   &  -   & $\,\,\,\,\,\,8.758^{+0.179}_{-0.176}$  & $\,\,\,\,\,\,8.764^{+0.178}_{-0.174}$  & 5 \\[+0.10cm]
E 3  & $\,\,\,\,\,\,7.326^{+0.765}_{-0.633}$  &  -   &  -   &  -   &  -   & $\,\,\,\,\,\,7.940^{+0.268}_{-0.259}$  & $\,\,\,\,\,\,7.876^{+0.254}_{-0.245}$  & 3 \\[+0.10cm]
ESO 280  &  -   &  -   &  -   &  -   &  -   & $\,\,\,20.941^{+0.656}_{-0.636}$  & $\,\,\,20.945^{+0.656}_{-0.636}$  & 3 \\[+0.10cm]
ESO 452  &  -   &  -   &  -   &  -   &  -   & $\,\,\,\,\,\,7.440^{+0.208}_{-0.203}$  & $\,\,\,\,\,\,7.389^{+0.202}_{-0.196}$  & 4 \\[+0.10cm]
Eridanus  &  -   &  -   &  -   &  -   &  -   & $\,\,\,84.723^{+2.937}_{-2.839}$  & $\,\,\,84.684^{+2.936}_{-2.838}$  & 1 \\[+0.10cm]
FSR 1716  &  -   &  -   &  -   &  -   &  -   & $\,\,\,\,\,\,7.321^{+0.275}_{-0.265}$  & $\,\,\,\,\,\,7.431^{+0.270}_{-0.260}$  & 4 \\[+0.10cm]
FSR 1735  &  -   &  -   &  -   &  -   &  -   & $\,\,\,\,\,\,9.082^{+0.543}_{-0.512}$  & $\,\,\,\,\,\,9.082^{+0.543}_{-0.512}$  & 4 \\[+0.10cm]
FSR 1758  &  -   &  -   &  -   &  -   &  -   & $\,\,\,11.482^{+0.832}_{-0.776}$  & $\,\,\,11.085^{+0.764}_{-0.710}$  & 2 \\[+0.10cm]
HP 1  &  -   &  -   &  -   &  -   &  -   & $\,\,\,\,\,\,6.944^{+0.145}_{-0.142}$  & $\,\,\,\,\,\,6.995^{+0.144}_{-0.141}$  & 10 \\[+0.10cm]
IC 1257  &  -   &  -   &  -   &  -   &  -   & $\,\,\,26.632^{+1.474}_{-1.397}$  & $\,\,\,26.587^{+1.469}_{-1.392}$  & 2 \\[+0.10cm]
IC 1276  & $\,\,\,\,\,\,5.023^{+0.491}_{-0.411}$  &  -   &  -   &  -   &  -   & $\,\,\,\,\,\,4.159^{+0.345}_{-0.318}$  & $\,\,\,\,\,\,4.554^{+0.263}_{-0.243}$  & 3 \\[+0.10cm]
IC 4499  &  -   &  -   &  -   & $\,\,\,17.378^{+0.652}_{-0.629}$  &  -   & $\,\,\,18.880^{+0.254}_{-0.250}$  & $\,\,\,18.891^{+0.253}_{-0.250}$  & 11 \\[+0.10cm]
Laevens 3  &  -   &  -   &  -   &  -   &  -   & $\,\,\,61.773^{+1.672}_{-1.628}$  & $\,\,\,61.767^{+1.672}_{-1.628}$  & 3 \\[+0.10cm]
Liller 1  &  -   &  -   &  -   &  -   &  -   & $\,\,\,\,\,\,8.061^{+0.353}_{-0.338}$  & $\,\,\,\,\,\,8.061^{+0.353}_{-0.338}$  & 5 \\[+0.10cm]
Lynga 7  &  -   &  -   &  -   & $\,\,\,\,\,\,7.798^{+0.368}_{-0.351}$  &  -   & $\,\,\,\,\,\,7.867^{+0.165}_{-0.161}$  & $\,\,\,\,\,\,7.899^{+0.163}_{-0.159}$  & 8 \\[+0.10cm]
Mercer 5  &  -   &  -   &  -   &  -   &  -   & $\,\,\,\,\,\,5.495^{+0.533}_{-0.486}$  & $\,\,\,\,\,\,5.466^{+0.523}_{-0.476}$  & 1 \\[+0.10cm]
NGC 104  & $\,\,\,\,\,\,4.367^{+0.189}_{-0.174}$  & $\,\,\,\,\,\,4.533 \pm 0.067$  & $\,\,\,\,\,\,4.522 \pm 0.085$  & $\,\,\,\,\,\,4.406^{+0.165}_{-0.159}$  & $\,\,\,\,\,\,4.779 \pm 0.197$  & $\,\,\,\,\,\,4.512^{+0.040}_{-0.039}$  & $\,\,\,\,\,\,4.521^{+0.031}_{-0.031}$  & 33 \\[+0.10cm]
NGC 288  & $\,\,\,\,\,\,7.692^{+0.731}_{-0.615}$  & $\,\,\,\,\,\,9.814 \pm 0.576$  & $\,\,\,\,\,\,9.098 \pm 0.348$  & $\,\,\,\,\,\,9.333^{+0.350}_{-0.338}$  & $\,\,\,\,\,\,8.798 \pm 0.446$  & $\,\,\,\,\,\,8.983^{+0.096}_{-0.095}$  & $\,\,\,\,\,\,8.988^{+0.089}_{-0.088}$  & 23 \\[+0.10cm]
NGC 362  & $\,\,\,\,\,\,9.099^{+0.317}_{-0.296}$  & $\,\,\,\,\,\,8.267 \pm 0.646$  & $\,\,\,\,\,\,9.202 \pm 0.321$  & $\,\,\,\,\,\,8.790^{+0.330}_{-0.318}$  & $\,\,\,\,\,\,8.707 \pm 0.374$  & $\,\,\,\,\,\,8.770^{+0.114}_{-0.112}$  & $\,\,\,\,\,\,8.829^{+0.096}_{-0.096}$  & 21 \\[+0.10cm]
NGC 1261  &  -   &  -   & $\,\,\,16.775 \pm 0.872$  & $\,\,\,16.293^{+0.611}_{-0.589}$  & $\,\,\,16.680 \pm 0.726$  & $\,\,\,16.353^{+0.205}_{-0.202}$  & $\,\,\,16.400^{+0.192}_{-0.190}$  & 15 \\[+0.10cm]
NGC 1851  &  -   &  -   & $\,\,\,11.440 \pm 0.320$  & $\,\,\,12.303^{+0.462}_{-0.445}$  & $\,\,\,12.376 \pm 0.528$  & $\,\,\,12.017^{+0.156}_{-0.154}$  & $\,\,\,11.951^{+0.134}_{-0.133}$  & 18 \\[+0.10cm]
NGC 1904  &  -   &  -   &  -   &  -   &  -   & $\,\,\,13.080^{+0.182}_{-0.179}$  & $\,\,\,13.078^{+0.181}_{-0.179}$  & 13 \\[+0.10cm]
NGC 2298  &  -   &  -   &  -   & $\,\,\,10.000^{+0.471}_{-0.450}$  & $\,\,\,\,\,\,7.471 \pm 1.208$  & $\,\,\,\,\,\,9.890^{+0.175}_{-0.172}$  & $\,\,\,\,\,\,9.828^{+0.170}_{-0.167}$  & 10 \\[+0.10cm]
NGC 2419  &  -   &  -   &  -   &  -   &  -   & $\,\,\,88.471^{+2.437}_{-2.371}$  & $\,\,\,88.471^{+2.437}_{-2.371}$  & 6 \\[+0.10cm]
NGC 2808  &  -   & $\,\,\,\,\,\,9.688 \pm 0.428$  & $\,\,\,\,\,\,9.837 \pm 0.233$  & $\,\,\,10.280^{+0.435}_{-0.417}$  & $\,\,\,10.603 \pm 0.494$  & $\,\,\,10.139^{+0.141}_{-0.139}$  & $\,\,\,10.060^{+0.112}_{-0.111}$  & 20 \\[+0.10cm]
NGC 3201  & $\,\,\,\,\,\,4.819^{+0.239}_{-0.217}$  & $\,\,\,\,\,\,4.745 \pm 0.176$  &  -   & $\,\,\,\,\,\,4.571^{+0.172}_{-0.165}$  & $\,\,\,\,\,\,4.416 \pm 0.202$  & $\,\,\,\,\,\,4.749^{+0.046}_{-0.046}$  & $\,\,\,\,\,\,4.737^{+0.043}_{-0.042}$  & 29 \\[+0.10cm]
NGC 4147  &  -   &  -   &  -   & $\,\,\,17.947^{+0.674}_{-0.649}$  &  -   & $\,\,\,18.510^{+0.214}_{-0.212}$  & $\,\,\,18.535^{+0.214}_{-0.212}$  & 16 \\[+0.10cm]
NGC 4372  & $\,\,\,\,\,\,5.688^{+0.425}_{-0.370}$  & $\,\,\,\,\,\,5.933 \pm 0.423$  &  -   &  -   & $\,\,\,\,\,\,4.593 \pm 0.614$  & $\,\,\,\,\,\,5.959^{+0.385}_{-0.362}$  & $\,\,\,\,\,\,5.713^{+0.214}_{-0.207}$  & 5 \\[+0.10cm]
NGC 4590  &  -   &  -   &  -   & $\,\,\,10.328^{+0.388}_{-0.374}$  & $\,\,\,10.396 \pm 0.841$  & $\,\,\,10.409^{+0.101}_{-0.100}$  & $\,\,\,10.404^{+0.100}_{-0.099}$  & 24 \\[+0.10cm]
NGC 4833  & $\,\,\,\,\,\,6.557^{+0.490}_{-0.426}$  & $\,\,\,\,\,\,5.822 \pm 0.360$  &  -   & $\,\,\,\,\,\,6.546^{+0.246}_{-0.237}$  & $\,\,\,\,\,\,6.453 \pm 0.395$  & $\,\,\,\,\,\,6.519^{+0.091}_{-0.089}$  & $\,\,\,\,\,\,6.480^{+0.084}_{-0.083}$  & 12 \\[+0.10cm]
NGC 5024  &  -   & $\,\,\,17.313 \pm 1.353$  &  -   & $\,\,\,18.535^{+0.696}_{-0.670}$  & $\,\,\,18.803 \pm 1.446$  & $\,\,\,18.518^{+0.189}_{-0.187}$  & $\,\,\,18.498^{+0.185}_{-0.183}$  & 20 \\[+0.10cm]
NGC 5053  &  -   &  -   &  -   & $\,\,\,18.030^{+0.677}_{-0.652}$  &  -   & $\,\,\,17.515^{+0.235}_{-0.232}$  & $\,\,\,17.537^{+0.235}_{-0.232}$  & 20 \\[+0.10cm]
NGC 5139  & $\,\,\,\,\,\,5.485^{+0.302}_{-0.272}$  & $\,\,\,\,\,\,5.359 \pm 0.141$  & $\,\,\,\,\,\,5.264 \pm 0.121$  &  -   &  -   & $\,\,\,\,\,\,5.468^{+0.056}_{-0.055}$  & $\,\,\,\,\,\,5.426^{+0.047}_{-0.047}$  & 31 \\[+0.10cm]
NGC 5272  &  -   & $\,\,\,10.116 \pm 0.384$  &  -   & $\,\,\,\,\,\,9.638^{+0.362}_{-0.349}$  & $\,\,\,10.770 \pm 0.610$  & $\,\,\,10.167^{+0.085}_{-0.084}$  & $\,\,\,10.175^{+0.082}_{-0.081}$  & 38 \\[+0.10cm]
NGC 5286  &  -   & $\,\,\,11.181 \pm 0.689$  &  -   & $\,\,\,10.864^{+0.408}_{-0.393}$  &  -   & $\,\,\,11.071^{+0.149}_{-0.147}$  & $\,\,\,11.096^{+0.145}_{-0.143}$  & 13 \\[+0.10cm]
NGC 5466  &  -   &  -   &  -   & $\,\,\,15.776^{+0.592}_{-0.571}$  &  -   & $\,\,\,16.106^{+0.164}_{-0.162}$  & $\,\,\,16.120^{+0.164}_{-0.162}$  & 21 \\[+0.10cm]
NGC 5634  &  -   &  -   &  -   &  -   &  -   & $\,\,\,25.990^{+0.630}_{-0.615}$  & $\,\,\,25.959^{+0.628}_{-0.613}$  & 4 \\[+0.10cm]
NGC 5694  &  -   &  -   &  -   &  -   &  -   & $\,\,\,34.834^{+0.746}_{-0.730}$  & $\,\,\,34.840^{+0.745}_{-0.730}$  & 6 \\[+0.10cm]
\hline
\end{tabular}
\label{disresult}
\end{table*}
\begin{table*}
\contcaption{}
\begin{tabular}{@{}l@{\hspace*{0.1cm}}c@{\hspace*{0.1cm}}c@{\hspace*{0.1cm}}c@{\hspace*{0.1cm}}c@{\hspace*{0.1cm}}c@{\hspace*{0.1cm}}c@{\hspace*{0.1cm}}c@{\hspace*{0.1cm}}c}
\hline
\multirow{2}{*}{Name} & GEDR3 parallax dist. &   GEDR3 kin. dist. & HST kin. dist. & Subdwarf dist. & Star Count dist. & Lit. dist. & Mean distance & \multirow{2}{*}{N$_{\mbox{Tot}}$}\\
 & [kpc] & [kpc] & [kpc] & [kpc] & [kpc] & [kpc] & [kpc]  & \\
\hline
NGC 5824  &  -   &  -   &  -   &  -   &  -   & $\,\,\,31.754^{+0.605}_{-0.594}$  & $\,\,\,31.713^{+0.604}_{-0.593}$  & 5 \\[+0.10cm]
NGC 5897  &  -   &  -   &  -   & $\,\,\,12.359^{+0.464}_{-0.447}$  & $\,\,\,12.252 \pm 1.141$  & $\,\,\,12.607^{+0.246}_{-0.241}$  & $\,\,\,12.549^{+0.238}_{-0.233}$  & 8 \\[+0.10cm]
NGC 5904  & $\,\,\,\,\,\,7.651^{+0.677}_{-0.575}$  & $\,\,\,\,\,\,7.467 \pm 0.357$  & $\,\,\,\,\,\,7.456 \pm 0.201$  & $\,\,\,\,\,\,7.551^{+0.283}_{-0.273}$  & $\,\,\,\,\,\,7.721 \pm 0.350$  & $\,\,\,\,\,\,7.471^{+0.066}_{-0.065}$  & $\,\,\,\,\,\,7.479^{+0.060}_{-0.060}$  & 38 \\[+0.10cm]
NGC 5927  & $\,\,\,\,\,\,8.606^{+0.900}_{-0.744}$  & $\,\,\,\,\,\,8.209 \pm 0.647$  &  -   & $\,\,\,\,\,\,8.472^{+0.318}_{-0.306}$  & $\,\,\,\,\,\,8.757 \pm 0.371$  & $\,\,\,\,\,\,8.215^{+0.118}_{-0.116}$  & $\,\,\,\,\,\,8.270^{+0.111}_{-0.109}$  & 14 \\[+0.10cm]
NGC 5946  &  -   &  -   &  -   &  -   &  -   & $\,\,\,\,\,\,9.554^{+0.589}_{-0.555}$  & $\,\,\,\,\,\,9.642^{+0.529}_{-0.496}$  & 4 \\[+0.10cm]
NGC 5986  &  -   & $\,\,\,10.269 \pm 0.668$  &  -   & $\,\,\,10.471^{+0.393}_{-0.379}$  & $\,\,\,10.885 \pm 0.837$  & $\,\,\,10.549^{+0.137}_{-0.135}$  & $\,\,\,10.540^{+0.132}_{-0.130}$  & 14 \\[+0.10cm]
NGC 6093  &  -   &  -   &  -   & $\,\,\,10.375^{+0.389}_{-0.375}$  & $\,\,\,10.193 \pm 0.489$  & $\,\,\,10.342^{+0.120}_{-0.118}$  & $\,\,\,10.339^{+0.116}_{-0.115}$  & 20 \\[+0.10cm]
NGC 6101  &  -   &  -   &  -   & $\,\,\,14.322^{+0.537}_{-0.518}$  &  -   & $\,\,\,14.454^{+0.188}_{-0.185}$  & $\,\,\,14.449^{+0.187}_{-0.184}$  & 13 \\[+0.10cm]
NGC 6121  & $\,\,\,\,\,\,1.830^{+0.035}_{-0.034}$  & $\,\,\,\,\,\,1.878 \pm 0.033$  &  -   & $\,\,\,\,\,\,1.888^{+0.071}_{-0.068}$  & $\,\,\,\,\,\,2.016 \pm 0.090$  & $\,\,\,\,\,\,1.839^{+0.020}_{-0.020}$  & $\,\,\,\,\,\,1.851^{+0.015}_{-0.016}$  & 23 \\[+0.10cm]
NGC 6139  &  -   &  -   &  -   &  -   &  -   & $\,\,\,10.046^{+0.488}_{-0.465}$  & $\,\,\,10.035^{+0.463}_{-0.441}$  & 6 \\[+0.10cm]
NGC 6144  & $\,\,\,\,\,\,8.292^{+0.866}_{-0.716}$  &  -   &  -   & $\,\,\,\,\,\,8.511^{+0.319}_{-0.308}$  &  -   & $\,\,\,\,\,\,8.147^{+0.129}_{-0.127}$  & $\,\,\,\,\,\,8.151^{+0.127}_{-0.125}$  & 8 \\[+0.10cm]
NGC 6171  & $\,\,\,\,\,\,5.464^{+0.411}_{-0.357}$  & $\,\,\,\,\,\,6.017 \pm 0.407$  &  -   & $\,\,\,\,\,\,5.998^{+0.254}_{-0.244}$  & $\,\,\,\,\,\,5.479 \pm 0.407$  & $\,\,\,\,\,\,5.629^{+0.081}_{-0.080}$  & $\,\,\,\,\,\,5.631^{+0.077}_{-0.076}$  & 18 \\[+0.10cm]
NGC 6205  & $\,\,\,\,\,\,8.621^{+0.822}_{-0.690}$  & $\,\,\,\,\,\,6.906 \pm 0.336$  &  -   & $\,\,\,\,\,\,7.379^{+0.277}_{-0.267}$  &  -   & $\,\,\,\,\,\,7.427^{+0.079}_{-0.078}$  & $\,\,\,\,\,\,7.419^{+0.076}_{-0.075}$  & 25 \\[+0.10cm]
NGC 6218  & $\,\,\,\,\,\,5.068^{+0.282}_{-0.254}$  & $\,\,\,\,\,\,4.632 \pm 0.194$  &  -   & $\,\,\,\,\,\,5.321^{+0.200}_{-0.192}$  & $\,\,\,\,\,\,5.352 \pm 0.265$  & $\,\,\,\,\,\,5.136^{+0.052}_{-0.052}$  & $\,\,\,\,\,\,5.109^{+0.049}_{-0.048}$  & 24 \\[+0.10cm]
NGC 6229  &  -   &  -   &  -   &  -   &  -   & $\,\,\,30.102^{+0.475}_{-0.468}$  & $\,\,\,30.106^{+0.475}_{-0.467}$  & 7 \\[+0.10cm]
NGC 6235  &  -   &  -   &  -   &  -   &  -   & $\,\,\,11.896^{+0.395}_{-0.383}$  & $\,\,\,11.937^{+0.387}_{-0.374}$  & 5 \\[+0.10cm]
NGC 6254  & $\,\,\,\,\,\,5.394^{+0.412}_{-0.358}$  & $\,\,\,\,\,\,4.976 \pm 0.199$  &  -   & $\,\,\,\,\,\,4.989^{+0.235}_{-0.225}$  & $\,\,\,\,\,\,5.346 \pm 0.310$  & $\,\,\,\,\,\,5.051^{+0.070}_{-0.069}$  & $\,\,\,\,\,\,5.067^{+0.064}_{-0.063}$  & 19 \\[+0.10cm]
NGC 6256  &  -   &  -   &  -   & $\,\,\,\,\,\,7.516^{+0.650}_{-0.598}$  & $\,\,\,\,\,\,7.680 \pm 0.804$  & $\,\,\,\,\,\,7.204^{+0.326}_{-0.312}$  & $\,\,\,\,\,\,7.242^{+0.299}_{-0.288}$  & 9 \\[+0.10cm]
NGC 6266  & $\,\,\,\,\,\,5.764^{+0.426}_{-0.371}$  & $\,\,\,\,\,\,6.395 \pm 0.327$  & $\,\,\,\,\,\,6.502 \pm 0.163$  &  -   & $\,\,\,\,\,\,6.956 \pm 0.325$  & $\,\,\,\,\,\,6.249^{+0.172}_{-0.167}$  & $\,\,\,\,\,\,6.412^{+0.105}_{-0.104}$  & 13 \\[+0.10cm]
NGC 6273  & $\,\,\,\,\,\,7.657^{+0.683}_{-0.580}$  & $\,\,\,\,\,\,7.658 \pm 0.424$  &  -   &  -   & $\,\,\,\,\,\,8.463 \pm 0.536$  & $\,\,\,\,\,\,8.527^{+0.203}_{-0.198}$  & $\,\,\,\,\,\,8.343^{+0.165}_{-0.163}$  & 10 \\[+0.10cm]
NGC 6284  &  -   &  -   &  -   &  -   &  -   & $\,\,\,14.309^{+0.435}_{-0.422}$  & $\,\,\,14.208^{+0.427}_{-0.415}$  & 6 \\[+0.10cm]
NGC 6287  & $\,\,\,\,\,\,7.184^{+0.721}_{-0.601}$  &  -   &  -   &  -   &  -   & $\,\,\,\,\,\,8.143^{+0.427}_{-0.406}$  & $\,\,\,\,\,\,7.929^{+0.376}_{-0.356}$  & 6 \\[+0.10cm]
NGC 6293  &  -   &  -   &  -   &  -   & $\,\,\,\,\,\,9.273 \pm 1.131$  & $\,\,\,\,\,\,9.209^{+0.297}_{-0.288}$  & $\,\,\,\,\,\,9.192^{+0.282}_{-0.274}$  & 7 \\[+0.10cm]
NGC 6304  & $\,\,\,\,\,\,6.373^{+0.504}_{-0.435}$  & $\,\,\,\,\,\,7.304 \pm 0.684$  &  -   & $\,\,\,\,\,\,5.702^{+0.214}_{-0.206}$  &  -   & $\,\,\,\,\,\,6.059^{+0.161}_{-0.157}$  & $\,\,\,\,\,\,6.152^{+0.150}_{-0.146}$  & 13 \\[+0.10cm]
NGC 6316  &  -   & $\,\,\,\,\,\,9.631 \pm 0.901$  &  -   &  -   & $\,\,\,\,\,\,9.131 \pm 1.785$  & $\,\,\,11.792^{+0.499}_{-0.479}$  & $\,\,\,11.152^{+0.393}_{-0.382}$  & 7 \\[+0.10cm]
NGC 6325  & $\,\,\,\,\,\,6.627^{+0.646}_{-0.540}$  &  -   &  -   & $\,\,\,\,\,\,8.241^{+0.795}_{-0.725}$  &  -   & $\,\,\,\,\,\,7.755^{+0.369}_{-0.352}$  & $\,\,\,\,\,\,7.533^{+0.330}_{-0.314}$  & 7 \\[+0.10cm]
NGC 6333  & $\,\,\,\,\,\,8.052^{+0.775}_{-0.650}$  &  -   &  -   &  -   &  -   & $\,\,\,\,\,\,8.310^{+0.147}_{-0.144}$  & $\,\,\,\,\,\,8.300^{+0.144}_{-0.142}$  & 8 \\[+0.10cm]
NGC 6341  &  -   & $\,\,\,\,\,\,8.627 \pm 0.473$  & $\,\,\,\,\,\,8.231 \pm 0.374$  & $\,\,\,\,\,\,8.872^{+0.333}_{-0.321}$  & $\,\,\,\,\,\,8.188 \pm 0.345$  & $\,\,\,\,\,\,8.511^{+0.075}_{-0.074}$  & $\,\,\,\,\,\,8.501^{+0.071}_{-0.070}$  & 35 \\[+0.10cm]
NGC 6342  & $\,\,\,\,\,\,7.716^{+0.787}_{-0.654}$  &  -   &  -   & $\,\,\,\,\,\,8.054^{+0.302}_{-0.291}$  &  -   & $\,\,\,\,\,\,8.043^{+0.244}_{-0.237}$  & $\,\,\,\,\,\,8.013^{+0.233}_{-0.226}$  & 6 \\[+0.10cm]
NGC 6352  & $\,\,\,\,\,\,5.583^{+0.378}_{-0.333}$  &  -   & $\,\,\,\,\,\,6.260 \pm 0.816$  & $\,\,\,\,\,\,5.702^{+0.214}_{-0.206}$  & $\,\,\,\,\,\,5.564 \pm 0.295$  & $\,\,\,\,\,\,5.534^{+0.077}_{-0.076}$  & $\,\,\,\,\,\,5.543^{+0.073}_{-0.072}$  & 16 \\[+0.10cm]
NGC 6355  & $\,\,\,\,\,\,7.236^{+0.707}_{-0.591}$  &  -   &  -   & $\,\,\,\,\,\,8.472^{+0.318}_{-0.306}$  & $\,\,\,\,\,\,9.444 \pm 0.449$  & $\,\,\,\,\,\,8.527^{+0.259}_{-0.251}$  & $\,\,\,\,\,\,8.655^{+0.224}_{-0.220}$  & 8 \\[+0.10cm]
NGC 6356  &  -   &  -   &  -   &  -   &  -   & $\,\,\,15.900^{+0.973}_{-0.917}$  & $\,\,\,15.656^{+0.946}_{-0.890}$  & 4 \\[+0.10cm]
NGC 6362  & $\,\,\,\,\,\,8.026^{+0.731}_{-0.618}$  & $\,\,\,\,\,\,7.717 \pm 0.440$  & $\,\,\,\,\,\,7.720 \pm 0.341$  & $\,\,\,\,\,\,7.656^{+0.287}_{-0.277}$  & $\,\,\,\,\,\,6.849 \pm 0.326$  & $\,\,\,\,\,\,7.677^{+0.071}_{-0.070}$  & $\,\,\,\,\,\,7.649^{+0.067}_{-0.066}$  & 28 \\[+0.10cm]
NGC 6366  & $\,\,\,\,\,\,3.643^{+0.181}_{-0.165}$  & $\,\,\,\,\,\,3.404 \pm 0.256$  &  -   & $\,\,\,\,\,\,3.548^{+0.167}_{-0.160}$  & $\,\,\,\,\,\,3.861 \pm 0.297$  & $\,\,\,\,\,\,3.407^{+0.055}_{-0.054}$  & $\,\,\,\,\,\,3.444^{+0.051}_{-0.050}$  & 14 \\[+0.10cm]
NGC 6380  &  -   &  -   &  -   & $\,\,\,\,\,\,9.550^{+0.358}_{-0.345}$  &  -   & $\,\,\,\,\,\,9.532^{+0.312}_{-0.302}$  & $\,\,\,\,\,\,9.607^{+0.306}_{-0.296}$  & 4 \\[+0.10cm]
NGC 6388  &  -   &  -   & $\,\,\,10.894 \pm 0.230$  & $\,\,\,10.765^{+0.404}_{-0.389}$  & $\,\,\,11.196 \pm 0.466$  & $\,\,\,11.545^{+0.280}_{-0.273}$  & $\,\,\,11.171^{+0.162}_{-0.161}$  & 15 \\[+0.10cm]
NGC 6397  & $\,\,\,\,\,\,2.458^{+0.061}_{-0.058}$  & $\,\,\,\,\,\,2.467 \pm 0.056$  & $\,\,\,\,\,\,2.371 \pm 0.051$  & $\,\,\,\,\,\,2.410^{+0.090}_{-0.087}$  & $\,\,\,\,\,\,2.531 \pm 0.107$  & $\,\,\,\,\,\,2.512^{+0.024}_{-0.024}$  & $\,\,\,\,\,\,2.482^{+0.019}_{-0.019}$  & 29 \\[+0.10cm]
NGC 6401  & $\,\,\,\,\,\,7.283^{+0.781}_{-0.643}$  &  -   &  -   & $\,\,\,\,\,\,8.204^{+0.308}_{-0.297}$  & $\,\,\,\,\,\,8.448 \pm 0.376$  & $\,\,\,\,\,\,7.910^{+0.316}_{-0.304}$  & $\,\,\,\,\,\,8.064^{+0.238}_{-0.234}$  & 7 \\[+0.10cm]
NGC 6402  & $\,\,\,\,\,\,8.482^{+0.847}_{-0.706}$  &  -   &  -   &  -   &  -   & $\,\,\,\,\,\,9.204^{+0.267}_{-0.259}$  & $\,\,\,\,\,\,9.144^{+0.255}_{-0.248}$  & 6 \\[+0.10cm]
NGC 6426  &  -   &  -   &  -   & $\,\,\,21.677^{+1.022}_{-0.976}$  &  -   & $\,\,\,20.701^{+0.356}_{-0.350}$  & $\,\,\,20.710^{+0.355}_{-0.349}$  & 8 \\[+0.10cm]
NGC 6440  & $\,\,\,\,\,\,7.283^{+0.742}_{-0.617}$  & $\,\,\,\,\,\,7.394 \pm 0.776$  &  -   &  -   &  -   & $\,\,\,\,\,\,8.461^{+0.281}_{-0.272}$  & $\,\,\,\,\,\,8.248^{+0.248}_{-0.241}$  & 7 \\[+0.10cm]
NGC 6441  &  -   &  -   & $\,\,\,12.190 \pm 0.244$  &  -   & $\,\,\,13.095 \pm 0.674$  & $\,\,\,13.176^{+0.245}_{-0.240}$  & $\,\,\,12.728^{+0.163}_{-0.162}$  & 16 \\[+0.10cm]
NGC 6453  &  -   &  -   &  -   &  -   & $\,\,\,11.149 \pm 0.523$  & $\,\,\,\,\,\,9.804^{+0.238}_{-0.232}$  & $\,\,\,10.070^{+0.220}_{-0.216}$  & 8 \\[+0.10cm]
NGC 6496  &  -   & $\,\,\,\,\,\,9.823 \pm 0.612$  &  -   & $\,\,\,\,\,\,9.204^{+0.345}_{-0.333}$  & $\,\,\,\,\,\,9.326 \pm 0.988$  & $\,\,\,\,\,\,9.647^{+0.161}_{-0.159}$  & $\,\,\,\,\,\,9.641^{+0.153}_{-0.151}$  & 11 \\[+0.10cm]
NGC 6517  &  -   & $\,\,\,10.202 \pm 1.704$  &  -   &  -   &  -   & $\,\,\,\,\,\,8.913^{+0.717}_{-0.664}$  & $\,\,\,\,\,\,9.227^{+0.578}_{-0.538}$  & 4 \\[+0.10cm]
NGC 6522  &  -   & $\,\,\,\,\,\,8.120 \pm 0.929$  &  -   &  -   &  -   & $\,\,\,\,\,\,7.158^{+0.221}_{-0.214}$  & $\,\,\,\,\,\,7.295^{+0.211}_{-0.205}$  & 8 \\[+0.10cm]
NGC 6528  &  -   &  -   &  -   &  -   &  -   & $\,\,\,\,\,\,7.791^{+0.244}_{-0.237}$  & $\,\,\,\,\,\,7.829^{+0.239}_{-0.232}$  & 8 \\[+0.10cm]
\hline
\end{tabular}
\end{table*}
\begin{table*}
\contcaption{}
\begin{tabular}{@{}l@{\hspace*{0.1cm}}c@{\hspace*{0.1cm}}c@{\hspace*{0.1cm}}c@{\hspace*{0.1cm}}c@{\hspace*{0.1cm}}c@{\hspace*{0.1cm}}c@{\hspace*{0.1cm}}c@{\hspace*{0.1cm}}c}
\hline
\multirow{2}{*}{Name} & GEDR3 parallax dist. &   GEDR3 kin. dist. & HST kin. dist. & Subdwarf dist. & Star Count dist. & Lit. dist. & Mean distance & \multirow{2}{*}{N$_{\mbox{Tot}}$}\\
 & [kpc] & [kpc] & [kpc] & [kpc] & [kpc] & [kpc] & [kpc]  & \\
\hline
NGC 6535  & $\,\,\,\,\,\,6.570^{+0.669}_{-0.556}$  &  -   &  -   & $\,\,\,\,\,\,6.486^{+0.274}_{-0.263}$  &  -   & $\,\,\,\,\,\,6.353^{+0.127}_{-0.125}$  & $\,\,\,\,\,\,6.363^{+0.124}_{-0.122}$  & 8 \\[+0.10cm]
NGC 6539  & $\,\,\,\,\,\,8.292^{+0.866}_{-0.716}$  & $\,\,\,\,\,\,8.686 \pm 0.746$  &  -   &  -   & $\,\,\,\,\,\,8.694 \pm 1.140$  & $\,\,\,\,\,\,7.565^{+0.609}_{-0.563}$  & $\,\,\,\,\,\,8.165^{+0.395}_{-0.379}$  & 5 \\[+0.10cm]
NGC 6540  &  -   &  -   &  -   &  -   &  -   & $\,\,\,\,\,\,5.897^{+0.301}_{-0.286}$  & $\,\,\,\,\,\,5.909^{+0.279}_{-0.265}$  & 4 \\[+0.10cm]
NGC 6541  & $\,\,\,\,\,\,7.429^{+0.648}_{-0.552}$  & $\,\,\,\,\,\,7.056 \pm 0.324$  &  -   & $\,\,\,\,\,\,7.311^{+0.274}_{-0.264}$  & $\,\,\,\,\,\,8.493 \pm 0.416$  & $\,\,\,\,\,\,7.617^{+0.113}_{-0.111}$  & $\,\,\,\,\,\,7.609^{+0.102}_{-0.101}$  & 11 \\[+0.10cm]
NGC 6544  & $\,\,\,\,\,\,2.585^{+0.074}_{-0.070}$  & $\,\,\,\,\,\,2.516 \pm 0.239$  &  -   &  -   &  -   & $\,\,\,\,\,\,2.589^{+0.113}_{-0.109}$  & $\,\,\,\,\,\,2.582^{+0.059}_{-0.057}$  & 6 \\[+0.10cm]
NGC 6553  & $\,\,\,\,\,\,5.488^{+0.370}_{-0.326}$  & $\,\,\,\,\,\,6.211 \pm 0.457$  &  -   &  -   &  -   & $\,\,\,\,\,\,5.219^{+0.141}_{-0.138}$  & $\,\,\,\,\,\,5.332^{+0.128}_{-0.125}$  & 12 \\[+0.10cm]
NGC 6558  &  -   &  -   &  -   & $\,\,\,\,\,\,7.943^{+0.298}_{-0.287}$  &  -   & $\,\,\,\,\,\,7.495^{+0.306}_{-0.294}$  & $\,\,\,\,\,\,7.474^{+0.294}_{-0.282}$  & 6 \\[+0.10cm]
NGC 6569  &  -   &  -   &  -   &  -   & $\,\,\,10.961 \pm 1.031$  & $\,\,\,10.529^{+0.275}_{-0.268}$  & $\,\,\,10.534^{+0.261}_{-0.255}$  & 9 \\[+0.10cm]
NGC 6584  &  -   &  -   &  -   & $\,\,\,12.647^{+0.475}_{-0.457}$  &  -   & $\,\,\,13.602^{+0.177}_{-0.174}$  & $\,\,\,13.611^{+0.176}_{-0.174}$  & 13 \\[+0.10cm]
NGC 6624  &  -   & $\,\,\,\,\,\,8.027 \pm 0.530$  & $\,\,\,\,\,\,7.972 \pm 0.308$  & $\,\,\,\,\,\,7.621^{+0.286}_{-0.276}$  &  -   & $\,\,\,\,\,\,8.017^{+0.119}_{-0.117}$  & $\,\,\,\,\,\,8.019^{+0.108}_{-0.107}$  & 15 \\[+0.10cm]
NGC 6626  & $\,\,\,\,\,\,5.297^{+0.325}_{-0.289}$  & $\,\,\,\,\,\,5.734 \pm 0.306$  &  -   &  -   & $\,\,\,\,\,\,5.571 \pm 0.403$  & $\,\,\,\,\,\,5.311^{+0.114}_{-0.111}$  & $\,\,\,\,\,\,5.368^{+0.099}_{-0.098}$  & 10 \\[+0.10cm]
NGC 6637  &  -   & $\,\,\,\,\,\,9.960 \pm 1.149$  &  -   & $\,\,\,\,\,\,9.204^{+0.345}_{-0.333}$  & $\,\,\,\,\,\,9.139 \pm 1.054$  & $\,\,\,\,\,\,8.880^{+0.107}_{-0.106}$  & $\,\,\,\,\,\,8.900^{+0.106}_{-0.104}$  & 17 \\[+0.10cm]
NGC 6638  &  -   &  -   &  -   &  -   &  -   & $\,\,\,\,\,\,9.786^{+0.363}_{-0.350}$  & $\,\,\,\,\,\,9.775^{+0.347}_{-0.334}$  & 5 \\[+0.10cm]
NGC 6642  &  -   &  -   &  -   & $\,\,\,\,\,\,8.054^{+0.302}_{-0.291}$  &  -   & $\,\,\,\,\,\,8.009^{+0.209}_{-0.204}$  & $\,\,\,\,\,\,8.049^{+0.204}_{-0.198}$  & 7 \\[+0.10cm]
NGC 6652  &  -   &  -   &  -   & $\,\,\,\,\,\,9.506^{+0.357}_{-0.344}$  &  -   & $\,\,\,\,\,\,9.467^{+0.141}_{-0.138}$  & $\,\,\,\,\,\,9.464^{+0.139}_{-0.137}$  & 13 \\[+0.10cm]
NGC 6656  & $\,\,\,\,\,\,3.368^{+0.119}_{-0.111}$  & $\,\,\,\,\,\,3.181 \pm 0.123$  & $\,\,\,\,\,\,3.161 \pm 0.088$  & $\,\,\,\,\,\,3.373^{+0.127}_{-0.122}$  & $\,\,\,\,\,\,3.556 \pm 0.149$  & $\,\,\,\,\,\,3.327^{+0.049}_{-0.049}$  & $\,\,\,\,\,\,3.303^{+0.037}_{-0.037}$  & 14 \\[+0.10cm]
NGC 6681  &  -   & $\,\,\,\,\,\,9.722 \pm 0.578$  & $\,\,\,\,\,\,9.260 \pm 0.228$  & $\,\,\,\,\,\,9.120^{+0.342}_{-0.330}$  & $\,\,\,\,\,\,9.634 \pm 0.414$  & $\,\,\,\,\,\,9.354^{+0.130}_{-0.128}$  & $\,\,\,\,\,\,9.362^{+0.107}_{-0.106}$  & 15 \\[+0.10cm]
NGC 6712  & $\,\,\,\,\,\,7.402^{+0.676}_{-0.571}$  & $\,\,\,\,\,\,7.463 \pm 0.546$  &  -   &  -   &  -   & $\,\,\,\,\,\,7.355^{+0.294}_{-0.282}$  & $\,\,\,\,\,\,7.382^{+0.240}_{-0.233}$  & 9 \\[+0.10cm]
NGC 6715  &  -   &  -   & $\,\,\,25.019 \pm 0.829$  &  -   & $\,\,\,25.386 \pm 1.185$  & $\,\,\,26.644^{+0.383}_{-0.378}$  & $\,\,\,26.283^{+0.328}_{-0.325}$  & 17 \\[+0.10cm]
NGC 6717  &  -   &  -   &  -   &  -   &  -   & $\,\,\,\,\,\,7.464^{+0.135}_{-0.133}$  & $\,\,\,\,\,\,7.524^{+0.133}_{-0.131}$  & 10 \\[+0.10cm]
NGC 6723  & $\,\,\,\,\,\,8.237^{+0.829}_{-0.690}$  & $\,\,\,\,\,\,8.708 \pm 0.586$  &  -   & $\,\,\,\,\,\,8.472^{+0.318}_{-0.306}$  & $\,\,\,\,\,\,8.691 \pm 0.601$  & $\,\,\,\,\,\,8.241^{+0.103}_{-0.102}$  & $\,\,\,\,\,\,8.267^{+0.100}_{-0.099}$  & 18 \\[+0.10cm]
NGC 6749  & $\,\,\,\,\,\,7.849^{+0.779}_{-0.650}$  &  -   &  -   &  -   &  -   & $\,\,\,\,\,\,7.561^{+0.230}_{-0.223}$  & $\,\,\,\,\,\,7.591^{+0.218}_{-0.212}$  & 5 \\[+0.10cm]
NGC 6752  & $\,\,\,\,\,\,4.092^{+0.171}_{-0.158}$  & $\,\,\,\,\,\,4.036 \pm 0.114$  & $\,\,\,\,\,\,4.005 \pm 0.147$  & $\,\,\,\,\,\,3.981^{+0.149}_{-0.144}$  & $\,\,\,\,\,\,3.917 \pm 0.167$  & $\,\,\,\,\,\,4.176^{+0.050}_{-0.050}$  & $\,\,\,\,\,\,4.125^{+0.041}_{-0.041}$  & 18 \\[+0.10cm]
NGC 6760  & $\,\,\,\,\,\,8.244^{+0.839}_{-0.697}$  & $\,\,\,\,\,\,9.322 \pm 0.900$  &  -   &  -   &  -   & $\,\,\,\,\,\,8.110^{+0.600}_{-0.559}$  & $\,\,\,\,\,\,8.411^{+0.441}_{-0.418}$  & 4 \\[+0.10cm]
NGC 6779  &  -   &  -   &  -   & $\,\,\,10.139^{+0.381}_{-0.367}$  &  -   & $\,\,\,10.404^{+0.145}_{-0.143}$  & $\,\,\,10.430^{+0.144}_{-0.142}$  & 12 \\[+0.10cm]
NGC 6809  & $\,\,\,\,\,\,5.020^{+0.285}_{-0.256}$  & $\,\,\,\,\,\,5.201 \pm 0.246$  &  -   & $\,\,\,\,\,\,5.200^{+0.195}_{-0.188}$  & $\,\,\,\,\,\,5.223 \pm 0.303$  & $\,\,\,\,\,\,5.370^{+0.055}_{-0.054}$  & $\,\,\,\,\,\,5.348^{+0.052}_{-0.051}$  & 27 \\[+0.10cm]
NGC 6838  & $\,\,\,\,\,\,4.160^{+0.190}_{-0.174}$  & $\,\,\,\,\,\,4.154 \pm 0.237$  &  -   & $\,\,\,\,\,\,4.036^{+0.151}_{-0.146}$  &  -   & $\,\,\,\,\,\,3.977^{+0.053}_{-0.053}$  & $\,\,\,\,\,\,4.001^{+0.050}_{-0.050}$  & 15 \\[+0.10cm]
NGC 6864  &  -   &  -   &  -   &  -   &  -   & $\,\,\,20.549^{+0.459}_{-0.449}$  & $\,\,\,20.517^{+0.457}_{-0.448}$  & 5 \\[+0.10cm]
NGC 6934  &  -   & $\,\,\,16.718 \pm 1.382$  &  -   & $\,\,\,14.791^{+0.555}_{-0.535}$  &  -   & $\,\,\,15.704^{+0.175}_{-0.173}$  & $\,\,\,15.716^{+0.173}_{-0.171}$  & 19 \\[+0.10cm]
NGC 6981  &  -   &  -   &  -   & $\,\,\,16.069^{+0.603}_{-0.581}$  &  -   & $\,\,\,16.672^{+0.185}_{-0.183}$  & $\,\,\,16.661^{+0.185}_{-0.183}$  & 15 \\[+0.10cm]
NGC 7006  &  -   &  -   &  -   & $\,\,\,39.264^{+1.474}_{-1.420}$  &  -   & $\,\,\,39.319^{+0.565}_{-0.557}$  & $\,\,\,39.318^{+0.565}_{-0.557}$  & 9 \\[+0.10cm]
NGC 7078  &  -   &  -   & $\,\,\,10.375 \pm 0.295$  & $\,\,\,10.617^{+0.398}_{-0.384}$  & $\,\,\,10.769 \pm 0.452$  & $\,\,\,10.740^{+0.104}_{-0.103}$  & $\,\,\,10.709^{+0.096}_{-0.095}$  & 28 \\[+0.10cm]
NGC 7089  &  -   & $\,\,\,11.940 \pm 0.703$  &  -   & $\,\,\,11.169^{+0.419}_{-0.404}$  & $\,\,\,12.267 \pm 0.543$  & $\,\,\,11.647^{+0.119}_{-0.117}$  & $\,\,\,11.693^{+0.115}_{-0.114}$  & 21 \\[+0.10cm]
NGC 7099  & $\,\,\,\,\,\,7.981^{+0.753}_{-0.633}$  & $\,\,\,\,\,\,8.682 \pm 0.496$  &  -   & $\,\,\,\,\,\,8.630^{+0.324}_{-0.312}$  & $\,\,\,\,\,\,7.959 \pm 0.435$  & $\,\,\,\,\,\,8.480^{+0.094}_{-0.093}$  & $\,\,\,\,\,\,8.458^{+0.090}_{-0.089}$  & 19 \\[+0.10cm]
NGC 7492  &  -   &  -   &  -   &  -   &  -   & $\,\,\,24.434^{+0.581}_{-0.567}$  & $\,\,\,24.390^{+0.579}_{-0.566}$  & 4 \\[+0.10cm]
Pal 1  &  -   &  -   &  -   & $\,\,\,10.520^{+0.547}_{-0.520}$  &  -   & $\,\,\,11.189^{+0.329}_{-0.320}$  & $\,\,\,11.176^{+0.328}_{-0.319}$  & 4 \\[+0.10cm]
Pal 2  &  -   &  -   &  -   &  -   &  -   & $\,\,\,26.158^{+1.321}_{-1.258}$  & $\,\,\,26.174^{+1.317}_{-1.253}$  & 2 \\[+0.10cm]
Pal 3  &  -   &  -   &  -   &  -   &  -   & $\,\,\,94.842^{+3.288}_{-3.178}$  & $\,\,\,94.842^{+3.288}_{-3.178}$  & 2 \\[+0.10cm]
Pal 4  &  -   &  -   &  -   &  -   &  -   & $ 101.391^{+2.601}_{-2.536}$  & $ 101.391^{+2.601}_{-2.536}$  & 2 \\[+0.10cm]
Pal 5  &  -   &  -   &  -   &  -   &  -   & $\,\,\,21.928^{+0.521}_{-0.509}$  & $\,\,\,21.941^{+0.520}_{-0.508}$  & 6 \\[+0.10cm]
Pal 6  &  -   &  -   &  -   &  -   &  -   & $\,\,\,\,\,\,6.887^{+0.472}_{-0.442}$  & $\,\,\,\,\,\,7.047^{+0.463}_{-0.433}$  & 6 \\[+0.10cm]
Pal 8  &  -   &  -   &  -   &  -   &  -   & $\,\,\,11.695^{+0.722}_{-0.680}$  & $\,\,\,11.316^{+0.652}_{-0.611}$  & 4 \\[+0.10cm]
Pal 10  &  -   &  -   &  -   &  -   &  -   & $\,\,\,\,\,\,8.402^{+1.621}_{-1.359}$  & $\,\,\,\,\,\,8.944^{+1.292}_{-1.072}$  & 3 \\[+0.10cm]
Pal 11  &  -   &  -   &  -   &  -   &  -   & $\,\,\,14.125^{+0.523}_{-0.505}$  & $\,\,\,14.024^{+0.517}_{-0.498}$  & 3 \\[+0.10cm]
Pal 12  &  -   &  -   &  -   & $\,\,\,18.281^{+0.686}_{-0.661}$  &  -   & $\,\,\,18.484^{+0.300}_{-0.296}$  & $\,\,\,18.494^{+0.300}_{-0.295}$  & 8 \\[+0.10cm]
Pal 13  &  -   &  -   &  -   &  -   &  -   & $\,\,\,23.475^{+0.403}_{-0.397}$  & $\,\,\,23.475^{+0.403}_{-0.397}$  & 6 \\[+0.10cm]
Pal 14  &  -   &  -   &  -   &  -   &  -   & $\,\,\,73.587^{+1.645}_{-1.609}$  & $\,\,\,73.579^{+1.645}_{-1.609}$  & 4 \\[+0.10cm]
\hline
\end{tabular}
\end{table*}
\begin{table*}
\contcaption{}
\begin{tabular}{@{}l@{\hspace*{0.1cm}}c@{\hspace*{0.1cm}}c@{\hspace*{0.1cm}}c@{\hspace*{0.1cm}}c@{\hspace*{0.1cm}}c@{\hspace*{0.1cm}}c@{\hspace*{0.1cm}}c@{\hspace*{0.1cm}}c}
\hline
\multirow{2}{*}{Name} & GEDR3 parallax dist. &   GEDR3 kin. dist. & HST kin. dist. & Subdwarf dist. & Star Count dist. & Lit. dist. & Mean distance & \multirow{2}{*}{N$_{\mbox{Tot}}$}\\
 & [kpc] & [kpc] & [kpc] & [kpc] & [kpc] & [kpc] & [kpc]  & \\
\hline
Pal 15  &  -   &  -   &  -   & $\,\,\,42.855^{+2.227}_{-2.117}$  &  -   & $\,\,\,44.096^{+1.152}_{-1.123}$  & $\,\,\,44.096^{+1.152}_{-1.123}$  & 4 \\[+0.10cm]
Pyxis  &  -   &  -   &  -   & $\,\,\,37.670^{+1.414}_{-1.363}$  &  -   & $\,\,\,36.526^{+0.662}_{-0.650}$  & $\,\,\,36.526^{+0.662}_{-0.650}$  & 6 \\[+0.10cm]
RLGC 1  &  -   &  -   &  -   &  -   &  -   & $\,\,\,28.840^{+4.288}_{-3.733}$  & $\,\,\,28.840^{+4.288}_{-3.733}$  & 1 \\[+0.10cm]
RLGC 2  &  -   &  -   &  -   &  -   &  -   & $\,\,\,15.849^{+2.356}_{-2.051}$  & $\,\,\,15.849^{+2.356}_{-2.051}$  & 1 \\[+0.10cm]
Rup 106  &  -   &  -   &  -   & $\,\,\,19.588^{+0.735}_{-0.709}$  &  -   & $\,\,\,20.730^{+0.366}_{-0.360}$  & $\,\,\,20.711^{+0.365}_{-0.359}$  & 6 \\[+0.10cm]
Sgr II  &  -   &  -   &  -   & $\,\,\,66.988^{+2.514}_{-2.423}$  &  -   & $\,\,\,66.527^{+1.581}_{-1.544}$  & $\,\,\,66.527^{+1.581}_{-1.544}$  & 5 \\[+0.10cm]
Ter 1  &  -   &  -   &  -   &  -   &  -   & $\,\,\,\,\,\,5.673^{+0.175}_{-0.170}$  & $\,\,\,\,\,\,5.673^{+0.175}_{-0.170}$  & 6 \\[+0.10cm]
Ter 2  &  -   &  -   &  -   &  -   &  -   & $\,\,\,\,\,\,7.681^{+0.336}_{-0.322}$  & $\,\,\,\,\,\,7.753^{+0.332}_{-0.318}$  & 4 \\[+0.10cm]
Ter 3  &  -   &  -   &  -   &  -   &  -   & $\,\,\,\,\,\,7.568^{+0.338}_{-0.324}$  & $\,\,\,\,\,\,7.644^{+0.318}_{-0.304}$  & 3 \\[+0.10cm]
Ter 4  &  -   &  -   &  -   &  -   &  -   & $\,\,\,\,\,\,7.558^{+0.316}_{-0.304}$  & $\,\,\,\,\,\,7.591^{+0.315}_{-0.302}$  & 5 \\[+0.10cm]
Ter 5  &  -   & $\,\,\,\,\,\,6.780 \pm 0.220$  &  -   &  -   & $\,\,\,\,\,\,6.346 \pm 0.503$  & $\,\,\,\,\,\,6.462^{+0.224}_{-0.217}$  & $\,\,\,\,\,\,6.617^{+0.150}_{-0.148}$  & 10 \\[+0.10cm]
Ter 6  &  -   &  -   &  -   &  -   &  -   & $\,\,\,\,\,\,7.271^{+0.360}_{-0.343}$  & $\,\,\,\,\,\,7.271^{+0.360}_{-0.343}$  & 4 \\[+0.10cm]
Ter 7  &  -   &  -   &  -   &  -   &  -   & $\,\,\,24.277^{+0.497}_{-0.487}$  & $\,\,\,24.278^{+0.496}_{-0.487}$  & 5 \\[+0.10cm]
Ter 8  &  -   &  -   &  -   & $\,\,\,27.669^{+1.038}_{-1.001}$  &  -   & $\,\,\,27.542^{+0.422}_{-0.415}$  & $\,\,\,27.535^{+0.421}_{-0.415}$  & 8 \\[+0.10cm]
Ter 9  &  -   & $\,\,\,\,\,\,5.030 \pm 1.037$  &  -   &  -   &  -   & $\,\,\,\,\,\,5.605^{+0.434}_{-0.403}$  & $\,\,\,\,\,\,5.770^{+0.356}_{-0.333}$  & 5 \\[+0.10cm]
Ter 10  &  -   &  -   &  -   &  -   &  -   & $\,\,\,10.247^{+0.414}_{-0.398}$  & $\,\,\,10.212^{+0.412}_{-0.396}$  & 3 \\[+0.10cm]
Ter 12  &  -   & $\,\,\,\,\,\,5.002 \pm 1.851$  &  -   &  -   &  -   & $\,\,\,\,\,\,5.107^{+0.444}_{-0.408}$  & $\,\,\,\,\,\,5.166^{+0.400}_{-0.369}$  & 4 \\[+0.10cm]
Ton 2  &  -   &  -   &  -   &  -   &  -   & $\,\,\,\,\,\,6.934^{+0.360}_{-0.343}$  & $\,\,\,\,\,\,6.987^{+0.344}_{-0.326}$  & 3 \\[+0.10cm]
UKS 1  &  -   &  -   &  -   &  -   &  -   & $\,\,\,15.581^{+0.570}_{-0.550}$  & $\,\,\,15.581^{+0.570}_{-0.550}$  & 2 \\[+0.10cm]
VVV-CL001  &  -   &  -   &  -   &  -   &  -   & $\,\,\,\,\,\,8.204^{+1.664}_{-1.383}$  & $\,\,\,\,\,\,8.082^{+1.622}_{-1.343}$  & 1 \\[+0.10cm]
Whiting 1  &  -   &  -   &  -   &  -   &  -   & $\,\,\,30.591^{+1.192}_{-1.147}$  & $\,\,\,30.591^{+1.192}_{-1.147}$  & 2 \\[+0.10cm]
Pal 6  &  -   &  -   &  -   &  -   &  -   & $\,\,\,\,\,\,6.887^{+0.472}_{-0.442}$  & $\,\,\,\,\,\,7.047^{+0.463}_{-0.433}$  & 6 \\[+0.10cm]
\hline
\end{tabular}
\end{table*}

\label{lastpage}

\newpage

\section*{Appendix A: Comparison of distance determinations for globular clusters}

\begin{figure*}
\begin{center}
\includegraphics[width=0.99\textwidth]{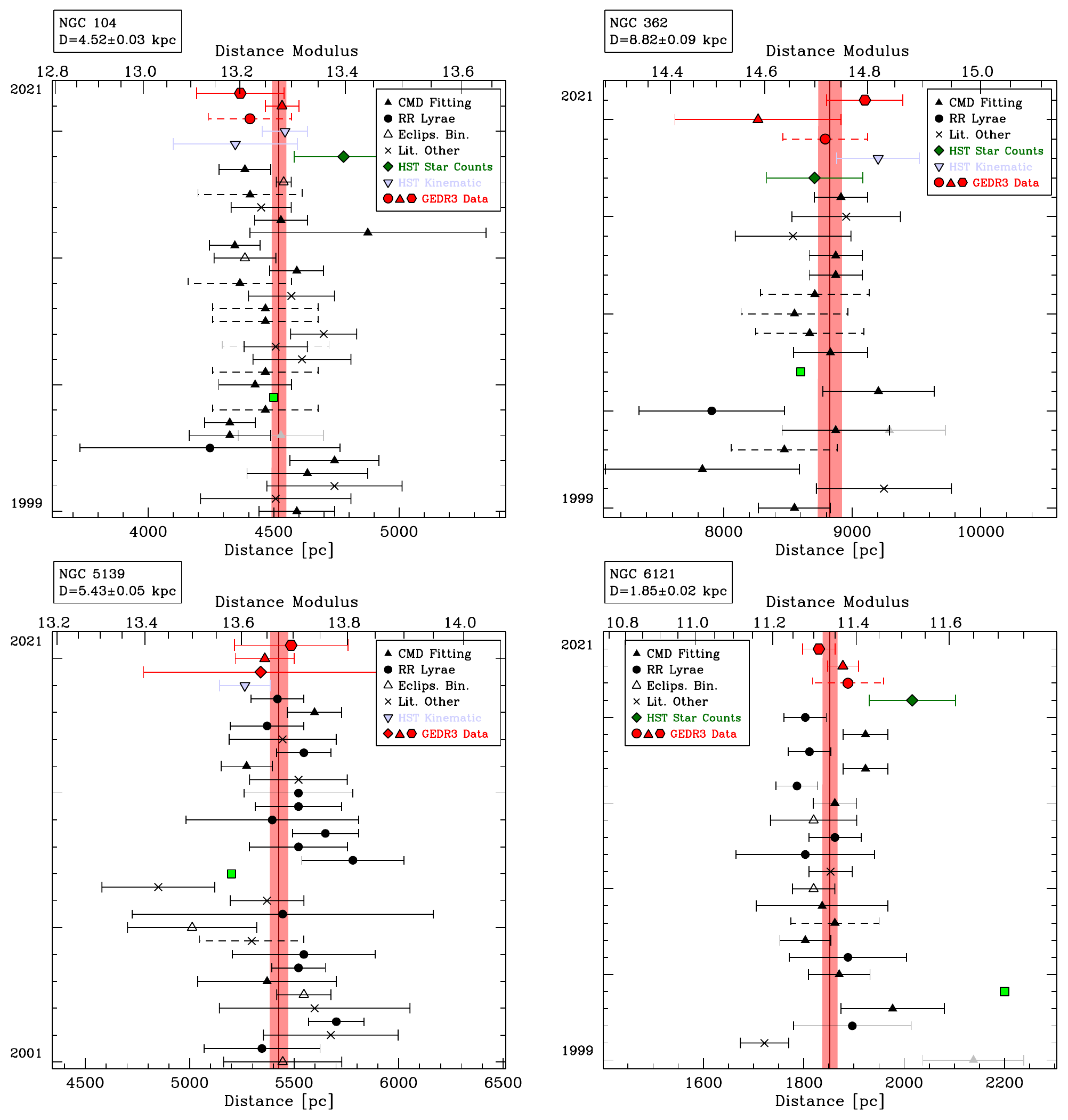}
\end{center}
\vspace*{-0.2cm}
\caption{Comparison of recent literature distance determinations for the globular clusters NGC~104, NGC~362, NGC~5139 and NGC~6121 with our data. Black symbols depict distance determinations taken from the literature, while red 
 symbols show our own measurements based on \gaia EDR3 data. Red triangles show \gaia EDR3 kinematic distances, red circles subdwarf distances, red squares  
 moving group distances and red sexagons show \gaia EDR3 parallax distances. Blue triangles show HST kinematic distances and dark green squares with error bars star count distances. The bright green square is 
  the distance given by Harris (2010) based on an averaging of literature values up to 2010. The distances are sorted according to publication year, with newest measurements at the top
   of each panel and the first and last year indicated. Distance determinations shown in grey have been neglected when calculating the mean cluster distance. For literature distances 
    without an error, the dashed error bars show our adopted error. Solid error bars are literature measurements with error bars. The red vertical line and shaded region marks the 
     average distance and its 1$\sigma$ error bar. The average distances are also given at the top of each panel.
}
\label{fig:appfirst}
\end{figure*}

Figs. \ref{fig:appfirst} to \ref{fig:applast} depict the individual distances determinations to globular clusters used in this work. Only clusters
with more than 5 measurements are shown.

\begin{figure*}
\begin{center}
\includegraphics[width=0.87\textwidth]{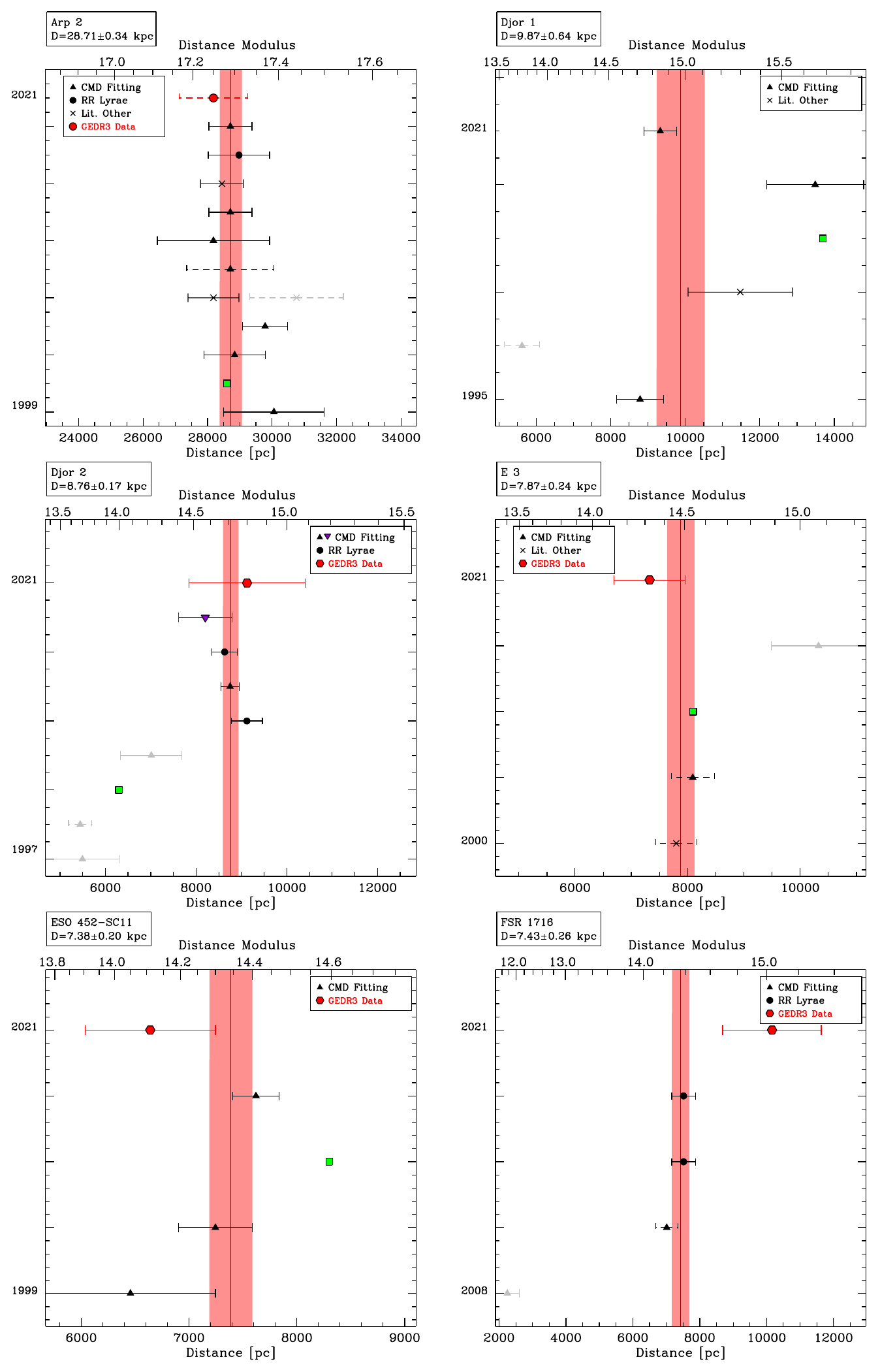}
\end{center}
\vspace*{-0.2cm}
\caption{Same as Fig.~\ref{fig:appfirst} for Arp~2, Djor~1, Djor~2 and E~3, ESO~452-SC11 and FSR 1716.\hspace{8cm}}
\end{figure*}

\begin{figure*}
\begin{center}
\includegraphics[width=0.87\textwidth]{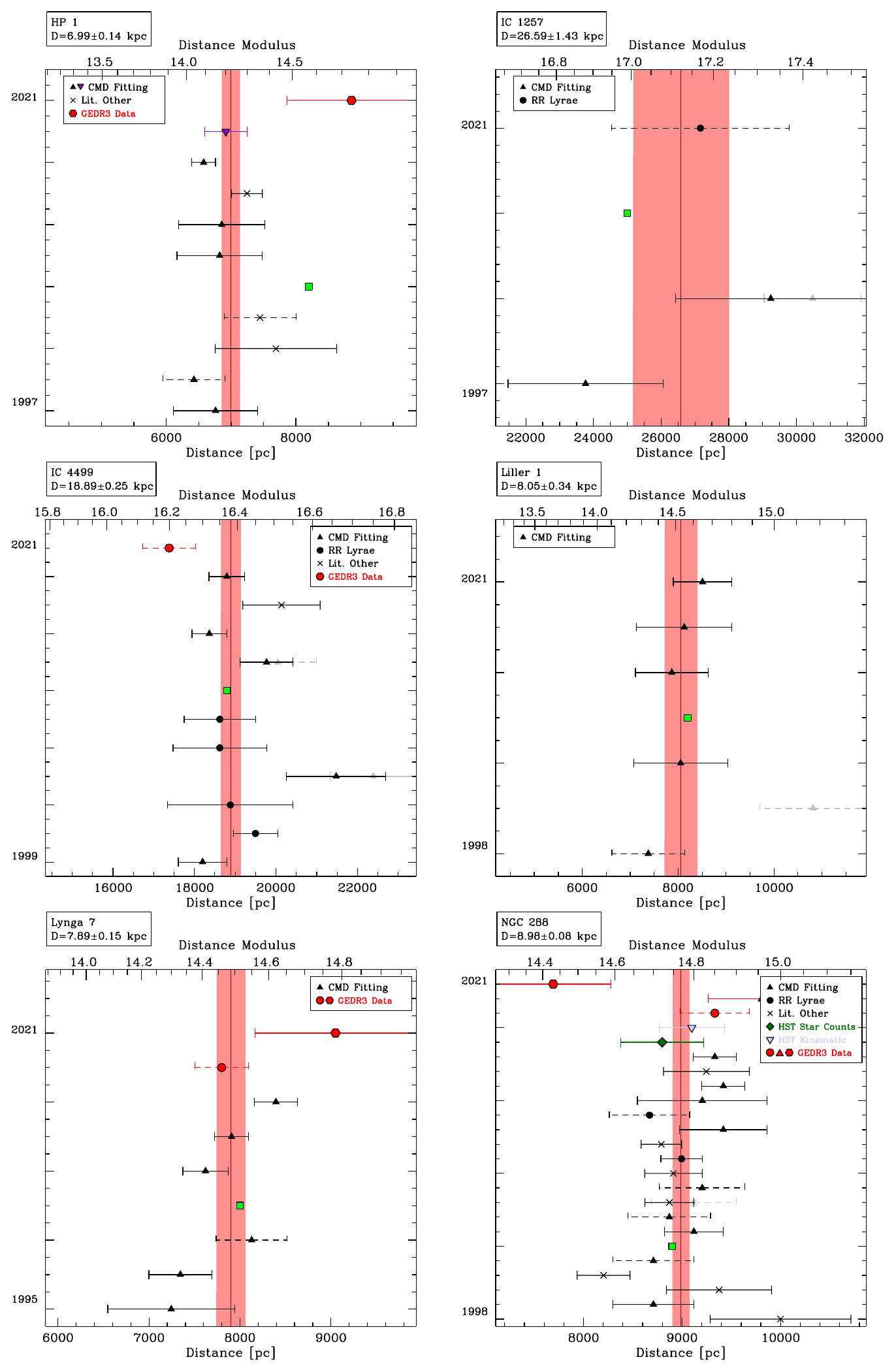}
\end{center}
\vspace*{-0.2cm}
\caption{Same as Fig.~\ref{fig:appfirst} for HP 1, IC 1257, IC 4499 and Liller 1, Lynga 7 and NGC 288.\hspace{7cm}}
\end{figure*}

\begin{figure*}
\begin{center}
\includegraphics[width=0.87\textwidth]{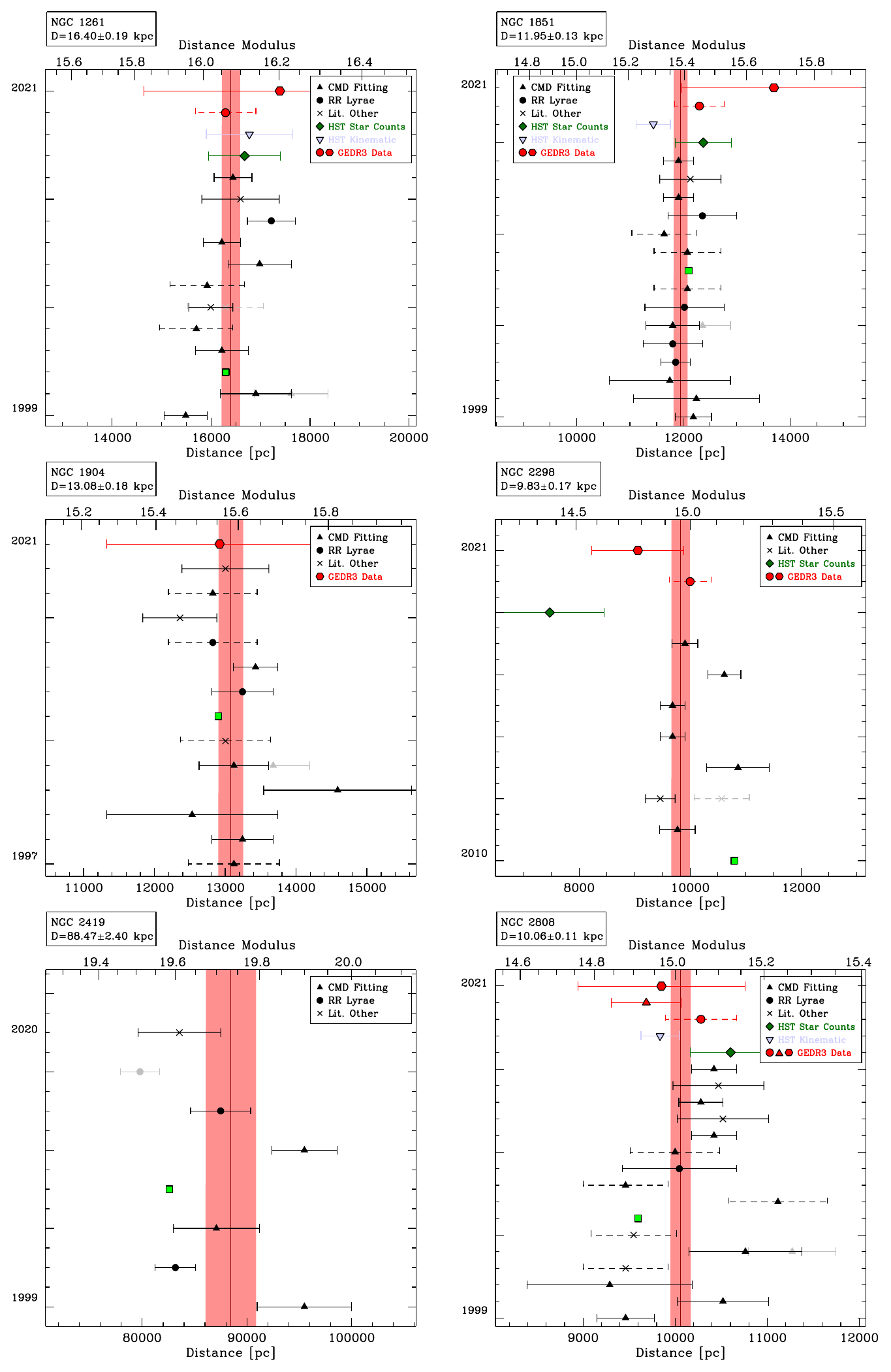}
\end{center}
\vspace*{-0.2cm}
\caption{Same as Fig.~\ref{fig:appfirst} for NGC 1261, NGC 1851, NGC 1904, NGC 2298, NGC 2419 and NGC 2808.\hspace{7cm}}
\end{figure*}

\begin{figure*}
\begin{center}
\includegraphics[width=0.87\textwidth]{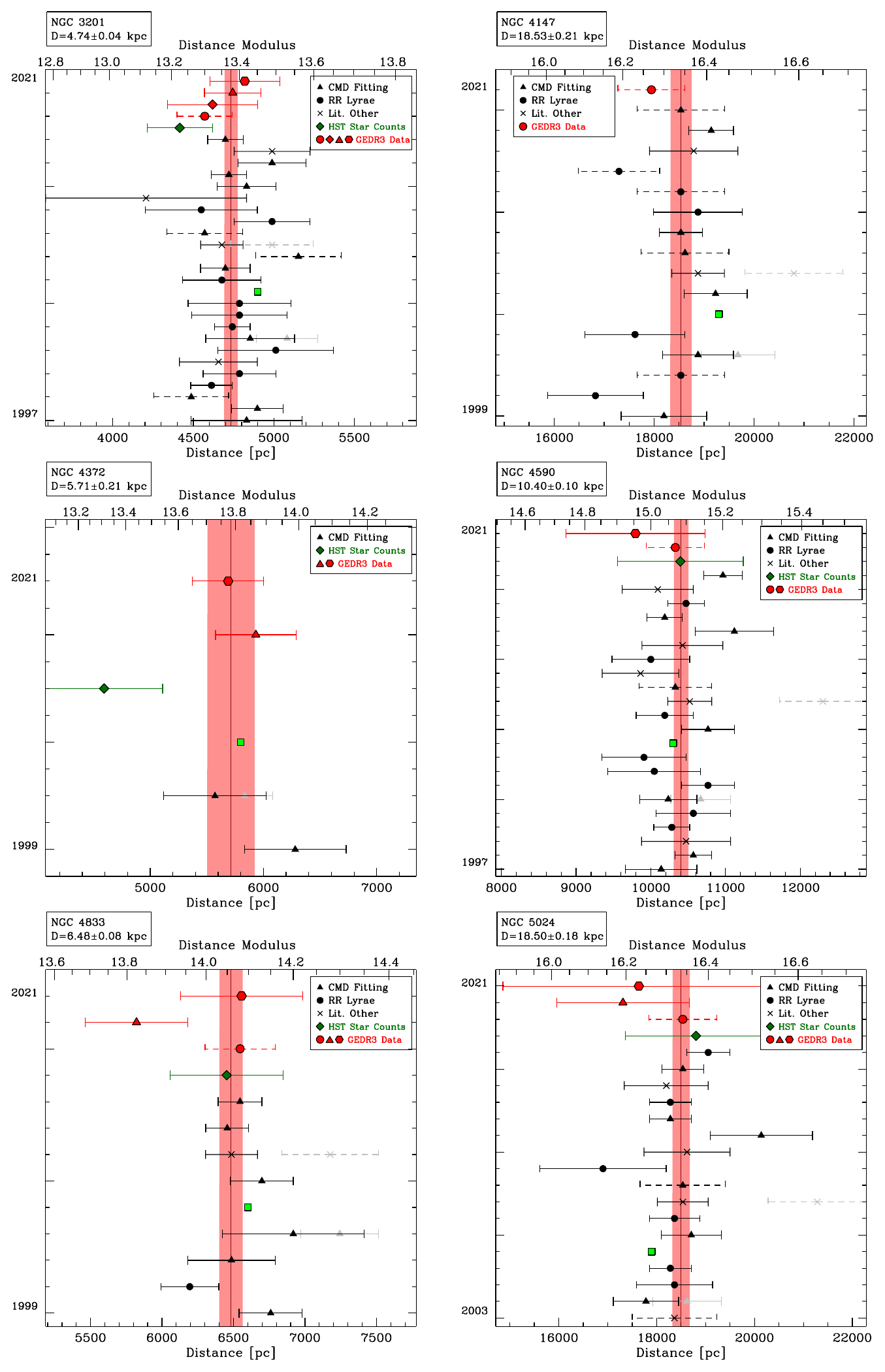}
\end{center}
\vspace*{-0.2cm}
\caption{Same as Fig.~\ref{fig:appfirst} for NGC 3201, NGC 4147, NGC 4372, NGC 4590, NGC 4833 and NGC 5024.\hspace{7cm}} 
\end{figure*}

\begin{figure*}
\begin{center}
\includegraphics[width=0.87\textwidth]{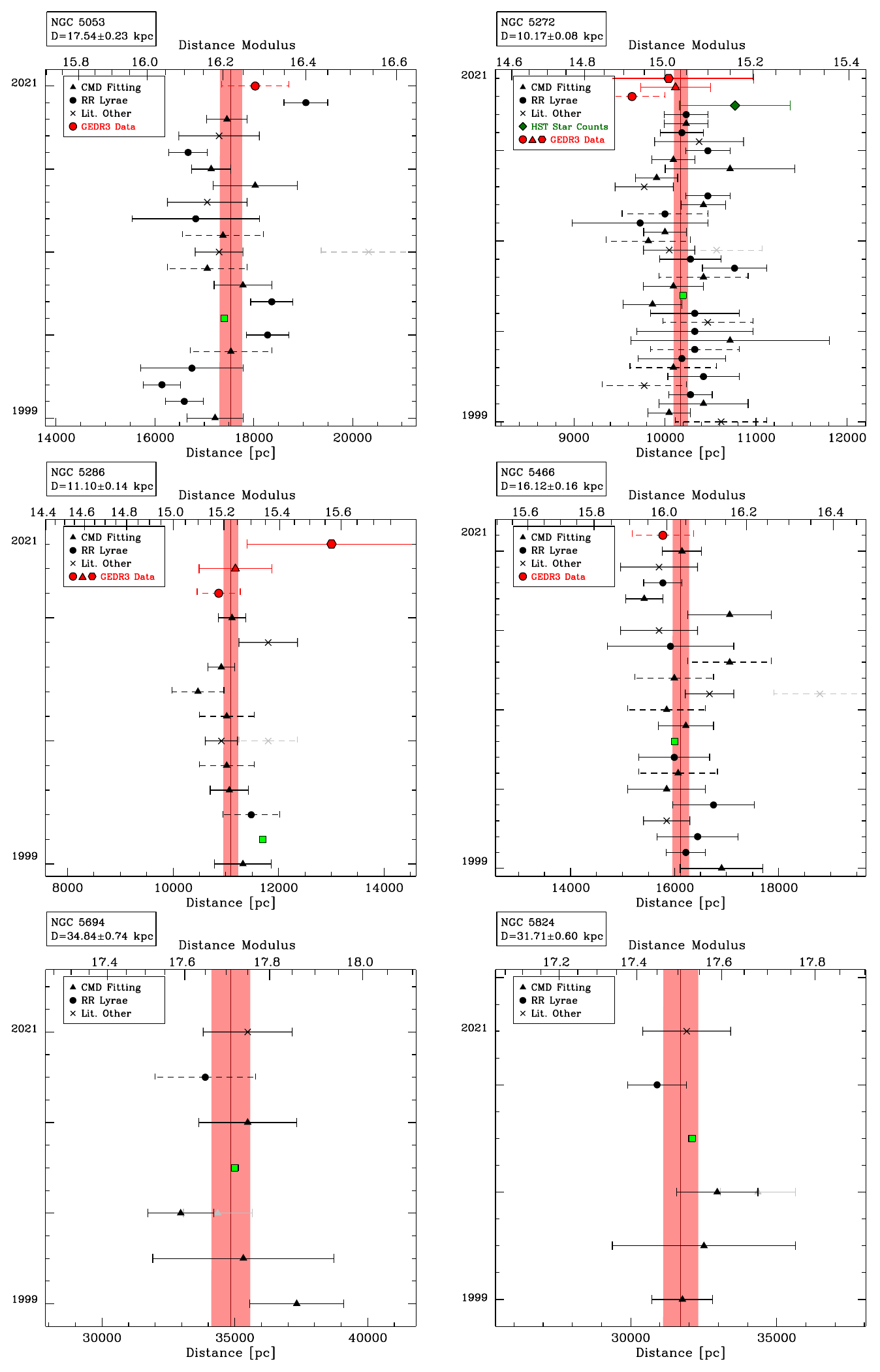}
\end{center}
\vspace*{-0.2cm}
\caption{Same as Fig.~\ref{fig:appfirst} for NGC 5053, NGC 5272, NGC 5286, NGC 5466, NGC 5694 and NGC 5824.\hspace{7cm}} 
\end{figure*}

\begin{figure*}
\begin{center}
\includegraphics[width=0.87\textwidth]{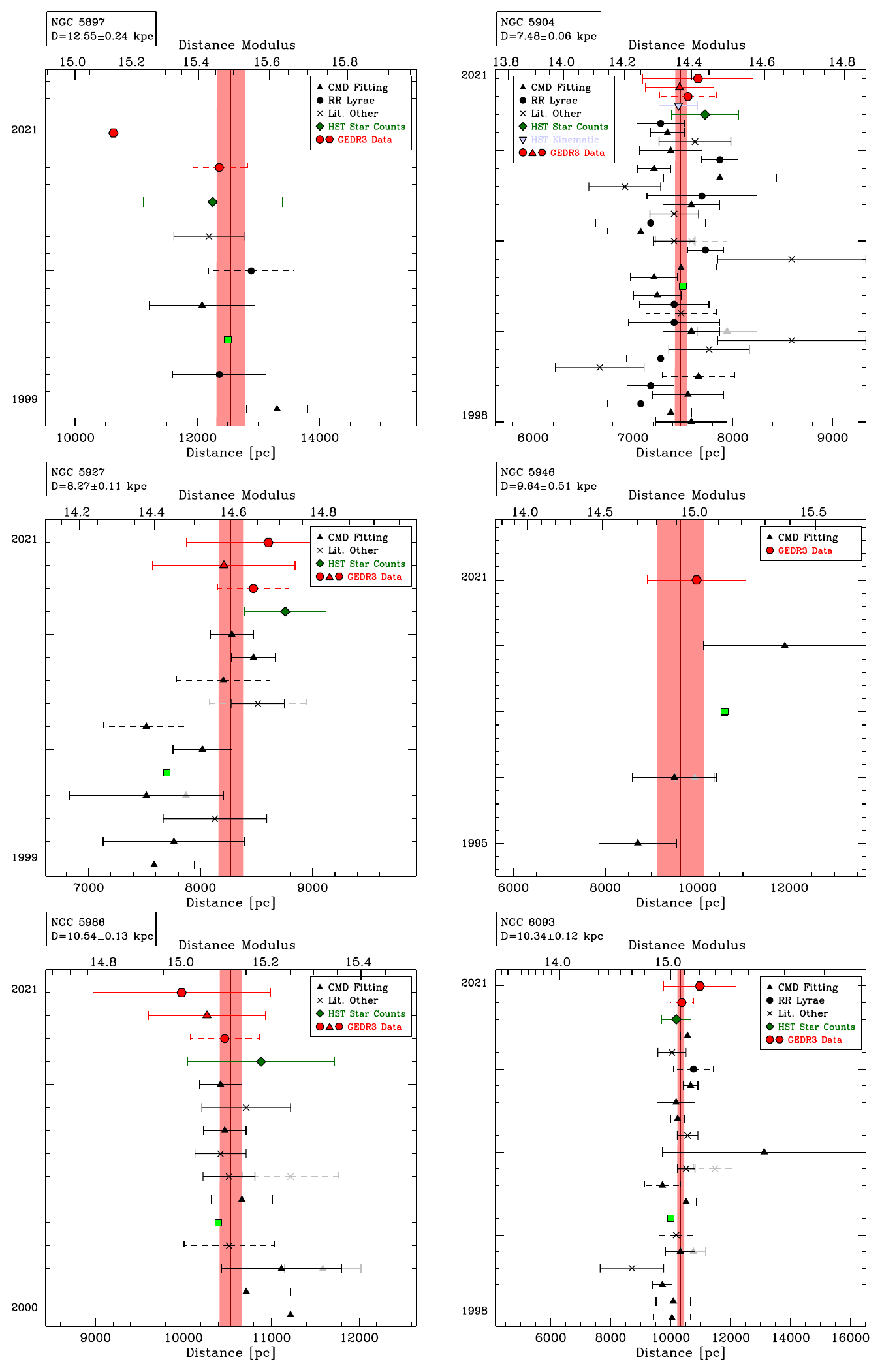}
\end{center}
\vspace*{-0.2cm}
\caption{Same as Fig.~\ref{fig:appfirst} for NGC 5897, NGC 5904, NGC 5927, NGC 5946, NGC 5986 and NGC 6093.\hspace{7cm}}
\end{figure*}

\begin{figure*}
\begin{center}
\includegraphics[width=0.87\textwidth]{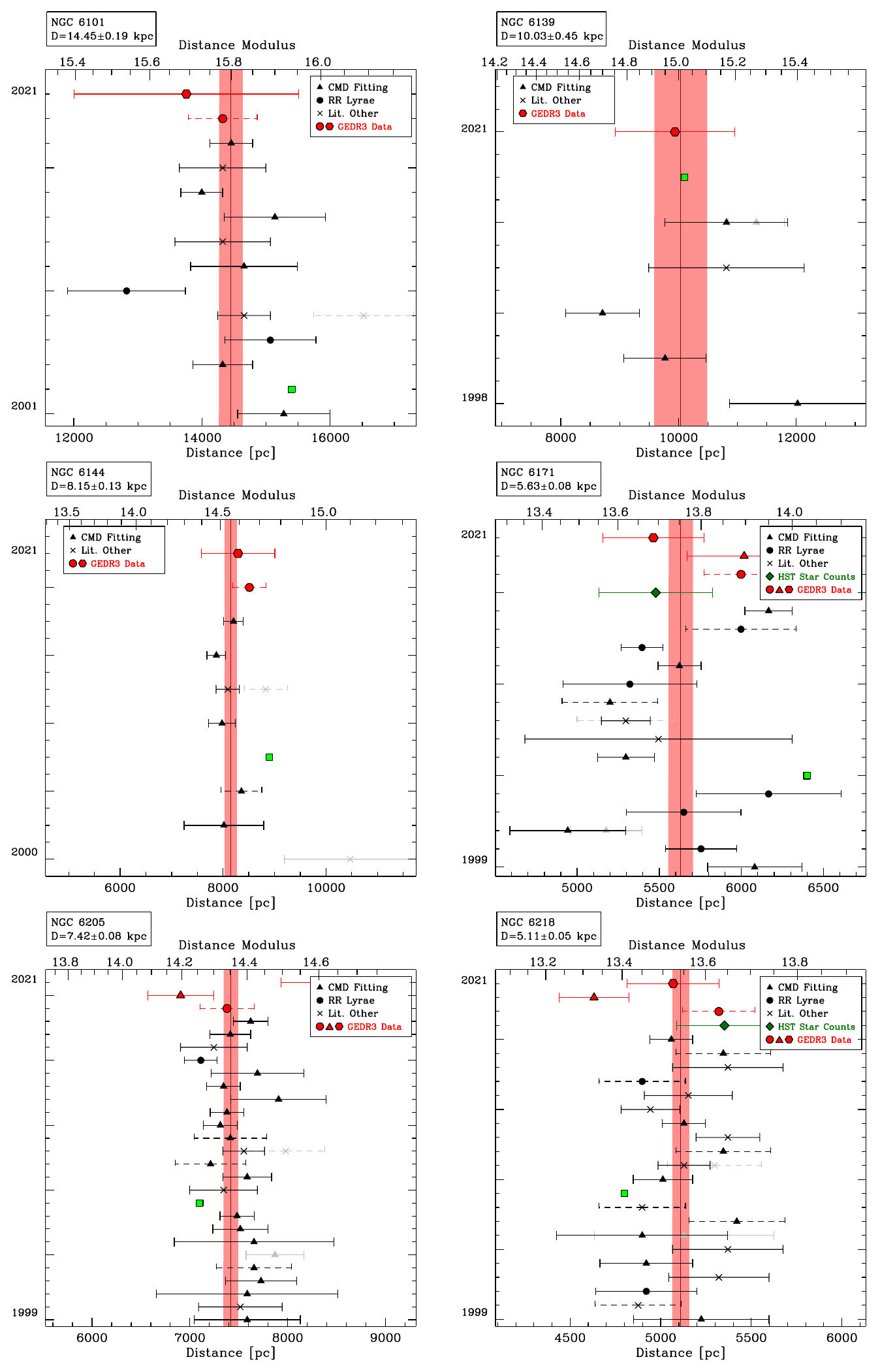}
\end{center}
\vspace*{-0.2cm}
\caption{Same as Fig.~\ref{fig:appfirst} for NGC 6101, NGC 6139, NGC 6144 and NGC 6171, NGC 6205 and NGC 6218.\hspace{7cm}}
\end{figure*}

\begin{figure*}
\begin{center}
\includegraphics[width=0.87\textwidth]{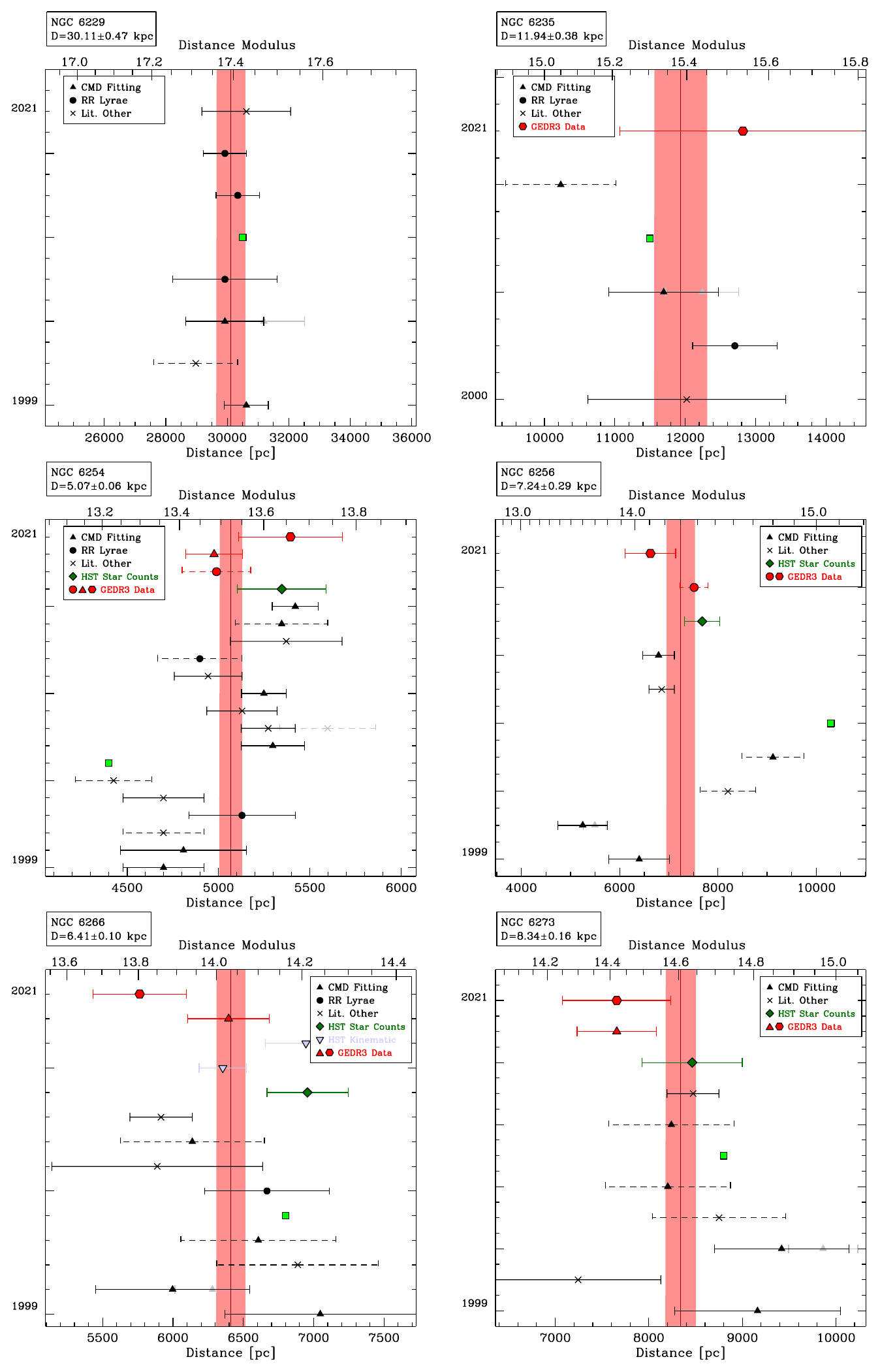}
\end{center}
\vspace*{-0.2cm}
\caption{Same as Fig.~\ref{fig:appfirst} for NGC 6229, NGC 6235, NGC 6254 and NGC 6256, NGC 6266 and NGC 6273.\hspace{7cm}}
\end{figure*}

\begin{figure*}
\begin{center}
\includegraphics[width=0.87\textwidth]{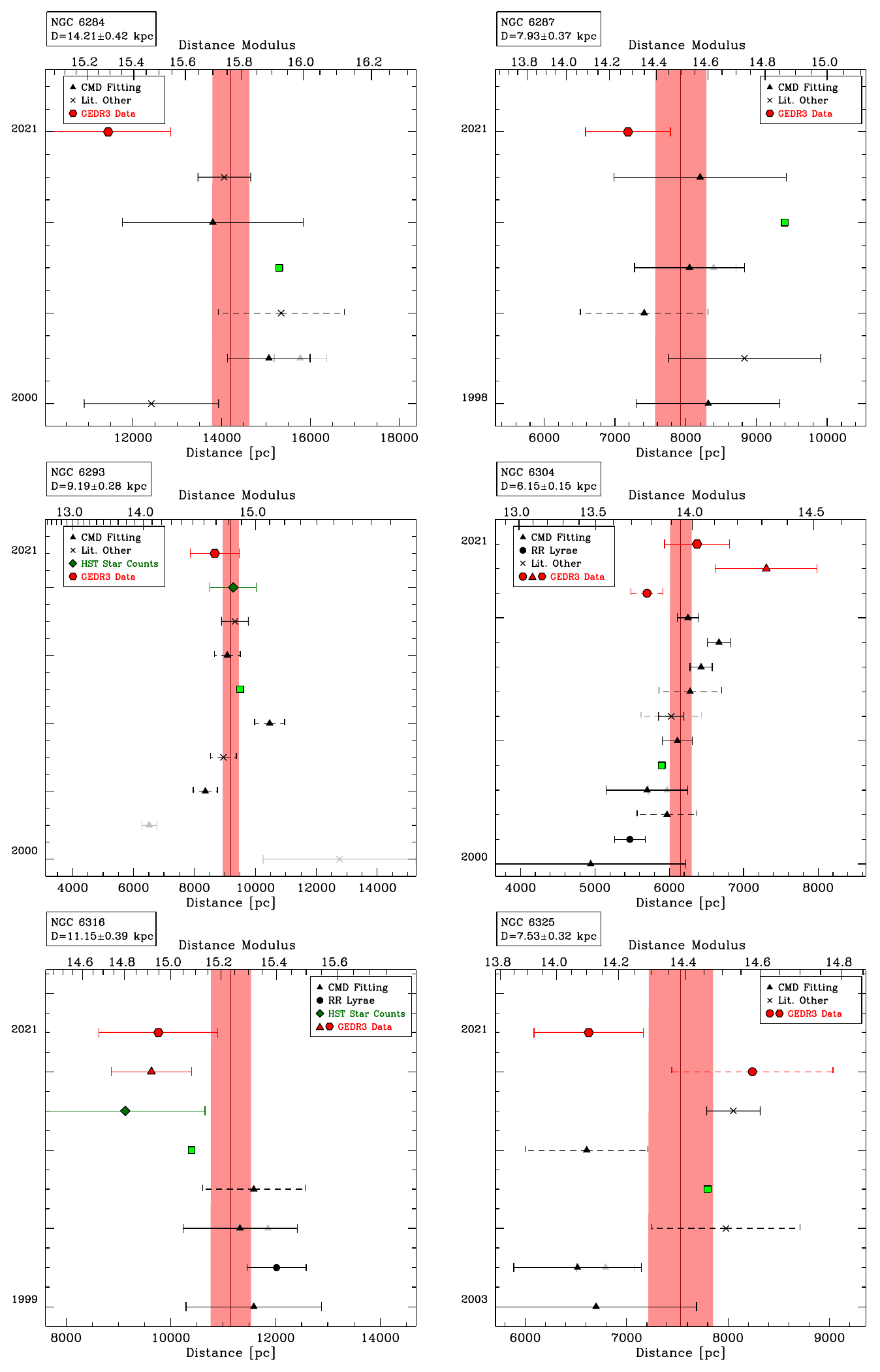}
\end{center}
\vspace*{-0.2cm}
\caption{Same as Fig.~\ref{fig:appfirst} for NGC 6284, NGC 6287, NGC 6293 and NGC 6304, NGC 6316 and NGC 6325.\hspace{7cm}}
\end{figure*}

\begin{figure*}
\begin{center}
\includegraphics[width=0.87\textwidth]{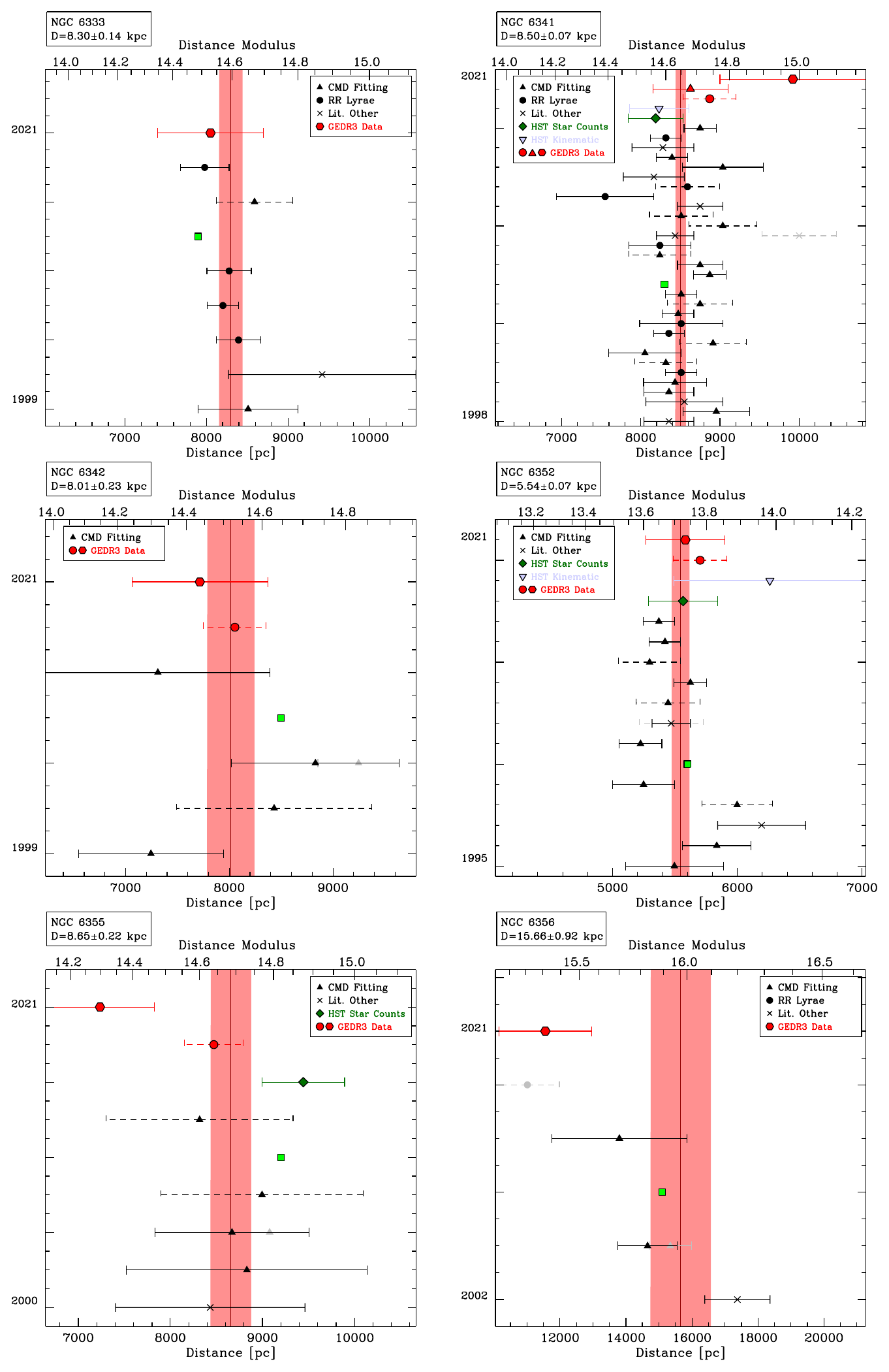}
\end{center}
\vspace*{-0.2cm}
\caption{Same as Fig.~\ref{fig:appfirst} for NGC 6333, NGC 6341, NGC 6342 and NGC 6352, NGC 6355 and NGC 6356.\hspace{7cm}}
\end{figure*}

\begin{figure*}
\begin{center}
\includegraphics[width=0.87\textwidth]{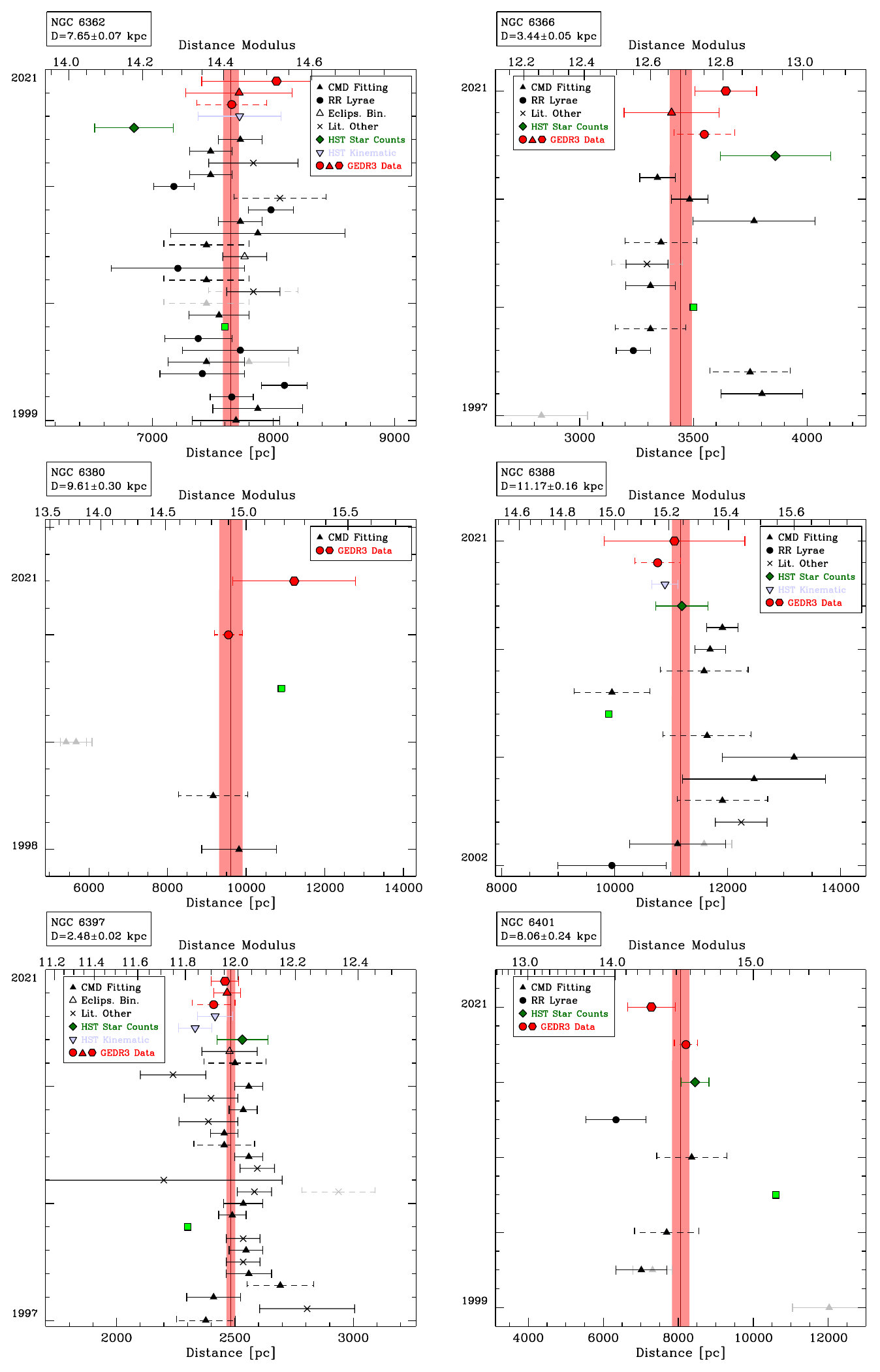}
\end{center}
\vspace*{-0.2cm}
\caption{Same as Fig.~\ref{fig:appfirst} for NGC 6362, NGC 6366, NGC 6380 and NGC 6388, NGC 6397 and NGC 6401.\hspace{7cm}}
\end{figure*}

\begin{figure*}
\begin{center}
\includegraphics[width=0.87\textwidth]{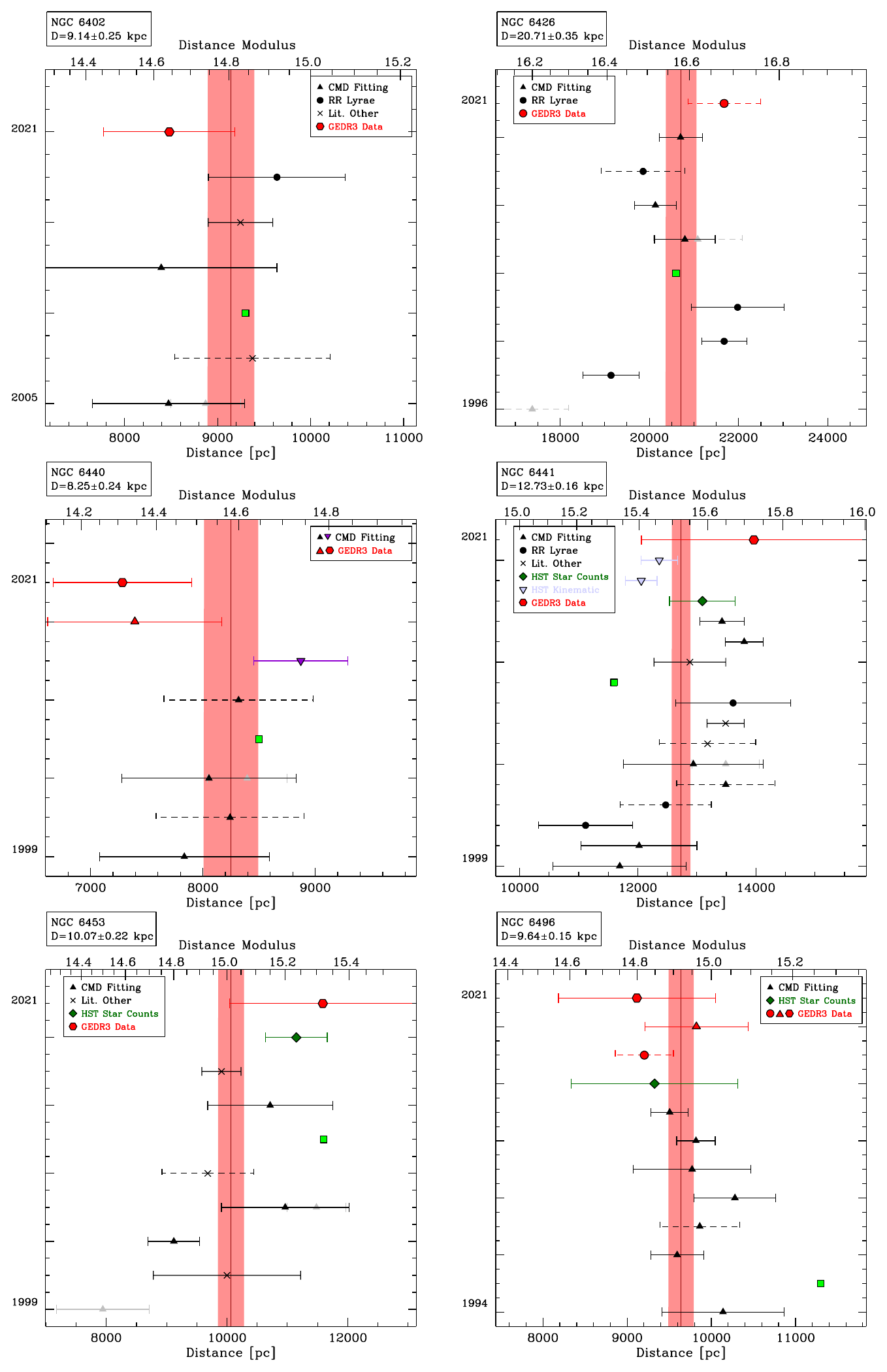}
\end{center}
\vspace*{-0.2cm}
\caption{Same as Fig.~\ref{fig:appfirst} for NGC 6402, NGC 6426, NGC 6440 and NGC 6441, NGC 6453 and NGC 6496.\hspace{7cm}}
\end{figure*}

\begin{figure*}
\begin{center}
\includegraphics[width=0.87\textwidth]{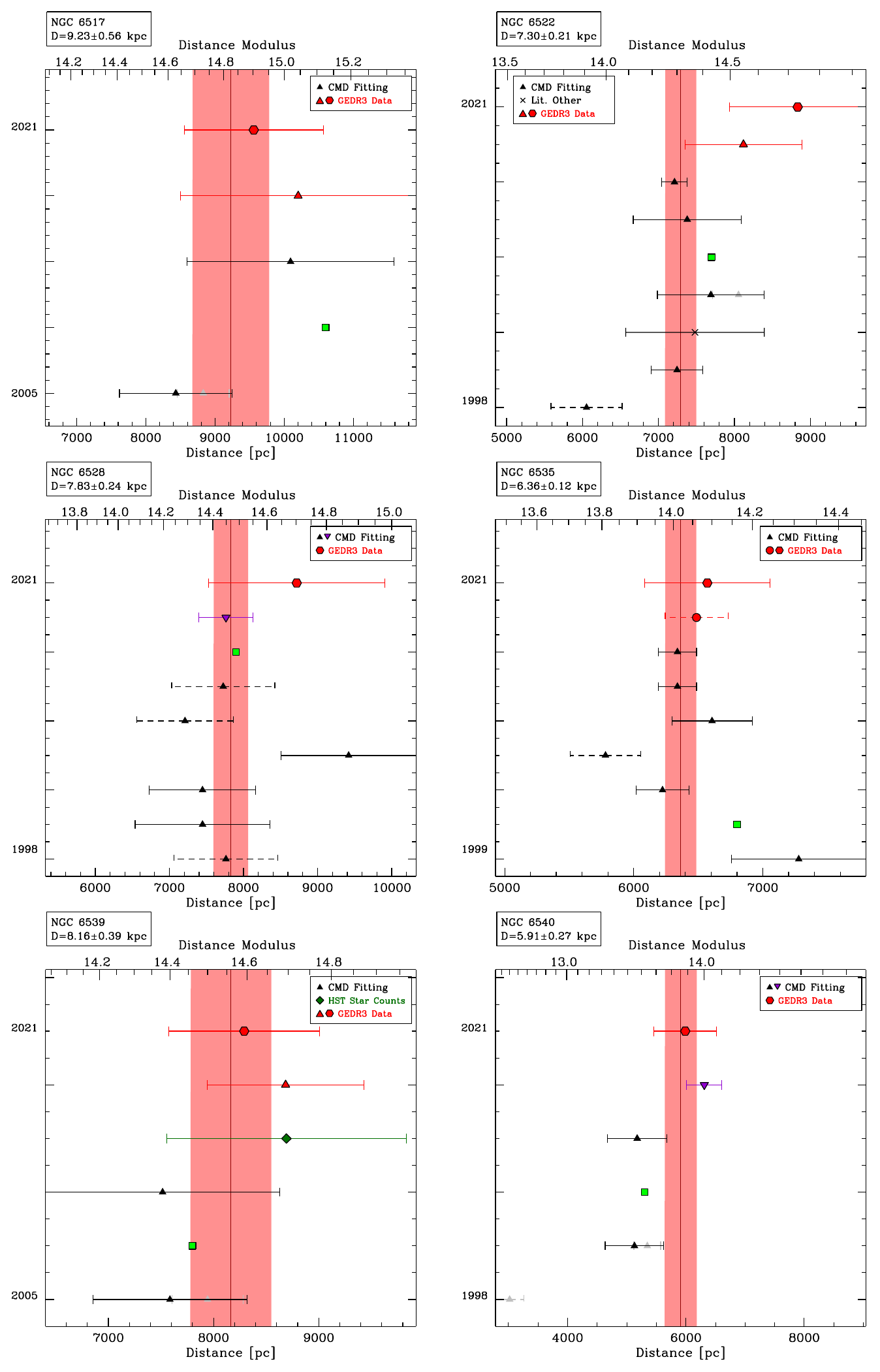}
\end{center}
\vspace*{-0.2cm}
\caption{Same as Fig.~\ref{fig:appfirst} for NGC 6517, NGC 6522, NGC 6528 and NGC 6535, NGC 6539 and NGC 6540.\hspace{7cm}}
\end{figure*}

\begin{figure*}
\begin{center}
\includegraphics[width=0.87\textwidth]{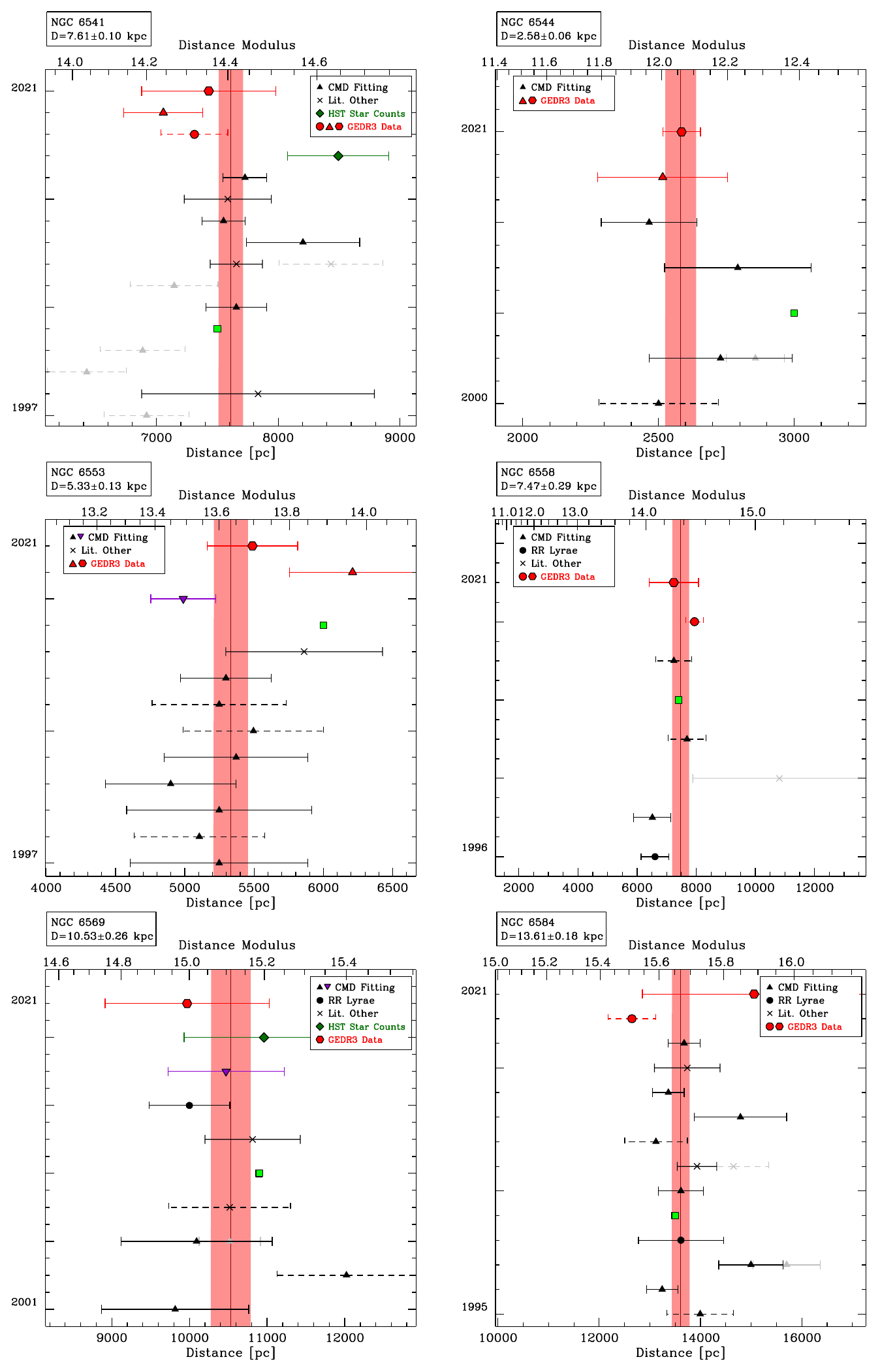}
\end{center}
\vspace*{-0.2cm}
\caption{Same as Fig.~\ref{fig:appfirst} for NGC 6541, NGC 6544, NGC 6553 and NGC 6558, NGC 6569 and NGC 6584.\hspace{7cm}}
\end{figure*}

\begin{figure*}
\begin{center}
\includegraphics[width=0.87\textwidth]{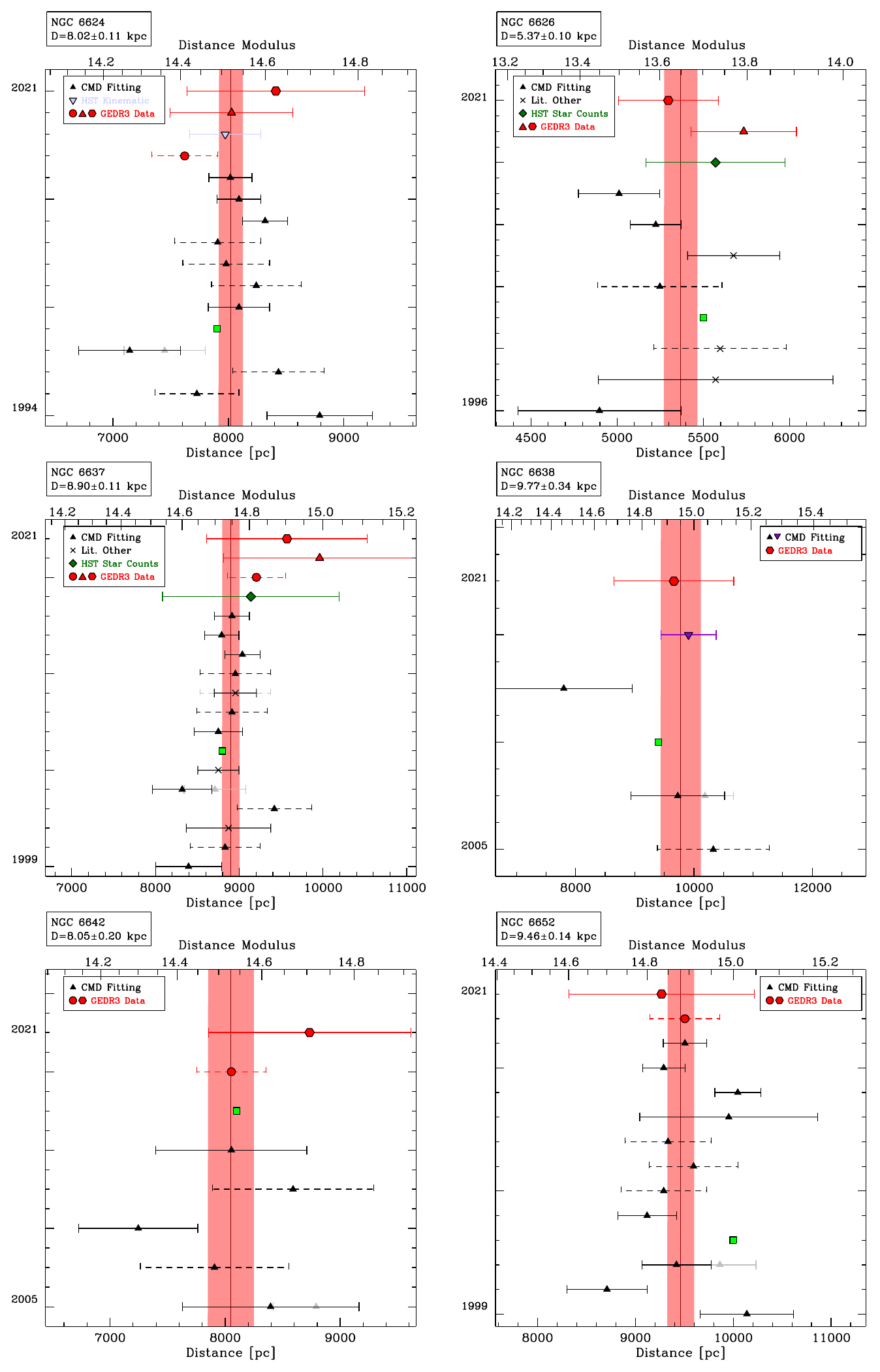}
\end{center}
\vspace*{-0.2cm}
\caption{Same as Fig.~\ref{fig:appfirst} for NGC 6624, NGC 66226 NGC 6637 and NGC 6638, NGC 6642 and NGC 6652.\hspace{7cm}}
\end{figure*}

\begin{figure*}
\begin{center}
\includegraphics[width=0.87\textwidth]{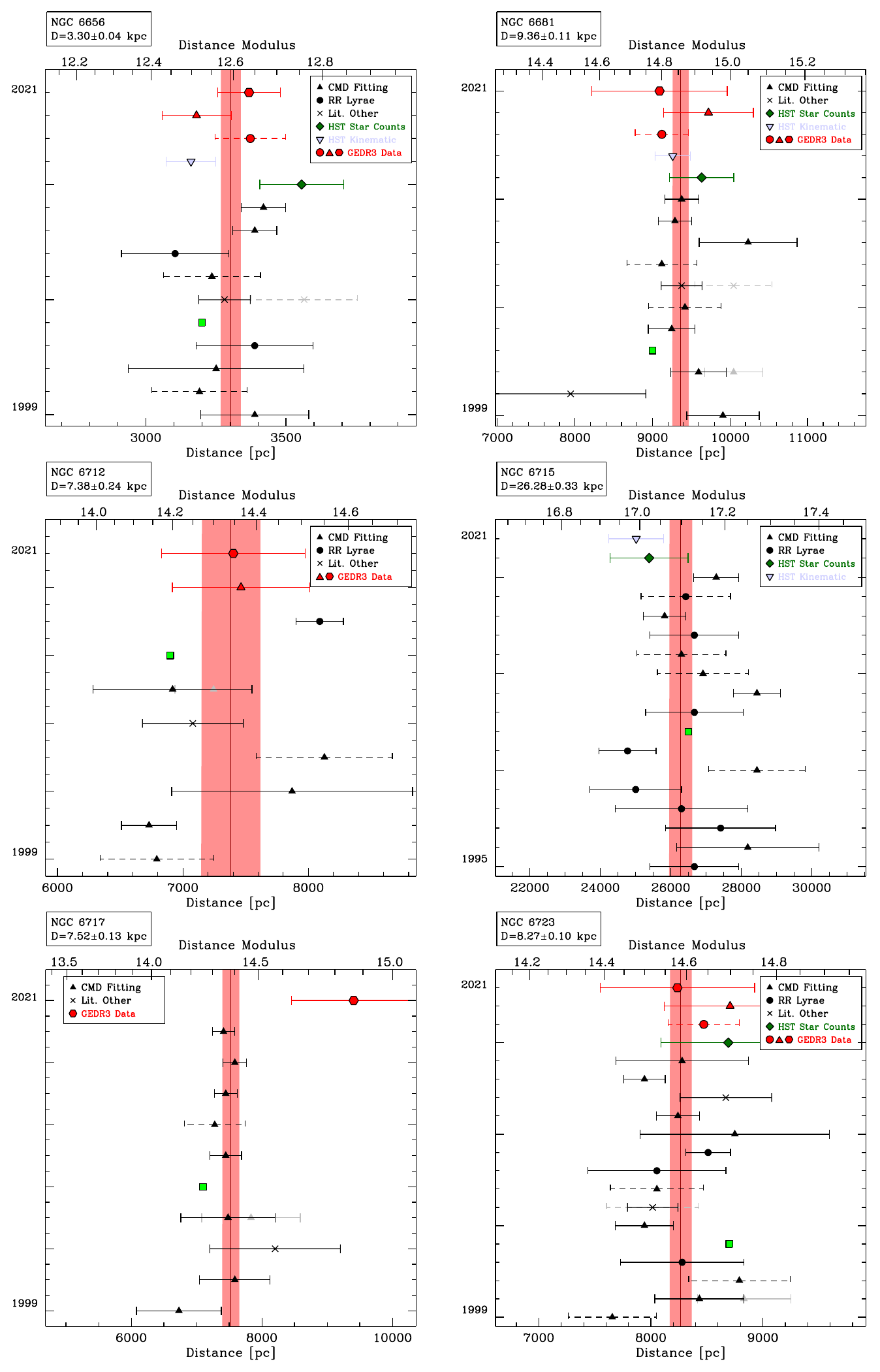}
\end{center}
\vspace*{-0.2cm}
\caption{Same as Fig.~\ref{fig:appfirst} for NGC 6656, NGC 6681, NGC 6712 and NGC 6715, NGC 6717 and NGC 6723.\hspace{7cm}}
\end{figure*}

\begin{figure*}
\begin{center}
\includegraphics[width=0.87\textwidth]{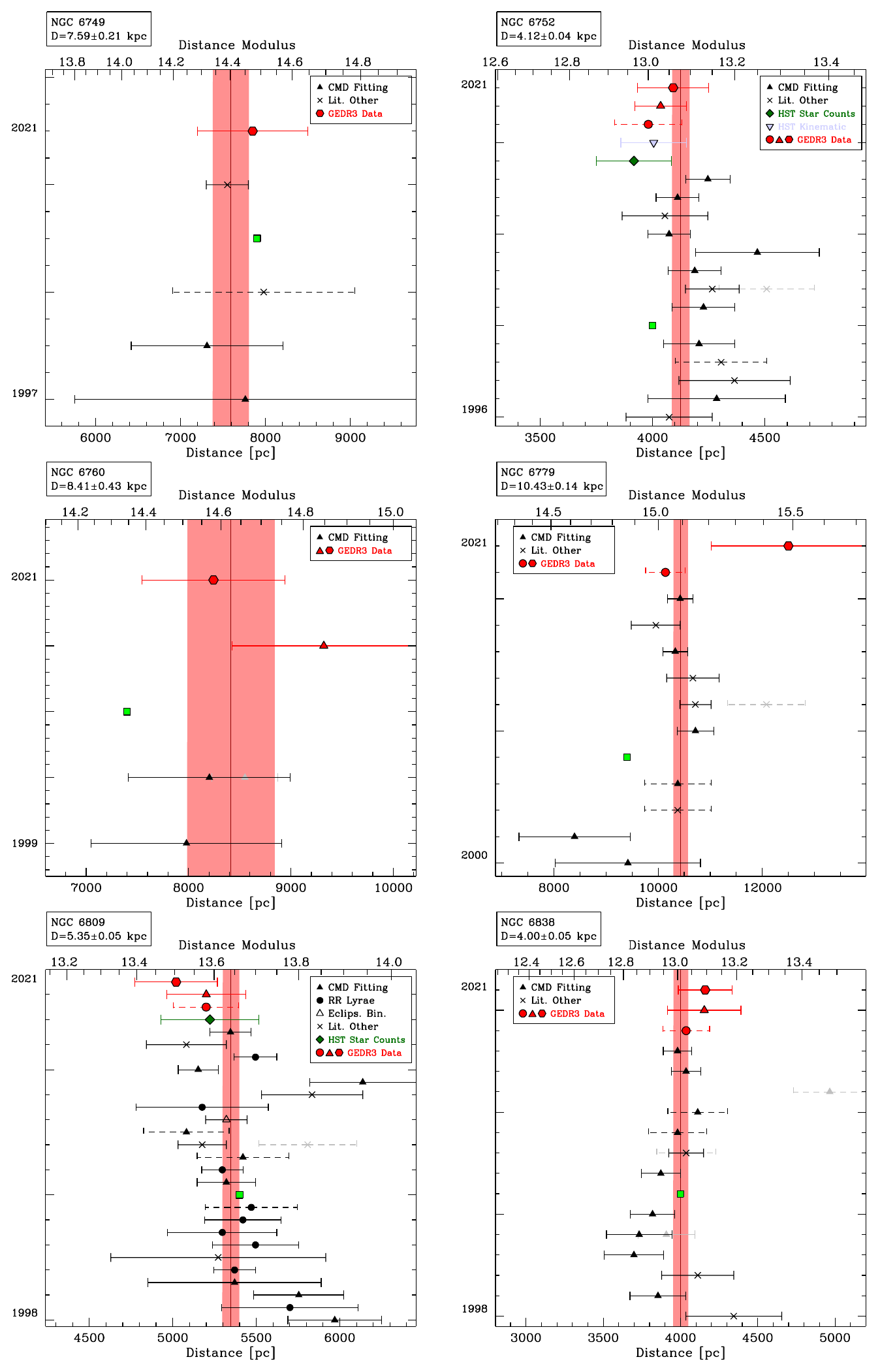}
\end{center}
\vspace*{-0.2cm}
\caption{Same as Fig.~\ref{fig:appfirst} for NGC 6749, NGC 6752, NGC 6760 and NGC 6779, NGC 6809 and NGC 6838.\hspace{7cm}}
\end{figure*}

\begin{figure*}
\begin{center}
\includegraphics[width=0.87\textwidth]{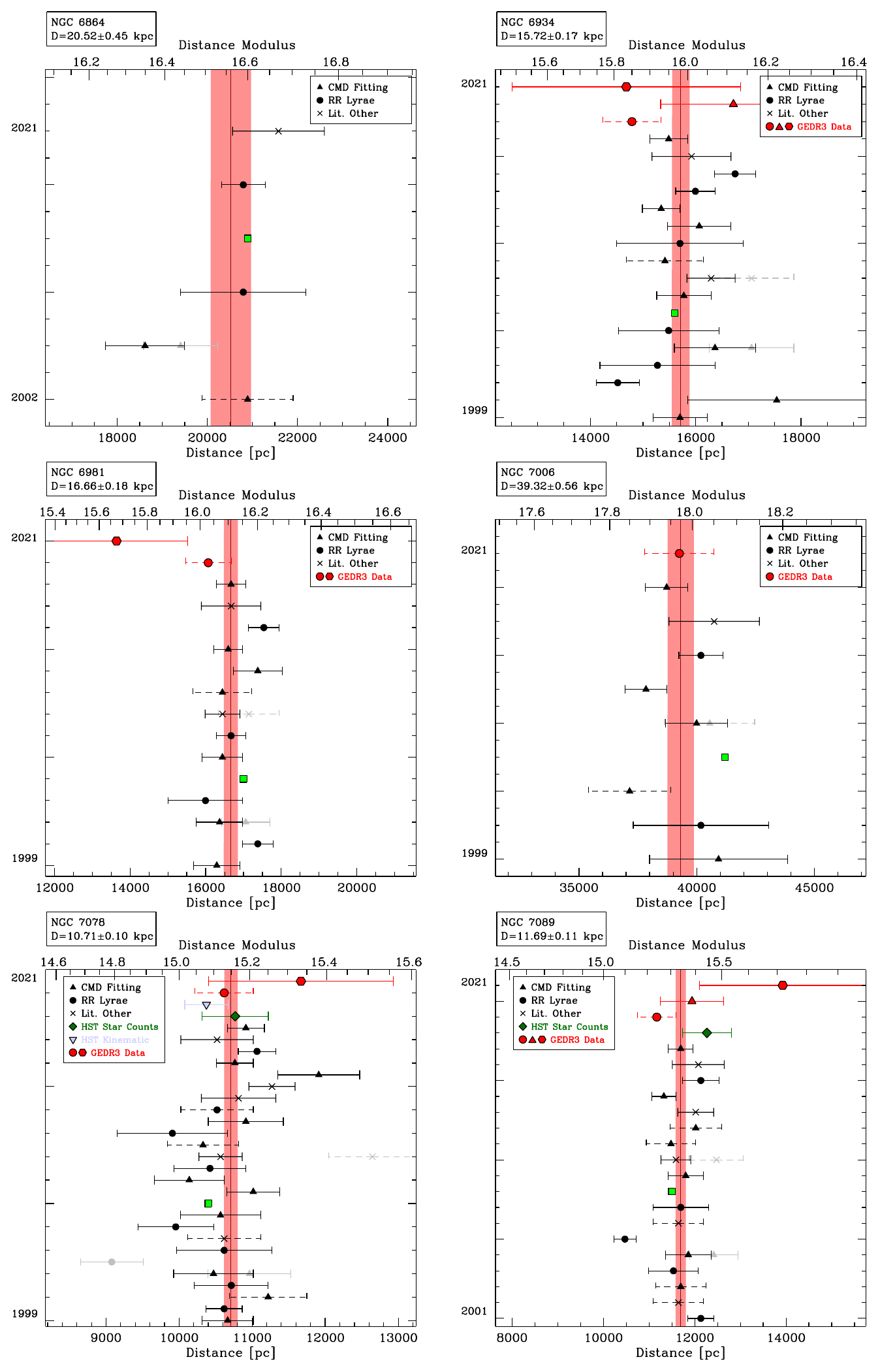}
\end{center}
\vspace*{-0.2cm}
\caption{Same as Fig.~\ref{fig:appfirst} for NGC 6864, NGC 69346 NGC 6981 and NGC 7006, NGC 7078 and NGC 7089.\hspace{7cm}}
\end{figure*}

\begin{figure*}
\begin{center}
\includegraphics[width=0.87\textwidth]{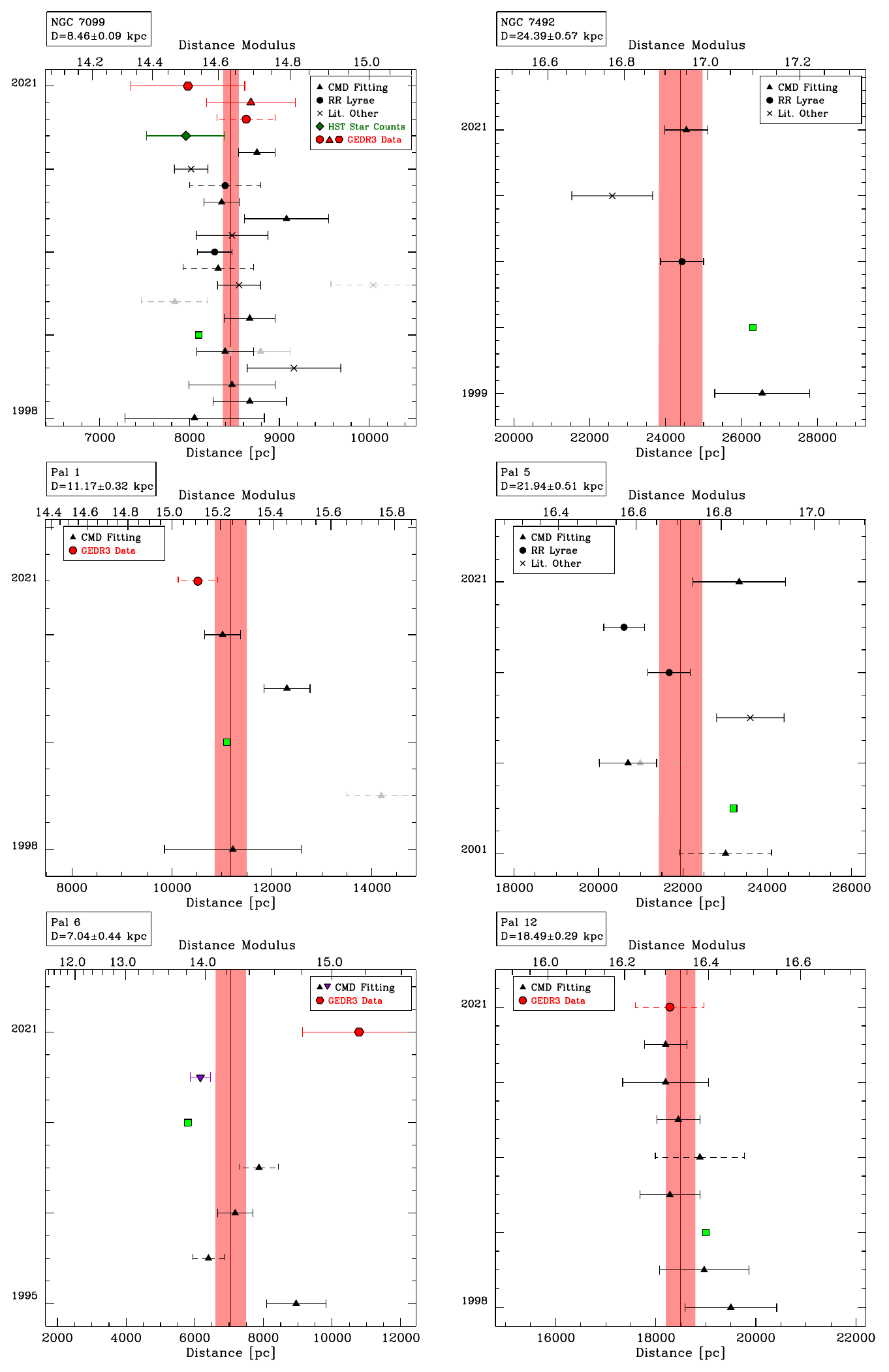}
\end{center}
\vspace*{-0.2cm}
\caption{Same as Fig.~\ref{fig:appfirst} for NGC 7099, NGC 7492, Pal 1, Pal 5, Pal 6 and Pal 12\hspace{8cm}}
\end{figure*}

\begin{figure*}
\begin{center}
\includegraphics[width=0.87\textwidth]{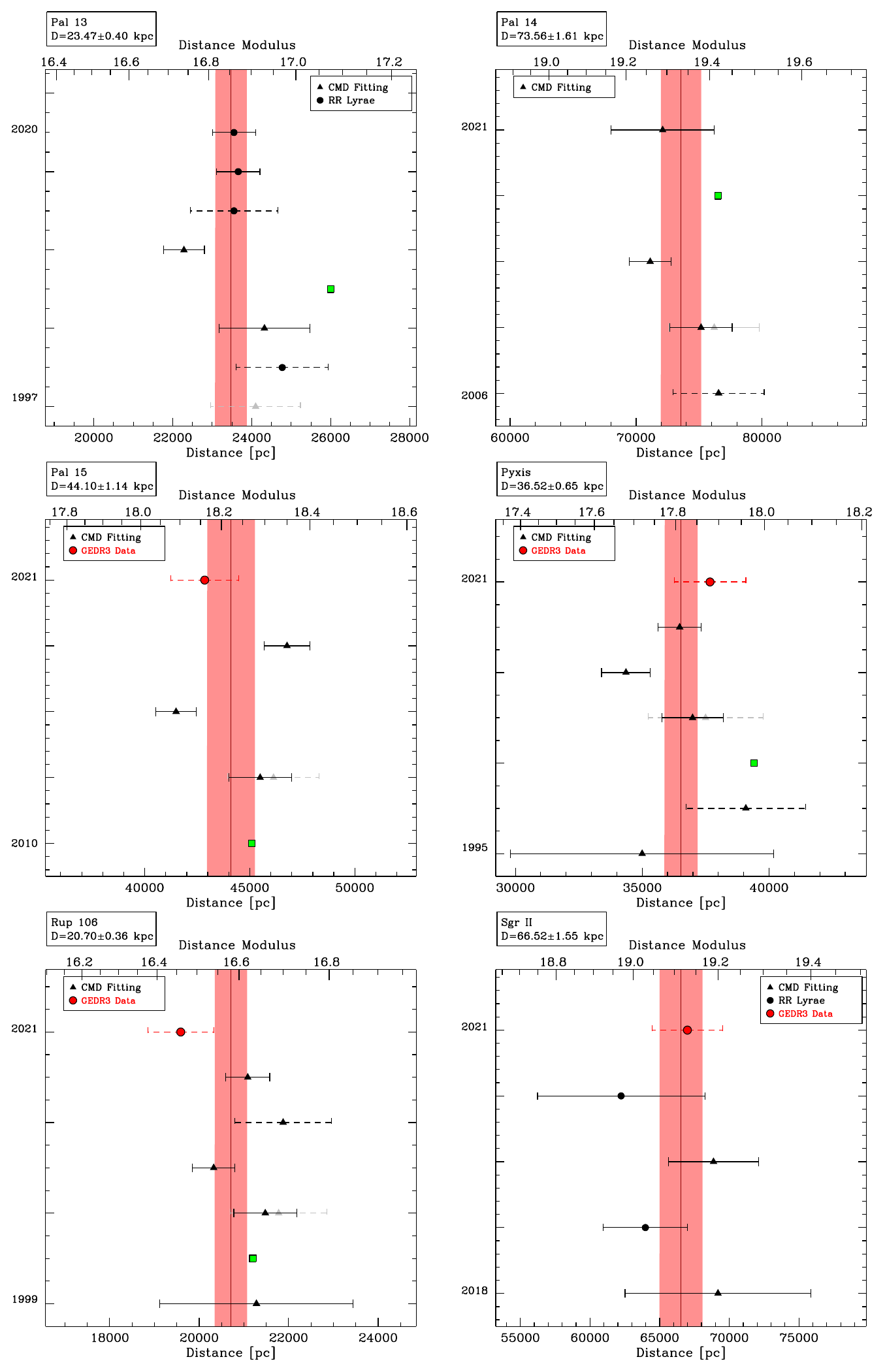}
\end{center}
\vspace*{-0.2cm}
\caption{Same as Fig.~\ref{fig:appfirst} for Pal 13, Pal 14, Pal 15, Pyxis, Rup 106 and Ter 1.\hspace{8cm}}
\end{figure*}

\begin{figure*}
\begin{center}
\includegraphics[width=0.87\textwidth]{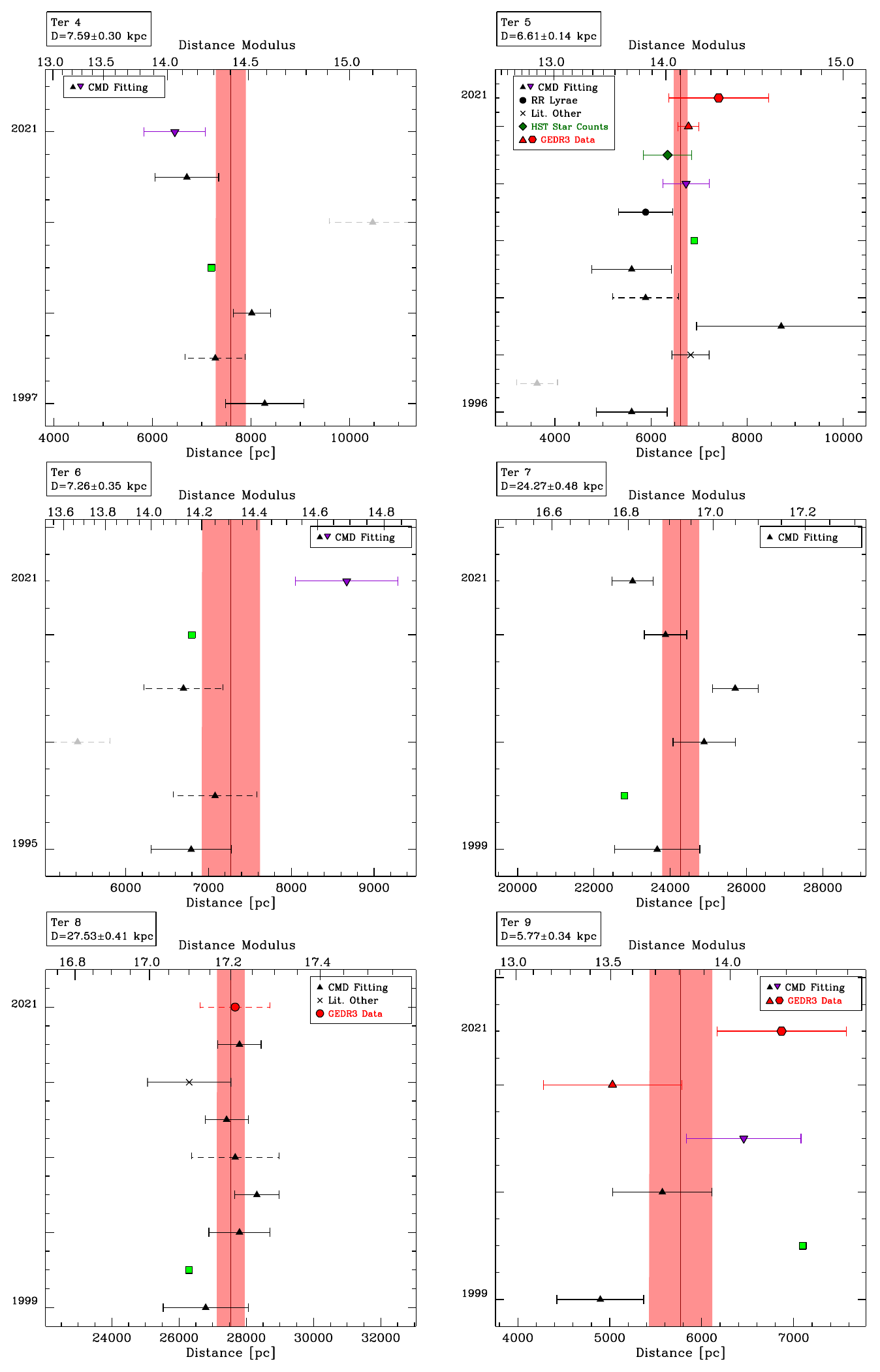}
\end{center}
\vspace*{-0.2cm}
\caption{Same as Fig.~\ref{fig:appfirst} for Ter 4, Ter 5, Ter 6, Ter 7, Ter 8 and Ter 9.\hspace{8cm}}
\label{fig:applast}
\end{figure*}

\section*{Appendix B: Literature distances used in this work}

\clearpage

\begin{table}
\caption{Globular cluster distances from the literature used in this paper. For each distance we give either the de-reddended distance modulus or the distance in kpc depending on what has been given in the original paper.}
\tiny
\begin{tabular}{l@{$\;$}l@{}ll}
Cluster & Source Paper & Method & DM/Distance \\ 
\hline
2MS-GC01 & \citet{ivanovetal2002} & TRGB &  $12.64 \pm 0.40$ \\[+0.10cm]
2MS-GC02 & \citet{borissovaetal2007} & CMD Fitting &  $13.55 \pm 0.18$ \\ 
2MS-GC02 & \citet{alonsogarcia2015} & RR Lyrae &  $14.28 \pm 0.35$ \\[+0.10cm]
AM 1 & \citet{hilker2006} & CMD Fitting &  $20.45$ \\ 
AM 1 & \citet{dotteretal2010} & CMD Fitting &  $20.37$ \\[+0.10cm]
AM 4 & \citet{carraro2009} & CMD Fitting &  $17.58 \pm 0.20$ \\ 
AM 4 & \citet{hamrenetal2013} & CMD Fitting &  $17.28 \pm 0.03$ \\
\hline
\end{tabular}
\end{table}

\begin{table}
\contcaption{}
\tiny
\begin{tabular}{l@{$\;$}l@{}ll}
Cluster & Source Paper & Method & DM/Distance \\ 
\hline
NGC 104 & \citet{grundahl2002} & CMD Fitting &  $13.33 \pm 0.11$ \\ 
NGC 104 & \citet{grattonetal2003} & CMD Fitting &  $13.38 \pm 0.08$ \\ 
NGC 104 & \citet{weldrake2004} & RR Lyrae &  $13.14 \pm 0.25$ \\ 
NGC 104 & \citet{recioblancoetal2005} & CMD Fitting &  $13.28 \pm 0.08$ \\ 
NGC 104 & \citet{salarisetal2007} & CMD Fitting &  $13.18 \pm 0.05$ \\ 
NGC 104 & \citet{bergbusch2009} & CMD Fitting &  $13.25$ \\ 
NGC 104 & \citet{dotteretal2010} & CMD Fitting &  $13.26$ \\
\hline
\end{tabular}
\end{table}

\begin{table}
\contcaption{}
\tiny
\begin{tabular}{l@{$\;$}l@{}ll}
Cluster & Source Paper & Method & DM/Distance \\ 
\hline
NGC 288 & \citet{valcinetal2020} & CMD Fitting &  $14.90 \pm 0.03$ \\[+0.10cm]
NGC 3201 & \citet{covino1997} & CMD Fitting &  $13.42 \pm 0.15$ \\ 
NGC 3201 & \citet{ferraroetal1999} & ZAHB &  $13.45 \pm 0.07$ \\ 
NGC 3201 & \citet{vonbraun2001} & CMD Fitting &  $13.26$ \\ 
NGC 3201 & \citet{piersimonietal2002} & RR Lyrae &  $13.32 \pm 0.06$ \\ 
NGC 3201 & \citet{layden2003} & RR Lyrae &  $13.40 \pm 0.10$ \\ 
NGC 3201 & \citet{mazur2003} & SX Phe &  $13.34 \pm 0.11$ \\
\hline
\end{tabular}
\end{table}

\begin{table}
\contcaption{}
\tiny
\begin{tabular}{l@{$\;$}l@{}ll}
Cluster & Source Paper & Method & DM/Distance \\ 
\hline
NGC 5053 & \citet{sarajedinietal2007} & CMD Fitting &  $16.22$ \\ 
NGC 5053 & \citet{dekany2009} & RR Lyrae &  $16.31 \pm 0.04$ \\ 
NGC 5053 & \citet{arellano2011} & RR Lyrae &  $16.32 \pm 0.05$ \\ 
NGC 5053 & \citet{dotteretal2010} & CMD Fitting &  $16.28$ \\ 
NGC 5053 & \citet{paustetal2010} & CMD Fitting &  $16.16$ \\ 
NGC 5053 & \citet{cohensarajedini2012} & Subdwarfs &  $16.54$ \\ 
NGC 5053 & \citet{vandenbergetal2013} & CMD Fitting &  $16.20$ \\
\hline
\end{tabular}
\end{table}

\begin{table}
\contcaption{}
\tiny
\begin{tabular}{l@{$\;$}l@{}ll}
Cluster & Source Paper & Method & DM/Distance \\ 
\hline
NGC 5694 & \citet{ferraroetal1999} & ZAHB &  $17.86 \pm 0.10$ \\ 
NGC 5694 & \citet{zoccali2000} & CMD Fitting &  $17.74 \pm 0.20$ \\ 
NGC 5694 & \citet{recioblancoetal2005} & CMD Fitting &  $17.68 \pm 0.08$ \\ 
NGC 5694 & \citet{correntietal2011} & CMD Fitting &  $17.75 \pm 0.11$ \\ 
NGC 5694 & \citet{hernitscheketal2019} & RR Lyrae &  $17.65$ \\ 
NGC 5694 & \citet{cernyetal2020} & TRGB &  $17.75 \pm 0.10$ \\[+0.10cm]
NGC 5824 & \citet{ferraroetal1999} & ZAHB &  $17.51 \pm 0.07$ \\
\hline
\end{tabular}
\end{table}

\begin{table}
\contcaption{}
\tiny
\begin{tabular}{l@{$\;$}l@{}ll}
Cluster & Source Paper & Method & DM/Distance \\ 
\hline
NGC 6121 & \citet{hendricksetal2012} & CMD Fitting &  $11.28 \pm 0.06$ \\ 
NGC 6121 & \citet{vandenbergetal2013} & CMD Fitting &  $11.35$ \\ 
NGC 6121 & \citet{kaluznyetal2013} & CMD Fitting &  $11.32 \pm 0.15$ \\ 
NGC 6121 & \citet{kaluznyetal2013} & Eclipsing Binary &  $11.30 \pm 0.05$ \\ 
NGC 6121 & \citet{richeretal2013} & WD Cooling &  $11.34 \pm 0.03$ \\ 
NGC 6121 & \citet{dambisetal2014} & RR Lyrae &  $11.28 \pm 0.16$ \\ 
NGC 6121 & \citet{bragaetal2015} & RR Lyrae &  $11.35 \pm 0.06$ \\
\hline
\end{tabular}
\end{table}

\begin{table}
\contcaption{}
\tiny
\begin{tabular}{l@{$\;$}l@{}ll}
Cluster & Source Paper & Method & DM/Distance \\ 
\hline
NGC 6254 & \citet{valcinetal2020} & CMD Fitting &  $13.78 \pm 0.03$ \\[+0.10cm]
NGC 6256 & \citet{ortolanietal1999} & CMD Fitting &  $14.03 \pm 0.20$ \\ 
NGC 6256 & \citet{recioblancoetal2005} & CMD Fitting &  $13.70 \pm 0.09$ \\ 
NGC 6256 & \citet{matsunagaetal2006} & TII Cepheids &  $14.57$ \\ 
NGC 6256 & \citet{valentietal2007} & CMD Fitting &  $14.80$ \\ 
NGC 6256 & \citet{bhardwajetal2017} & TII Cepheids &  $14.18 \pm 0.08$ \\ 
NGC 6256 & \citet{cadelanoetal2020} & CMD Fitting &  $14.16 \pm 0.10$ \\
\hline
\end{tabular}
\end{table}

\begin{table}
\contcaption{}
\tiny
\begin{tabular}{l@{$\;$}l@{}ll}
Cluster & Source Paper & Method & DM/Distance \\ 
\hline
NGC 6352 & \citet{dotteretal2010} & CMD Fitting &  $13.62$ \\ 
NGC 6352 & \citet{cohensarajedini2012} & Subdwarfs &  $13.69$ \\ 
NGC 6352 & \citet{vandenbergetal2013} & CMD Fitting &  $13.68$ \\ 
NGC 6352 & \citet{wagnerkaiseretal2017} & CMD Fitting &  $13.75 \pm 0.01$ \\ 
NGC 6352 & \citet{libralatoetal2019} & CMD Fitting &  $13.62$ \\ 
NGC 6352 & \citet{valcinetal2020} & CMD Fitting &  $13.76 \pm 0.04$ \\ 
NGC 6352 & \citet{oliveiraetal2020} & CMD Fitting &  $13.65 \pm 0.05$ \\
\hline
\end{tabular}
\end{table}

\begin{table}
\contcaption{}
\tiny
\begin{tabular}{l@{$\;$}l@{}ll}
Cluster & Source Paper & Method & DM/Distance \\ 
\hline
NGC 6440 & \citet{ferraroetal1999} & ZAHB &  $14.47 \pm 0.20$ \\ 
NGC 6440 & \citet{vfo2004} & CMD Fitting &  $14.58$ \\ 
NGC 6440 & \citet{recioblancoetal2005} & CMD Fitting &  $14.62 \pm 0.09$ \\ 
NGC 6440 & \citet{pallancaetal2021} & CMD Fitting &  $14.60$ \\[+0.10cm]
NGC 6441 & \citet{heitsch1999} & CMD Fitting &  $15.34 \pm 0.20$ \\ 
NGC 6441 & \citet{laydenetal1999} & CMD Fitting &  $15.40 \pm 0.17$ \\ 
NGC 6441 & \citet{pritzletal2001} & RR Lyrae &  $15.23 \pm 0.15$ \\
\hline
\end{tabular}
\end{table}

\begin{table}
\contcaption{}
\tiny
\begin{tabular}{l@{$\;$}l@{}ll}
Cluster & Source Paper & Method & DM/Distance \\ 
\hline
NGC 6624 & \citet{vfo2004} & CMD Fitting &  $14.63$ \\ 
NGC 6624 & \citet{recioblancoetal2005} & CMD Fitting &  $14.36 \pm 0.10$ \\ 
NGC 6624 & \citet{dotteretal2010} & CMD Fitting &  $14.57$ \\ 
NGC 6624 & \citet{siegeletal2011} & CMD Fitting &  $14.58$ \\ 
NGC 6624 & \citet{vandenbergetal2013} & CMD Fitting &  $14.51$ \\ 
NGC 6624 & \citet{saracinoetal2016} & CMD Fitting &  $14.49$ \\ 
NGC 6624 & \citet{wagnerkaiseretal2017} & CMD Fitting &  $14.60 \pm 0.02$ \\
\hline
\end{tabular}
\end{table}

\begin{table}
\contcaption{}
\tiny
\begin{tabular}{l@{$\;$}l@{}ll}
Cluster & Source Paper & Method & DM/Distance \\ 
\hline
NGC 6723 & \citet{dambisetal2014} & RR Lyrae &  $14.53 \pm 0.16$ \\ 
NGC 6723 & \citet{leeetal2014} & RR Lyrae &  $14.65 \pm 0.05$ \\ 
NGC 6723 & \citet{omalleyetal2017} & CMD Fitting &  $14.71 \pm 0.20$ \\ 
NGC 6723 & \citet{wagnerkaiseretal2017} & CMD Fitting &  $14.58 \pm 0.01$ \\ 
NGC 6723 & \citet{cernyetal2020} & TRGB &  $14.69 \pm 0.10$ \\ 
NGC 6723 & \citet{valcinetal2020} & CMD Fitting &  $14.55 \pm 0.03$ \\ 
NGC 6723 & \citet{oliveiraetal2020} & CMD Fitting &  $14.59 \pm 0.15$ \\
\hline
\end{tabular}
\end{table}

\begin{table}
\contcaption{}
\tiny
\begin{tabular}{l@{$\;$}l@{}ll}
Cluster & Source Paper & Method & DM/Distance \\ 
\hline
NGC 7006 & \citet{ferraroetal1999} & ZAHB &  $18.06 \pm 0.15$ \\ 
NGC 7006 & \citet{bonoetal2007} & RR Lyrae &  $18.02 \pm 0.15$ \\ 
NGC 7006 & \citet{gerashchenko2007} & ZAHB &  $17.85$ \\ 
NGC 7006 & \citet{dotteretal2011} & CMD Fitting &  $18.04$ \\ 
NGC 7006 & \citet{wagnerkaiseretal2017} & CMD Fitting &  $17.89 \pm 0.04$ \\ 
NGC 7006 & \citet{hernitscheketal2019} & RR Lyrae &  $18.02 \pm 0.01$ \\ 
NGC 7006 & \citet{cernyetal2020} & TRGB &  $18.05 \pm 0.10$ \\
\hline
\end{tabular}
\end{table}

\begin{table}
\contcaption{}
\tiny
\begin{tabular}{l@{$\;$}l@{}ll}
Cluster & Source Paper & Method & DM/Distance \\ 
\hline
Pal 3 & \citet{sharina2018} & CMD Fitting &  $19.82$ \\ 
Pal 3 & \citet{hernitscheketal2019} & RR Lyrae &  $19.65 \pm 0.01$ \\[+0.10cm]
Pal 4 & \citet{dotteretal2010} & CMD Fitting &  $20.03$ \\ 
Pal 4 & \citet{franketal2012} & CMD Fitting &  $20.06 \pm 0.05$ \\[+0.10cm]
Pal 5 & \citet{grillmairsmith2001} & CMD Fitting &  $16.81$ \\ 
Pal 5 & \citet{dotteretal2011} & CMD Fitting &  $16.61$ \\ 
Pal 5 & \citet{kuepperetal2015} & Tail Modeling &  $23.60 \pm  0.80$ kpc \\
\hline
\end{tabular}
\end{table}

\end{document}